\documentclass{jfm}
\usepackage{graphicx}
\usepackage{epstopdf, epsfig}
\usepackage{mathrsfs}
\usepackage{natbib}
\usepackage{bm}
\usepackage{amsmath}
\usepackage{amssymb}
\usepackage{lscape}
\usepackage{afterpage}
\usepackage{float}
\usepackage{multirow}
\usepackage{upgreek}
\usepackage{tikz}
\usetikzlibrary{arrows.meta, decorations.pathreplacing}

\shorttitle{Linear stability of nanofluid boundary-layer flow over a flat plate}
\shortauthor{C. Thomas, S. O. Stephen, J. S. B. Gajjar \& P. T. Griffiths}

\title{Linear stability of nanofluid boundary-layer flow over a flat plate}

\author{Christian Thomas\aff{1}
\corresp{\email{christian.thomas@mq.edu.au}},
Sharon O. Stephen\aff{2},
Jitesh S. B. Gajjar\aff{3}
\and Paul T. Griffiths\aff{4}}

\affiliation{\aff{1}School of Mathematical and Physical Sciences, Macquarie University, NSW 2109, Australia
\aff{2} School of Mathematics and Statistics, University of Sydney, Sydney, NSW 2006, Australia
\aff{3} Department of Mathematics, University of Manchester, Oxford Road, Manchester, M13 9PL, UK
\aff{4} Aston Fluids Group and School of Engineering \& Innovation, Aston University, Birmingham, B4 7ET, UK}

\begin{document}

\maketitle

\begin{abstract}
The linear stability of nanofluid boundary-layer flow over a flat plate is investigated using a two-phase model that incorporates Brownian motion and thermophoresis, building upon the earlier work of \citet{Buongiorno2006}. Solutions to the steady boundary-layer equations reveal a thin nanoparticle concentration layer near the plate surface, with a characteristic thickness of $O({\Rey}^{-1/2}{Sc}^{-1/3})$, for a Reynolds number $\Rey$ and Schmidt number $Sc$. When Brownian motion and thermophoresis are neglected, this nanoparticle concentration layer disappears, resulting in a uniform concentration across the boundary layer. Neutral stability curves and critical conditions for the onset of the Tollmien--Schlichting wave are computed for a range of nanoparticle materials and volume concentrations. Results indicate that while the effects of Brownian motion and thermophoresis are negligible, the impact of nanoparticle density is significant. Denser nanoparticles, such as silver (Ag) and copper (Cu), destabilise the Tollmien--Schlichting wave, whereas lighter nanoparticles, like aluminium (Al) and silicon (Si), establish a small stabilising effect. Additionally, stability characteristics are influenced by the viscosity model. Finally, a high-Reynolds number asymptotic analysis is undertaken for the lower branch of the neutral stability curve.
\end{abstract}


\section{Introduction}

This paper is concerned with the influence of nanofluids on the linear stability of disturbances in the boundary-layer flow over a flat plate. Nanofluids are fluids containing nanoscale particles ranging from 1 to 100 nanometres, dispersed in a base fluid like water. These nanoparticles, composed of metal-based or carbon-based materials, enhance the thermal properties of the base fluid.

Since the seminal work of~\citet{Choi1995}, nanofluids have received considerable interest, with a rapid growth in annual publications~\citep{Taylor2013}. Numerous studies have investigated the thermal benefits of nanofluids, including comprehensive reviews by~\citet{Das2006}, \citet{Wang2008a, Wang2008b}, \citet{Kakac2009}, \citet{Mahbubul2012}, and \citet{Mishra2014}. These thermal improvements have led to a wide range of heat transfer applications, including cooling systems for automotive engines~\citep{Sidik2015}, electronics~\citep{Bahiraei2018}, nuclear systems~\citep{Buongiorno2009}, solar thermal systems~\citep{Khullar2012}, biomedical processes~\citep{Sheikhpour2020}, and industrial applications~\citep{Wong2010}. 

Despite the ongoing interest in nanofluids for their thermal benefits, relatively few investigations have examined the impact of nanofluids on the hydrodynamic stability of flows. This study aims to address this knowledge gap by investigating the capabilities of nanofluids in controlling laminar-turbulent transition processes.


\subsection{Modelling nanofluid flows}

A key aspect of modelling nanofluid flows is how suspended nanoparticles modify the fluid's effective viscosity. For dilute suspensions of rigid, spherical particles,~\citet{Einstein1906} showed that the dynamic viscosity increases linearly with the nanoparticle volume concentration $\phi$. He defined the effective dynamic viscosity as
\begin{equation}
\mu^*=\mu_{bf}^*(1+2.5\phi),
\end{equation}
where $\mu_{bf}^*$ is the dynamic viscosity of the base fluid. Since Einstein’s work, many viscosity models have been proposed to account for additional factors, including particle shape, size distribution, and particle-particle interactions. \citet{Batchelor1977} extended Einstein's formula to include the effects of Brownian motion (that is, the random movement of nanoparticles in a base fluid), while~\citet{Brinkman1952} proposed a semi-empirical correlation valid for nanoparticle volume concentrations up to approximately $4\%$. (The formulas for the Batchelor and Brinkman models are given in the subsequent section.) Comprehensive reviews of nanofluid viscosity models, including experimental and theoretical developments, are provided by~\citet{Wang2008a} and~\citet{Mishra2014}.

Another key aspect of nanofluid modelling is the treatment of the fluid either as a single-phase or a two-phase flow. Single-phase models treat the nanofluid as a homogeneous mixture with effective properties, while two-phase models account for interactions between the base fluid and nanoparticles. The latter approach can capture additional effects such as particle migration, Brownian motion, and thermophoresis (that is, the movement of nanoparticles in a base fluid due to a temperature gradient). Moreover, two-phase flow models include a continuity equation for the nanoparticle volume concentration. 

The steady boundary-layer flow over a flat plate has been investigated by~\citet{Buongiorno2006},~\citet{Avramenko2011}, and~\citet{MacDevette2014}. These studies employed the~\citet{Brinkman1952} model to describe the nanofluid viscosity and incorporated Brownian motion and thermophoresis into the governing equations. To simplify the analysis,~\citet{Buongiorno2006} assumed the flow to be incompressible, even though modelling the nanofluid as a two-component mixture implies a non-constant density. Despite this apparent inconsistency,~\citet{Buongiorno2006} showed that the effects of Brownian motion and thermophoresis are negligible in nanofluids and attributed the observed heat transfer benefits to the improved thermophysical properties of the nanoparticles. 

While acknowledging that Brownian motion and thermophoresis effects are weak,~\citet{Avramenko2011} derived boundary-layer equations similar to those of~\citet{Blasius1908}. However, despite accounting for compressibility effects in the base flow, the study implemented several simplifying assumptions. Notably, the incompressible flow condition was applied to the nanoparticle continuity equation (see equations (1)–(4) of~\citet{Avramenko2011}). Additionally, the coefficients for Brownian motion and thermophoresis, defined below in equation~\eqref{BrownianThermoCoef}, were treated as constants, even though they depend on temperature and nanoparticle volume concentration, respectively. Yet despite these simplifications,~\citet{Avramenko2011} demonstrated that a thin concentration layer forms near the plate surface. This concentration layer modifies the velocity and temperature fields in the near-wall region, which may, in turn, influence instabilities within the boundary layer. 

Both~\citet{Buongiorno2006} and~\citet{Avramenko2011} confirmed that heat transfer, measured by the Nusselt number $Nu$, is enhanced as the nanoparticle volume concentration $\phi$ increases. In contrast,~\citet{MacDevette2014}, who also confirmed that Brownian motion and thermophoresis are negligible, observed a reduced heat transfer coefficient as $\phi$ increases. They attributed the discrepancy with earlier studies to differences in the definition of the heat transfer coefficient. 

The study of nanofluids in boundary-layer flows has been extended to include flows past vertical plates~\citep{Kuznetsov2010}, planar wall jets~\citep{Turkyilmazoglu2016}, and the flow due to a rotating-disk~\citep{Bachok2011, Turkyilmazoglu2014, Mehmood2018}, with these studies reporting enhanced heat transfer due to the introduction of nanoparticles. 

Using triple-deck theory,~\citet{Alruwaele2023} modelled a nanofluid boundary-layer flow past a hump, on an otherwise flat plate. The study demonstrated that a nanofluid can suppress the region of flow separation along the rear side of the bump. More recently,~\citet{Gandhi2025} examined thermosolutal instabilities in a nanofluid layer with a deformable surface, showing how the Soret effect and thermal properties influence instability characteristics.


\subsection{Linear stability studies}

The linear stability of the incompressible Blasius boundary layer has been extensively studied, beginning with the seminal investigations of~\citet{Tollmien1933} and~\citet{Schlichting1933}, which led to the Orr--Sommerfeld equation. These studies employed the parallel flow approximation, where the flow is assumed to be unidirectional and depends only on the wall-normal direction. The theoretical predictions for the Tollmien–Schlichting (TS) wave were subsequently confirmed experimentally by~\citet{Schubauer1947}. Further theoretical and experimental insights into the stability of TS waves were reported by~\citet{Jordinson1970},~\citet{Barry1970},~\citet{Ross1970}, and~\citet{Gaster1974} amongst many others.

Using triple-deck theory,~\citet{Smith1979} undertook an asymptotic, high-Reynolds number $\Rey$ analysis to describe the structure of the lower branch of the neutral stability curve in the Blasius boundary layer. The triple-deck framework consists of three layers: an upper deck, representing the inviscid outer flow and spans a thickness of $O({\Rey}^{-3/8})$; a main deck, corresponding to the boundary layer, with thickness $O({\Rey}^{-4/8})$; and a lower deck, a thin viscous sublayer of thickness $O({\Rey}^{-5/8})$, where viscous-inviscid interactions are dominant. (A formal definition for the Reynolds number $\Rey$ is given below in equation~(\ref{Parameters}\emph{a}).) A subsequent study by~\citet{BodonyiSmith1981} employed a multi-deck approach to derive the corresponding structure of the upper branch of the neutral stability curve. Later,~\citet{Smith1989} extended the asymptotic analysis of the lower branch to compressible boundary-layer flows. 

Building on earlier studies,~\citet{Bertolotti1992} employed parabolised stability equations to investigate both the linear and nonlinear development of TS waves in the Blasius boundary layer. \citet{Healey1995} compared the asymptotic scalings of the lower and upper branches with solutions from the Orr--Sommerfeld equation and experimental observations. More recently, both asymptotic and numerical approaches have been utilised to model the effects of non-Newtonian viscosity~\citep{Griffiths2016} and temperature-dependent viscosity~\citep{Miller2018} on the stability of the Blasius boundary layer.

To the authors' knowledge, there are only two previous studies concerning the linear stability of nanofluid boundary-layer flows. The first, by \citet{Turkyilmazoglu2020}, considered the application of nanofluids to several configurations, including the Kelvin–Helmholtz instability, Rayleigh–B\'{e}nard convection, instabilities in rotating disk flows, and instabilities in the boundary-layer flow over a flat plate. Turkyilmazoglu modelled the latter flow as a single-phase flow, with quantities scaled on nanofluid properties, i.e., the combined characteristics of the base fluid and nanoparticles. This approach led to a Reynolds number based on nanofluid characteristics and a base flow described by the Blasius equation. The findings suggest that the Reynolds number of the nanofluid can be predicted using the Reynolds number of the base fluid. Moreover, results indicate that denser nanoparticle materials, like silver (Ag), stabilise the flow, while less dense nanoparticle materials, such as alumina (Al$_2$O$_3$), destabilise the flow at sufficiently larger volume concentrations $\phi$. However, the rationale for scaling quantities on nanofluid characteristics is unclear, as the resulting Reynolds number changes as the nanoparticle volume concentration $\phi$ increases, making it difficult to compare solutions. In the following study, the nanofluid flow is modelled as a two-phase flow that includes diffusion effects due to Brownian motion and thermophoresis, with the Reynolds number based on the base fluid properties to facilitate comparisons across different nanoparticle materials and variable $\phi$.

A second study, by \citet{Laouer2024}, examined the linear stability of a nanofluid flow past stationary and moving wedges. Similar to \citet{Turkyilmazoglu2020}, \citet{Laouer2024} employed a single-phase flow approach, with the base flow based on the methodology of \citet{Yacob2011} and a linear stability analysis that simplifies to the standard Orr--Sommerfeld equation for a regular fluid. \citet{Laouer2024} showed that, for a nanofluid flow over a stationary wedge due to a favourable pressure gradient, increasing the volume concentration $\phi$ leads to a destabilising effect. Additionally, Laouer and co-workers suggest that heavier nanoparticle materials, such as copper (Cu), have a stabilising effect, while lighter materials, like titanium oxide (TiO$_2$) and alumina (Al$_2$O$_3$), destabilise the flow. However, this latter finding appears to contradict the results presented in figure 8 of their paper, which shows that copper (Cu) nanoparticles shift neutral stability curves to the left and smaller Reynolds numbers, while titanium oxide (TiO$_2$) and alumina (Al$_2$O$_3$) nanoparticles shift neutral stability curves to the right and higher Reynolds numbers.


\subsection{Outline of paper}

The following study investigates the linear stability of nanofluid flow over a flat plate using a two-phase flow model that accounts for Brownian motion and thermophoresis. This model addresses the inconsistencies in previous single-phase studies and provides a physically consistent method for analysing stability trends. Both numerical and asymptotic analyses are undertaken to compute neutral stability curves and examine the lower branch behaviour at high Reynolds numbers. The most amplified TS disturbances appear near the lower branch of the neutral curve, and this, combined with the need to validate our numerical solutions, motivates the analysis of the lower rather than the upper branch.

The remainder of this paper is outlined as follows. The governing equations are introduced in the next section, followed by the steady, two-dimensional boundary-layer equations and its solutions in \S3. Linear stability results for three-dimensional disturbances, including neutral stability curves and critical conditions, are presented in \S4. An asymptotic analysis of the lower branch is provided in \S5. Conclusions are given in \S6.


\section{Governing equations}


\subsection{Model}

Consider the flow of a nanofluid over a semi-infinite flat plate with free-stream velocity $U_{\infty}^*$. (Here, an asterisk denotes dimensional quantities.) The model is given in Cartesian coordinates $\boldsymbol{x}^*=(x^*,y^*,z^*)$, where $x^*$ measures the distance along the surface of the flat plate, $y^*$ denotes the direction normal to the plate, and $z^*$ the spanwise direction. Consequently, the governing system of equations comprise the continuity, momentum, and energy equations for fluid motion \citep{RubanGajjar2014}, along with a continuity equation for the nanoparticles \citep{Buongiorno2006, Avramenko2011, MacDevette2014}: 
\begin{subequations}\label{Gov1}
\begin{equation}\label{Continuity1}
\frac{\partial\rho^*}{\partial t^*} + \nabla^*\cdot(\rho^*\boldsymbol{u}^*) = 0,  
\end{equation}
\begin{equation}\label{Momentum1}
\rho^*\left(\frac{\partial\boldsymbol{u}^*}{\partial t^*} + (\boldsymbol{u}^*\cdot\nabla^*)\boldsymbol{u}^*\right) = -\nabla^*p^* + \nabla^*\cdot\left(\mu^*\left(\nabla^*\boldsymbol{u}^* + \left(\nabla^*\boldsymbol{u}^*\right)^T - \frac{2}{3}\nabla^*\cdot\boldsymbol{u}^*\mathsfbi{I}\right)\right),
\end{equation}
\begin{multline}\label{Energy1}
\rho^*\left(\frac{\partial \left(c^*T^*\right)}{\partial t^*} + (\boldsymbol{u}^*\cdot\nabla^*)\left(c^*T^*\right)\right) = 
\nabla^*\cdot\left(k^*\nabla^* T^*\right)  \\ + (\rho^*c^*)_{np}\left(D_B^*\nabla^*\phi+D_T^*\frac{\nabla^*T^*}{T^*}\right)\cdot\nabla^*T^*,
\end{multline}
\begin{equation}\label{Volume1}
\frac{\partial\phi}{\partial t^*}+\nabla^*\cdot(\phi\boldsymbol{u}^*) = \nabla^*\cdot\left(D_B^*\nabla^*\phi + D_T^*\frac{\nabla^*T^*}{T^*}\right),
\end{equation}
\end{subequations}
for a velocity $\boldsymbol{u}^*=(u^*,v^*,w^*)$, pressure $p^*$, temperature $T^*$, and dimensionless nanoparticle volume concentration $\phi$. Here, $\mathsfbi{I}$ is the identity matrix. 

The density of the nanofluid $\rho^*$ is defined using the law of mixtures as
\begin{equation}\label{DimDensity}
\rho^* = (1-\phi)\rho_{bf}^* + \phi\rho_{np}^*,
\end{equation}
where subscripts $bf$ and $np$ represent quantities associated with the base fluid and nanoparticles, respectively. In addition,
\begin{equation}
\rho^*c^* = (1-\phi)(\rho^*c^*)_{bf} + \phi(\rho^*c^*)_{np},
\end{equation}
where $c^*$ denotes the specific heat capacity of the nanofluid, while the thermal conductivity of the nanofluid $k^*$ is given by the \citet{Maxwell1881} model
\begin{equation}
k^* = \left(\frac{k_{np}^*+2k_{bf}^*+2\phi(k_{np}^*-k_{bf}^*)}{k_{np}^*+2k_{bf}^*-\phi(k_{np}^*-k_{bf}^*)}\right)k_{bf}^*.
\end{equation}
Alternative models for $k^*$ may be considered as discussed in \citet{Wang2008a}.

The dynamic viscosity of the nanofluid $\mu^*$, used throughout the subsequent study, is given by the \citet{Brinkman1952} model
\begin{equation}\label{Brinkman}
\mu^* = \frac{\mu_{bf}^*}{(1-\phi)^{2.5}},
\end{equation}
for a base fluid dynamic viscosity $\mu_{bf}^*$. The Brinkman relation is known to under predict the dynamic viscosity for $\phi>0.01$~\citep{MacDevette2014}. However, for theoretical purposes and to demonstrate trends, here we consider nanoparticle volume concentrations $\phi$ up to 10\% of the fluid volume. Similar to the thermal conductivity $k^*$, alternative models may be considered for the dynamic viscosity $\mu^*$, as listed in \citet{Wang2008a} and \citet{Mishra2014}, which encompass properties such as the size and distribution of nanoparticles. For instance, \citet{Batchelor1977} modelled the dynamic viscosity as
\begin{subequations}\label{Otherviscousmodels}
\begin{equation}\label{Batchelor}
\mu^* = \mu_{bf}^*(1+2.5\phi+6.2\phi^2), \tag{\theequation \emph{a}}
\end{equation}
whereas \citet{PakCho1998} and \citet{Maiga2004} obtained the correlations
\begin{equation}\label{PakCho}
\mu^* = \mu_{bf}^*(1+39.11\phi+533.9\phi^2) \quad \textrm{and} \quad 
\mu^* = \mu_{bf}^*(1+7.3\phi+123\phi^2), \tag{\theequation \emph{b,c}} 
\end{equation}
\end{subequations}
for nanofluids inside circular pipes and tubes, respectively. 

The latter two terms of \eqref{Energy1} and the two terms on the right-hand side of \eqref{Volume1} model the respective effects of Brownian motion and thermophoresis, with coefficients
\begin{subequations}\label{BrownianThermoCoef}
\begin{equation}
D_B^* = \frac{k_B^*T^*}{3\pi\mu_{bf}^*d_{np}^*}\equiv C_B^*T^* 
\quad \textrm{and} \quad 
D_T^* = \frac{\beta_T\mu_{bf}^*\phi}{\rho_{bf}^*}\equiv C_T^*\phi. 
\tag{\theequation \emph{a,b}}
\end{equation}
\end{subequations}
Here, $k_B^*$ denotes the Boltzmann constant, $d_{np}^*$ the diameter of the nanoparticles, and the proportionality constant
\begin{equation*}
\beta_T = 0.26\frac{k_{bf}^*}{2k_{bf}^*+k_{np}^*},  
\end{equation*}
as given in \citet{McNabMeisen1973}, \citet{Buongiorno2006}, and \citet{MacDevette2014}.

\begin{figure}
\centering
\begin{tikzpicture}[scale=1]
\tikzstyle{every node}=[font=\normalsize]
\draw [thick] (1,0) -- (10,0);
\node [below left] at (3,0) {$\textrm{Plate}$};
\node [below left] at (8,0) {$u^{*}=v^{*}=0,\quad T^{*}=T_{w}^{*}$};
\node [above left] at (8,3.5) {$u^{*}=U_{\infty}^{*},\quad T^{*}=T_{\infty}^{*},\quad \phi=\phi_{\infty}$};
\draw [] (0,1) -- (0,3.5) -- (5.5,3.5);
\draw [] (7,3.5) -- (10,3.5) -- (10,1.5);
\draw [>=latex,<->] (0,1) -- (0,0) -- (1,0);
\node [below] at (1,0) {$x^{*}$};
\node [left] at (0,1) {$y^{*}$};
%
\draw [blue] (-2.5,0)				-- (-2.5,3.5);
\draw [blue, >=latex,->] (-2.5,3.5)	-- (-1,3.5);
\draw [blue, >=latex,->] (-2.5,3) 	-- (-1,3);
\draw [blue, >=latex,->] (-2.5,2.5) 	-- (-1,2.5);
\draw [blue, >=latex,->] (-2.5,2) 	-- (-1,2);
\draw [blue, >=latex,->] (-2.5,1.5) 	-- (-1,1.5);
\draw [blue, >=latex,->] (-2.5,1) 	-- (-1,1);
\draw [blue, >=latex,->] (-2.5,0.5) 	-- (-1,0.5);
\draw [blue, >=latex,->] (-2.5,0) 	-- (-1,0);
\draw [blue] (-1,0)				-- (-1,3.5);
\node [right] at (-1,3.5) {$U_{\infty}^{*}$};
\draw [blue] (5.5,0) 	-- (5.5,3.5);
\draw [blue] (7,1.5) 	-- (7,3.5);
\draw [blue] (5.5,0) arc (-90:0:1.5);
\draw [blue, >=latex,->] (5.5,3.5) 	-- (7,3.5);
\draw [blue, >=latex,->] (5.5,3) 		-- (7,3);
\draw [blue, >=latex,->] (5.5,2.5) 	-- (7,2.5);
\draw [blue, >=latex,->] (5.5,2) 		-- (7,2);
\draw [blue, >=latex,->] (5.5,1.5) 	-- (7,1.5);
\draw [blue, >=latex,->] (5.5,1) 		-- ({5.5+sqrt(2)},1);
\draw [blue, >=latex,->] (5.5,0.5) 	-- ({5.5+sqrt(1.25)},0.5);
\draw [purple, >=latex,<->] (10,0)	-- (10,1.5);
\node [right] at (10,0.75) {$\delta^{*}$};
\clip (-2,0) rectangle (10,4);
\draw (10,0) -- (10,1.5);
\draw [red,dashed,fill opacity=0.125,fill=gray] (0,0) arc (104.3:0:50);
\node[fill=magenta, circle, inner sep=1pt] at (9.686, 2.74) {};
\node[fill=magenta, circle, inner sep=1pt] at (4.413, 3.435) {};
\node[fill=magenta, circle, inner sep=1pt] at (6.122, 0.319) {};
\node[fill=magenta, circle, inner sep=1pt] at (6.003, 3.25) {};
\node[fill=magenta, circle, inner sep=1pt] at (9.659, 0.56) {};
\node[fill=magenta, circle, inner sep=1pt] at (2.347, 2.059) {};
\node[fill=magenta, circle, inner sep=1pt] at (9.162, 2.987) {};
\node[fill=magenta, circle, inner sep=1pt] at (6.348, 1.19) {};
\node[fill=magenta, circle, inner sep=1pt] at (7.518, 1.649) {};
\node[fill=magenta, circle, inner sep=1pt] at (1.361, 1.75) {};
\node[fill=magenta, circle, inner sep=1pt] at (7.766, 0.362) {};
\node[fill=magenta, circle, inner sep=1pt] at (0.185, 3.338) {};
\node[fill=magenta, circle, inner sep=1pt] at (6.108, 3.228) {};
\node[fill=magenta, circle, inner sep=1pt] at (9.048, 3.156) {};
\node[fill=magenta, circle, inner sep=1pt] at (3.068, 2.838) {};
\node[fill=magenta, circle, inner sep=1pt] at (1.629, 1.082) {};
\node[fill=magenta, circle, inner sep=1pt] at (1.755, 0.458) {};
\node[fill=magenta, circle, inner sep=1pt] at (6.839, 1.25) {};
\node[fill=magenta, circle, inner sep=1pt] at (5.542, 1.355) {};
\node[fill=magenta, circle, inner sep=1pt] at (0.957, 0.121) {};
\node[fill=magenta, circle, inner sep=1pt] at (8.122, 1.27) {};
\node[fill=magenta, circle, inner sep=1pt] at (7.276, 3.278) {};
\node[fill=magenta, circle, inner sep=1pt] at (4.884, 1.504) {};
\node[fill=magenta, circle, inner sep=1pt] at (3.878, 3.081) {};
\node[fill=magenta, circle, inner sep=1pt] at (3.294, 2.378) {};
\node[fill=magenta, circle, inner sep=1pt] at (3.886, 2.111) {};
\node[fill=magenta, circle, inner sep=1pt] at (5.278, 1.821) {};
\node[fill=magenta, circle, inner sep=1pt] at (4.192, 1.949) {};
\node[fill=magenta, circle, inner sep=1pt] at (2.458, 1.641) {};
\node[fill=magenta, circle, inner sep=1pt] at (1.462, 2.749) {};
\node[fill=magenta, circle, inner sep=1pt] at (8.625, 0.891) {};
\node[fill=magenta, circle, inner sep=1pt] at (9.967, 0.453) {};
\node[fill=magenta, circle, inner sep=1pt] at (2.905, 0.719) {};
\node[fill=magenta, circle, inner sep=1pt] at (3.316, 0.129) {};
\node[fill=magenta, circle, inner sep=1pt] at (7.433, 0.671) {};
\node[fill=magenta, circle, inner sep=1pt] at (4.556, 2.578) {};
\node[fill=magenta, circle, inner sep=1pt] at (6.568, 2.456) {};
\node[fill=magenta, circle, inner sep=1pt] at (7.927, 2.879) {};
\node[fill=magenta, circle, inner sep=1pt] at (5.934, 0.989) {};
\node[fill=magenta, circle, inner sep=1pt] at (0.389, 0.33) {};
\node[fill=magenta, circle, inner sep=1pt] at (5.048, 2.255) {};
\node[fill=magenta, circle, inner sep=1pt] at (7.095, 2.692) {};
\node[fill=magenta, circle, inner sep=1pt] at (4.714, 0.893) {};
\node[fill=magenta, circle, inner sep=1pt] at (6.195, 1.73) {};
\node[fill=magenta, circle, inner sep=1pt] at (3.727, 1.63) {};
\node[fill=magenta, circle, inner sep=1pt] at (2.204, 3.14) {};
\node[fill=magenta, circle, inner sep=1pt] at (0.807, 2.09) {};
\node[fill=magenta, circle, inner sep=1pt] at (3.526, 1.016) {};
\node[fill=magenta, circle, inner sep=1pt] at (9.173, 1.761) {};
\node[fill=magenta, circle, inner sep=1pt] at (0.771, 1.345) {};
\node[fill=magenta, circle, inner sep=1pt] at (8.185, 3.19) {};
\node[fill=magenta, circle, inner sep=1pt] at (8.764, 1.072) {};
\node[fill=magenta, circle, inner sep=1pt] at (6.853, 3.4) {};
\node[fill=magenta, circle, inner sep=1pt] at (7.345, 1.066) {};
\node[fill=magenta, circle, inner sep=1pt] at (1.142, 2.354) {};
\node[fill=magenta, circle, inner sep=1pt] at (2.771, 2.946) {};
\node[fill=magenta, circle, inner sep=1pt] at (5.756, 2.103) {};
\node[fill=magenta, circle, inner sep=1pt] at (4.145, 0.783) {};
\node[fill=magenta, circle, inner sep=1pt] at (8.511, 2.318) {};
\node[fill=magenta, circle, inner sep=1pt] at (1.011, 3.202) {};
\draw[>=latex,->] (1.95,3.3) -- (1.08,3.2);
\draw[>=latex,->] (3.3,3.15) -- (3.3,2.45);
\draw[>=latex,->] (3.5,3.15) -- (3.85,2.2);
\node at (3.1,3.3) {Nanoparticles};
\end{tikzpicture}
\caption{Diagram of a nanofluid flow, composed of a base fluid ($bf$) and nanoparticles ($np$) over a flat plate. Here, $\delta^*$ represents the boundary-layer thickness.}\label{Fig1}
\end{figure}
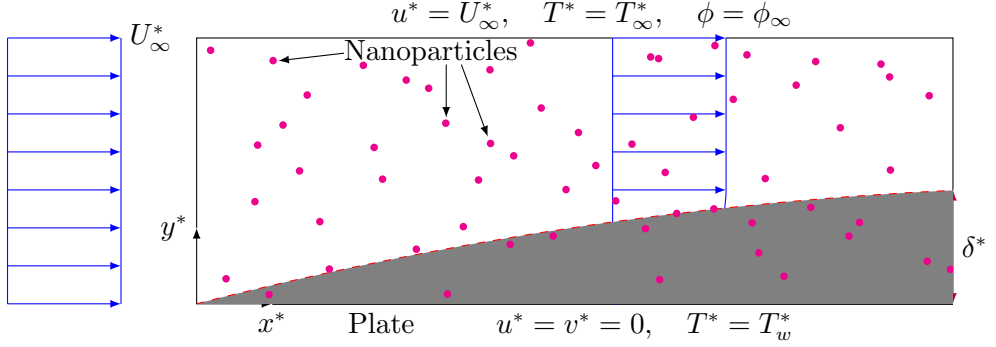

The nanofluid flow is subject to the no-slip condition and the fixed temperature condition on the plate surface
\begin{subequations}\label{BoundaryConditions1}
\begin{equation}
\boldsymbol{u}^*=0 \quad \textrm{and} \quad T^*=T_w^*
\quad \textrm{on} \quad y^*=0, \tag{\theequation \emph{a,b}}
\end{equation}
where $T_w^*$ denotes the constant wall temperature. (Here, a subscript $w$ references wall conditions.) In addition, 
\begin{equation}
D_B^*\frac{\partial\phi}{\partial y^*}+\frac{D_T^*}{T^*}{\frac{\partial T^*}{\partial y^*}}=0 
\quad \textrm{on} \quad y^*=0, 
\tag{\theequation \emph{c}}
\end{equation}
following \citet{Avramenko2011}, which imposes that the total flux of nanoparticles at the plate surface is zero.
\end{subequations}
Finally, in the far-field, the flow is subject to the free-stream conditions
\begin{subequations}\label{BoundaryConditions2}
\begin{equation}
\begin{alignedat}{3}
u^*&{}\rightarrow{} U_{\infty}^*, \qquad
&v^*{}\rightarrow{}& 0, \qquad 
&w^*{}\rightarrow{}& 0, \\
p^*&{}\rightarrow{} p_{\infty}^*, \qquad &T^*{}\rightarrow{}& T_{\infty}^*, \qquad 
&\phi{}\rightarrow{}&\phi_{\infty} \qquad 
\textrm{as} \quad y^*\rightarrow\infty,
\end{alignedat}
\tag{\theequation \emph{a-f}}
\end{equation}
\end{subequations}
where $p_{\infty}^*$, $T_{\infty}^*$, and $\phi_{\infty}$ denote the free-stream pressure, the free-stream temperature, and the dimensionless free-stream nanoparticle volume concentration, respectively. Figure~\ref{Fig1} shows a schematic diagram of the nanofluid flow over a flat plate.


%


\subsection{Non-dimensionalisation}

The governing system of equations \eqref{Gov1} are non-dimensionalised by setting 
\begin{subequations}\label{Scaling}
\begin{equation}
\begin{alignedat}{3}
\boldsymbol{x}^*&{}={}L^*\boldsymbol{x}, \qquad 
&\boldsymbol{u}^*{}={}&U_{\infty}^*\boldsymbol{u}, \qquad 
&t^*{}={}&L^*t/U_{\infty}^*, \\ 
p^*&{}={}p_{\infty}^*+\rho_{bf}^*U_{\infty}^{*2}p, \qquad
&T^*{}={}&T_{\infty}^*T, \qquad
&\rho^*{}={}&\rho_{bf}^*\rho, \\ 
\mu^*&{}={}\mu_{bf}^*\mu, \qquad 
&c^*{}={}&c_{bf}^*c, \qquad 
&k^*{}={}&k_{bf}^*k,
\end{alignedat}
\tag{\theequation \emph{a-i}}
\end{equation}
\end{subequations}
for a characteristic length scale $L^*$. Consequently, \eqref{Gov1} becomes 
\begin{subequations}\label{Gov2}
\begin{equation}\label{Continuity2}
\frac{\partial\rho}{\partial t} + \nabla\cdot\left(\rho\boldsymbol{u}\right) = 0,    
\end{equation}
\begin{equation}\label{Momentum2}
\rho\left(\frac{\partial\boldsymbol{u}}{\partial t} + (\boldsymbol{u}\cdot\nabla)\boldsymbol{u}\right) = -\nabla p + \frac{1}{\Rey}\nabla\cdot\left(\mu\left(\nabla\boldsymbol{u} + \left(\nabla\boldsymbol{u}\right)^T - \frac{2}{3}\nabla\cdot\boldsymbol{u}\mathsfbi{I}\right)\right),
\end{equation}
\begin{equation}\label{Energy2}
\rho\left(\frac{\partial \left(cT\right)}{\partial t} + (\boldsymbol{u}\cdot\nabla)\left(cT\right)\right) = 
\frac{1}{\Rey \Pran}\nabla\cdot(k\nabla T) + \frac{1}{\Rey \Pran Le} \left( T\nabla\phi + \frac{\phi\nabla T}{N_{\textrm{BT}}T} \right)\cdot\nabla T,
\end{equation}
\begin{equation}\label{Volume2}
\frac{\partial \phi}{\partial t} + \nabla\cdot(\phi\boldsymbol{u}) = \frac{1}{\Rey Sc}\nabla\cdot\left(T\nabla\phi+\frac{\phi\nabla T}{N_{\textrm{BT}}T}\right),
\end{equation}
\end{subequations}
where 
\begin{subequations}
\begin{alignat}{2}
\rho &= 1+(\hat{\rho}-1)\phi \quad &\textrm{for}& \quad \hat{\rho} = \frac{\rho_{np}^*}{\rho_{bf}^*}, \label{Density} \\
\rho c &= 1+(\hat{\rho}\hat{c}-1)\phi \quad &\textrm{for}& \quad \hat{\rho}\hat{c} = \frac{(\rho^*c^*)_{np}}{(\rho^*c^*)_{bf}}, \\
k &= \left(\frac{\hat{k}+2+2(\hat{k}-1)\phi}{\hat{k}+2-(\hat{k}-1)\phi}\right) \quad &\textrm{for}& \quad \hat{k} = \frac{k_{np}^*}{k_{bf}^*}.
\end{alignat}
\end{subequations}
%
Moreover, in the case of the \citet{Brinkman1952} viscosity model, given by equation \eqref{Brinkman}, the non-dimensional dynamic viscosity is given as
\begin{equation}\label{NondimBrinkman}
\mu = \frac{1}{(1-\phi)^{2.5}}.
\end{equation}
Similar representations for $\mu$ are given for the \citet{Batchelor1977}, \citet{PakCho1998}, and \citet{Maiga2004} models. 

\begin{figure*}
\centerline{\includegraphics[width=0.75\textwidth]{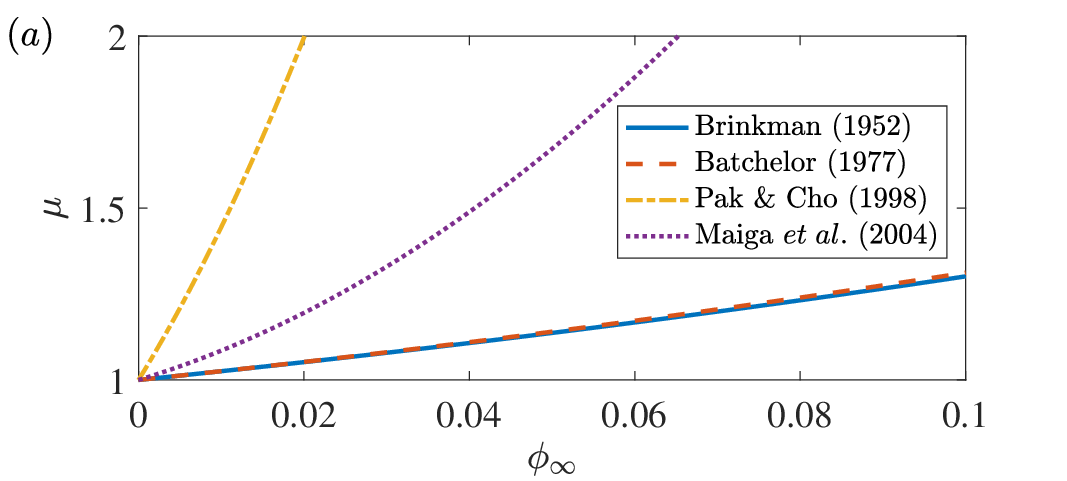}}
\centerline{\includegraphics[width=0.75\textwidth]{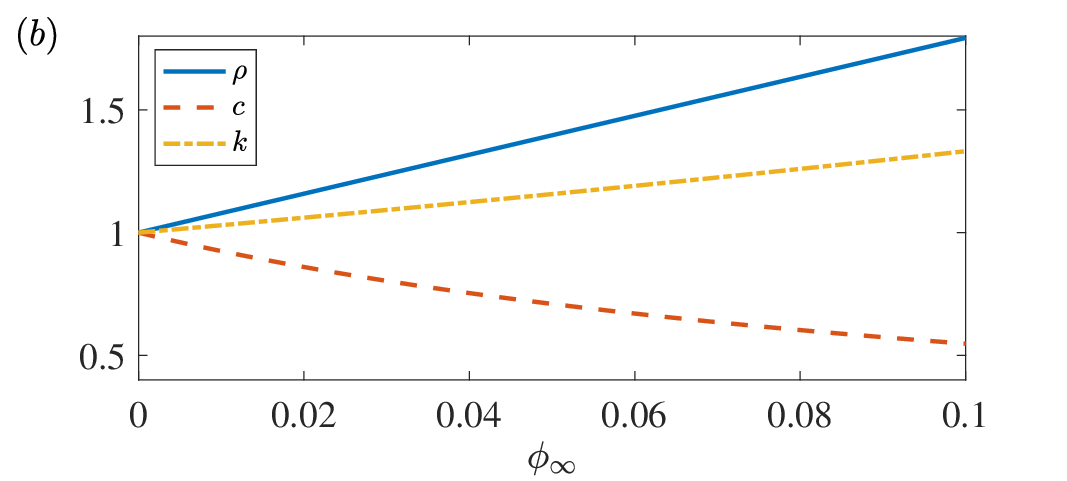}}
\caption{($a$) Non-dimensional dynamic viscosity $\mu$ as a function of $\phi_{\infty}$, for the \citet{Brinkman1952}, \citet{Batchelor1977}, \citet{PakCho1998}, and \citet{Maiga2004} models. ($b$) Non-dimensional density $\rho$, specific heat capacity $c$, and thermal conductivity $k$ as a function of $\phi_{\infty}$, for copper (Cu) nanoparticles in water. Refer to table~\ref{Table1} for fluid and nanoparticle properties.}\label{Fig2}
\end{figure*}

Figure~\ref{Fig2} compares the four models of the non-dimensional dynamic viscosity $\mu$ along with the non-dimensional density $\rho$, thermal conductivity $k$, and specific heat capacity $c$ for copper (Cu) nanoparticles in water (see table~\ref{Table1} for thermophysical properties). These quantities are plotted as functions of the free-stream nanoparticle volume concentration $\phi_{\infty}$. As $\phi_{\infty}$ increases, the Brinkman and Batchelor viscosity models show a similar rate of increase, while the Pak \& Cho and Maiga viscosity models exhibit a more rapid increase. In addition, $\rho$ also increases with $\phi_{\infty}$. Furthermore, $k$ increases, improving the flows heat transfer capability, while $c$ exhibits a reduction, causing temperature changes within the flow to occur more rapidly.

The dimensionless Reynolds, Prandtl, Lewis, and Schmidt numbers are defined as 
\begin{subequations}\label{Parameters}
\begin{alignat}{2}
\Rey &{}={} \frac{U_{\infty}^*L^*\rho_{bf}^*}{\mu_{bf}^*}, \quad 
&\Pran {}={}& \frac{\mu_{bf}^* c_{bf}^*}{k_{bf}^*}, \tag{\theequation \emph{a,b}} 
\\ 
Le &{}={} \frac{k_{bf}^*}{(\rho^*c^*)_{np}C_B^*T_{\infty}^*}, 
\quad 
&{} Sc ={}& \frac{\mu_{bf}^*}{\rho_{bf}^*C_B^*T_{\infty}^*}, \tag{\theequation \emph{c,d}} 
\end{alignat}
\end{subequations}
%
%
%
%
%
while the ratio of Brownian motion to thermophoresis is given as
\begin{equation}
 \quad N_{\textrm{BT}}=\frac{C_B^*T_{\infty}^*}{C_T^*}. 
\end{equation}
%

Finally, the boundary conditions~\eqref{BoundaryConditions1} on the plate surface are recast as
\begin{subequations}\label{BoundaryConditions3}
\begin{equation}
\boldsymbol{u}=0 \quad \textrm{and} \quad T = T_w \left(\equiv\tfrac{T_w^*}{T_{\infty}^*}\right) \quad \textrm{on} \quad y=0, \tag{\theequation \emph{a,b}}
\end{equation}
and
\begin{equation}\label{NonDimTempPhiCondition}
T\frac{\partial\phi}{\partial y}+\frac{\phi}{N_{\textrm{BT}}T}{\frac{\partial T}{\partial y}}=0 \quad \textrm{on} \quad y=0, \tag{\theequation \emph{c}}
\end{equation}
\end{subequations}
while the boundary conditions~\eqref{BoundaryConditions2} in the free-stream are given as
\begin{subequations}\label{BoundaryConditions4}
\begin{equation}
\begin{alignedat}{3}
u&{}\rightarrow{} 1, \qquad
&v{}\rightarrow{}& 0, \qquad 
&w{}\rightarrow{}& 0, \\
p&{}\rightarrow{} 0, \qquad 
&T{}\rightarrow{}& 1, \qquad 
&\phi{}\rightarrow{}&\phi_{\infty} 
\qquad \textrm{as} \quad y\rightarrow\infty.
\end{alignedat}
\tag{\theequation \emph{a-f}}
\end{equation}
\end{subequations}

Table~\ref{Table1} presents the thermophysical properties of various materials used for nanoparticles. The non-dimensional ratios $\hat{\rho}$, $\hat{k}$, and $\hat{c}$ are based on water as the base fluid, where the Prandtl number $\Pran=6.85$, while the Lewis number $Le$, the Schmidt number $Sc$, and the ratio $N_{\text{BT}}$ are given for the free-stream temperature $T_{\infty}^*=300$ K and the nanoparticle diameter $d_{np}^*=20$ nm. Both $Le$ and $Sc$ are of the order $10^4$ for all materials listed in table~\ref{Table1}. 

\afterpage{
\begin{landscape}
\begin{table}
\centering
    \begin{tabular}{l|ccccccccccc}
        \hline
         & $\rho^*$ (kg/m$^{3}$) & $\mu^*$ (kg/ms) & $k^*$ (W/mK) & $c^*$ (J/kgK) & $\hat{\rho}$ & $\hat{k}$ & $\hat{c}$ & $\beta_T$ & $Le$ & $Sc$ & $N_\text{BT}$ \\ \hline
        Water & $1000$ & $0.001$ & $0.61$ & $4180$ & - & - & - & - & - & - & - \\
        Silver (Ag) & $10 \; 500$ & - & $430$ & $235$ & $10.50$ & $704.9$ & $0.0562$ & $0.00037$ & $11 \; 250$ & $45 \; 509$ & $0.0597$ \\
        Copper (Cu) & $8933$ & - & $400$ & $385$ & $8.93$ & $655.7$ & $0.0921$ & $0.00040$ & $8072$ & $45 \; 509$ & $0.0556$ \\
        Copper Oxide (CuO) & $6320$ & - & $77$ & $532$ & $6.32$ & $126.2$ & $0.1273$ & $0.00203$ & $8257$ & $45 \; 509$ & $0.0108$ \\
        Alumina (Al$_2$O$_3$) & $3950$ & - & $35$ & $800$ & $3.95$ & $57.4$ & $0.1914$ & $0.00438$ & $8785$ & $45 \; 509$ & $0.0050$ \\
        Titanium Oxide (TiO$_2$) & $4250$ & - & $8.95$ & $686$ & $4.25$ & $14.7$ & $0.1641$ & $0.01559$ & $9522$ & $45 \; 509$ & $0.0014$ \\
        Aluminium (Al) & $2710$ & - & $235$ & $904$ & $2.71$ & $385.2$ & $0.2163$ & $0.00067$ & $11 \; 332$ & $45 \; 509$ & $0.0327$ \\
        Silicon (Si) & $2330$ & - & $150$ & $710$ & $2.33$ & $57.4$ & $0.1699$ & $0.00105$ & $16 \; 781$ & $45 \; 509$ & $0.0210$
    \end{tabular}
    \caption{Thermophysical properties of water and various materials used for nanoparticles, as given in \citet{Buongiorno2006}, \citet{Wang2008a}, \citet{Bachok2011}, \citet{MacDevette2014}, \citet{Turkyilmazoglu2014, Turkyilmazoglu2020}, and at https://periodictable.com/Elements. Here, the free-stream temperature $T_{\infty}^*=300$ K, the nanoparticle diameter $d_{np}^*=20$ nm, and the Prandtl number $\Pran=6.85$. The ratios $\hat{\rho}$, $\hat{k}$, and $\hat{c}$ are based on water as the base fluid.}
    \label{Table1}
\end{table}
\end{landscape}
}


\section{Steady boundary-layer flow}

\subsection{Boundary-layer equations}\label{BLEquations}

Following the derivation of \citet{Ruban2017}, the steady, two-dimensional boundary-layer equations are obtained by assuming a zero pressure gradient, setting $w=0$, and considering solutions that are independent of the $z$-direction and time $t$. On introducing the Prandtl boundary-layer transformation 
\begin{equation}
y = {\Rey}^{-1/2}Y, 
\end{equation}
with
\begin{subequations}
\begin{equation}
\begin{alignedat}{2}
u(x,y) &= U_B(x,Y), \qquad
&v(x,y) {}={}& {\Rey}^{-1/2}V_B(x,Y), \\
T(x,y) &= T_B(x,Y), \qquad
&\phi(x,y) {}={}& \phi_B(x,Y), \\
\mu(x,y) &= \mu_B(x,Y), \qquad 
&\rho(x,y) {}={}& \rho_B(x,Y), \\ 
c(x,y) &= c_B(x,Y), \qquad
&k(x,y) {}={}& k_B(x,Y),
\end{alignedat}
\tag{\theequation \emph{a-h}}
\end{equation}
\end{subequations}
and letting $\Rey\rightarrow\infty$, the non-dimensional governing equations \eqref{Gov2} become
\begin{subequations}\label{Gov3}
\begin{equation}\label{Continuity3}
\frac{\partial(\rho_B U_B)}{\partial x} + \frac{\partial(\rho_B V_B)}{\partial Y} = 0, 
\end{equation}
\begin{equation}\label{Momentum3}
\rho_B\left(U_B\frac{\partial U_B}{\partial x} + V_B\frac{\partial U_B}{\partial Y}\right) = \frac{\partial}{\partial Y}\left(\mu_B\frac{\partial U_B}{\partial Y}\right),  
\end{equation}
\begin{multline}\label{Energy3}
\rho_B \left(U_B\frac{\partial (c_BT_B)}{\partial x}+V_B\frac{\partial (c_BT_B)}{\partial Y}\right) = 
\frac{1}{\Pran}\frac{\partial}{\partial Y}\left(k_B\frac{\partial T_B}{\partial Y}\right) \\ 
+ \frac{1}{\Pran Le} \left( T_B\frac{\partial\phi_B}{\partial Y}\frac{\partial T_B}{\partial Y} +\frac{\phi_B}{N_{\textrm{BT}}T_B}\left(\frac{\partial T_B}{\partial Y}\right)^2\right),    
\end{multline}
\begin{equation}\label{Volume3}
\frac{\partial(\phi_B U_B)}{\partial x} + \frac{\partial(\phi_B V_B)}{\partial Y} = \frac{1}{Sc}\frac{\partial}{\partial Y}\left(T_B\frac{\partial\phi_B}{\partial Y}+\frac{\phi_B}{N_{\textrm{BT}}T_B}\frac{\partial T_B}{\partial Y}\right).
\end{equation}
\end{subequations}
A self-similar solution is then sought using the similarity variable $\eta=Y/\sqrt{x}$, coupled with the Dorodnitsyn–Howarth transformation
\begin{equation}\label{Dorodnitsyn–Howarth}
\xi=\int_0^{\eta}\rho(\grave{\eta}) \; \textrm{d}\grave{\eta},    
\end{equation}
with
\begin{subequations}
\begin{equation}\label{Baseflowsimilarity}
\begin{alignedat}{2}
U_B(x,Y)&{}={}f'(\xi), \quad 
&V_B(x,Y)&{}={}\frac{1}{2\sqrt{x}}\left(\eta f'-\frac{f}{\rho}\right), \\ 
T_B(x,Y)&{}={}\theta(\xi), \quad 
&\phi_B(x,Y)&{}={}\varphi(\xi), \\
\mu_B(x,Y)&{}={}\mu(\xi), \quad 
&\rho_B(x,Y)&{}={}\rho(\xi), \\ 
c_B(x,Y)&{}={}c(\xi), \quad 
&k_B(x,Y)&{}={}k(\xi).
\end{alignedat}
\tag{\theequation \emph{a-h}}
\end{equation}
\end{subequations}
(For notational simplicity, $\mu$, $\rho$, $c$, and $k$ are re-used to denote their similarity profiles.)
Consequently, the following boundary-layer equations are derived
\begin{subequations}\label{Gov4}
\begin{equation}\label{Momentum4}
2\left(\rho\mu f''\right)'+ff''=0, \tag{\theequation \emph{a}}
\end{equation}
\begin{equation}\label{Energy4}
2\left(\rho k \theta'\right)' +
\Pran f\left(c\theta\right)' + \frac{2\rho \theta'}{Le} \left( \theta\varphi' + \frac{\varphi \theta'}{N_{\textrm{BT}}\theta}\right) = 0, \tag{\theequation \emph{b}}
\end{equation}
\begin{equation}\label{Volume4}
\frac{2\rho^2}{Sc}\left(\rho\left(\theta\varphi' + \frac{\varphi \theta'}{N_{\textrm{BT}}\theta}\right)\right)'+f\varphi'=0, \tag{\theequation \emph{c}}
\end{equation}
%
%
subject to the boundary conditions 
%
%
\begin{equation}
f=f'=0, \; \theta=T_w
\quad \textrm{on} \quad \xi=0,
\tag{\theequation \emph{d-f}}
\end{equation}
\begin{equation}
\theta\varphi'+\frac{\varphi \theta'}{N_{\textrm{BT}}\theta}=0
\quad \textrm{on} \quad \xi=0, 
\tag{\theequation \emph{g}}
\end{equation}
and
\begin{equation}
f'\rightarrow 1, \quad 
\theta\rightarrow 1, \quad 
\varphi\rightarrow\phi_{\infty} \quad 
\textrm{as} \quad \xi\rightarrow\infty,  \tag{\theequation \emph{h-j}}
\end{equation}
\end{subequations}
where a prime denotes differentiation with respect to $\xi$.


\subsection{Boundary-layer simplifications}

In the limits $Le\rightarrow\infty$ and $Sc\rightarrow\infty$, equation~\eqref{Volume4} simplifies to $\varphi'=0$, implying $\varphi=\phi_{\infty}$ everywhere. Consequently, $\mu$, $\rho$, $c$, and $k$ are constants, and the boundary-layer equations~\eqref{Momentum4} and~\eqref{Energy4} reduce to
\begin{subequations}\label{Gov5}
\begin{equation}\label{Momentum5}
2\rho\mu f'''+ ff''=0 
\quad \textrm{and} \quad
2\rho\mu\theta'' + \widehat{\Pran}f\theta' = 0,  
\tag{\theequation \emph{a,b}}
\end{equation}
\end{subequations}
where $\widehat{\Pran}=\mu c\Pran/k$. 

A further simplification of the boundary-layer equations is obtained by introducing
\begin{subequations}
\begin{equation}
p=\rho\hat{p}, \quad T=1+(T_w-1)\widehat{T}, \quad \widehat{\Rey} = \frac{\rho}{\mu}\Rey, \tag{\theequation \emph{a-c}}
\end{equation}
\end{subequations}
into the governing equations~\eqref{Gov2} and following the procedure outlined in \S\ref{BLEquations} with $\widehat{\Rey}\rightarrow\infty$, to give
\begin{subequations}\label{Gov6}
\begin{equation}\label{Momentum6}
2 f'''+ff''=0 
\quad \textrm{and} \quad
2\widehat{\theta}'' + \widehat{\Pran}f\widehat{\theta}' = 0, 
\tag{\theequation \emph{a,b}}
\end{equation}
subject to the boundary conditions
\begin{equation}\label{BoundaryConditions6}
f=f'=0, \; \widehat{\theta}=1
\quad \textrm{on} \quad \xi=0,
\tag{\theequation \emph{c-e}}
\end{equation}
and
\begin{equation}
f'\rightarrow 1, \quad 
\widehat{\theta}\rightarrow 0, \quad 
\textrm{as} \quad \xi\rightarrow\infty.  \tag{\theequation \emph{f,g}}
\end{equation}
\end{subequations}
Thus, the model of the nanofluid flow simplifies to the standard Blasius equation with a modified Prandtl number $\widehat{\Pran}$. Consequently, in the absence of Brownian motion and thermophoresis, the Reynolds number of the nanofluid flow $\Rey$ is given in terms of $\widehat{\Rey}$ as $\Rey=\mu\widehat{\Rey}/\rho$. A detailed description of the Navier--Stokes equations in the absence of Brownian motion and thermophoresis, leading to the derivation of \eqref{Gov6}, is given in appendix~\ref{AppendixA}.


\subsection{Boundary-layer solutions}

\begin{figure*}
\centerline{
\includegraphics[width=0.5\textwidth]{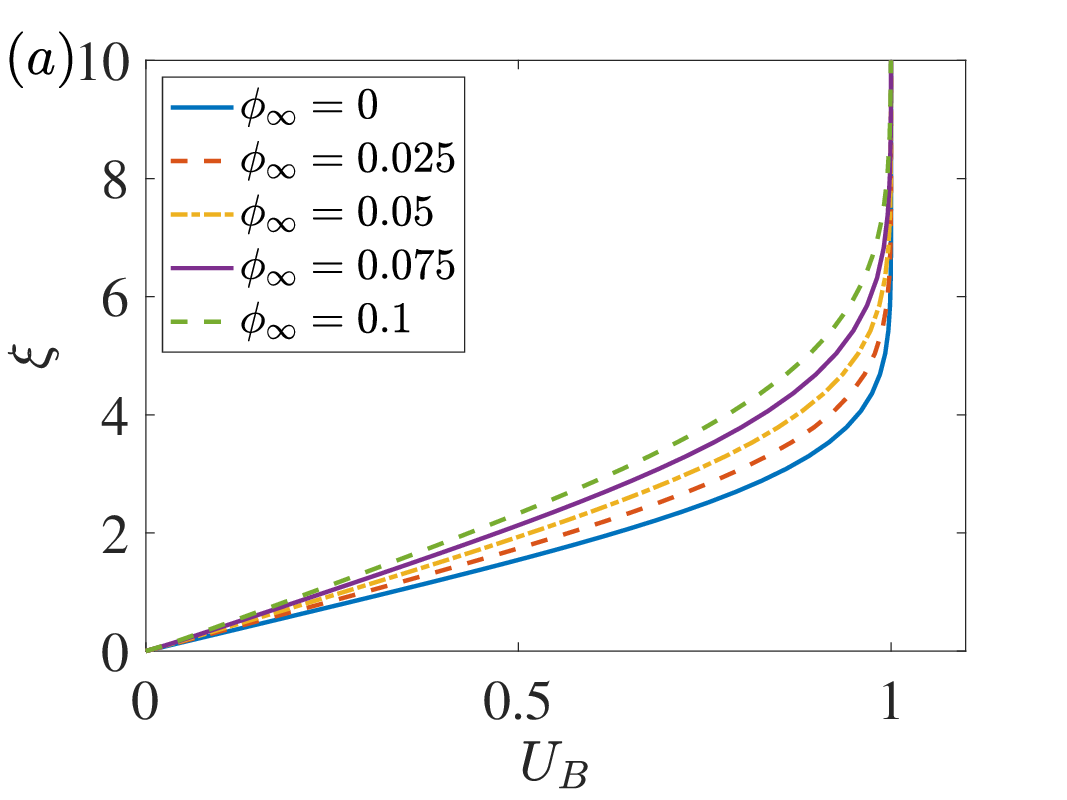}
\includegraphics[width=0.5\textwidth]{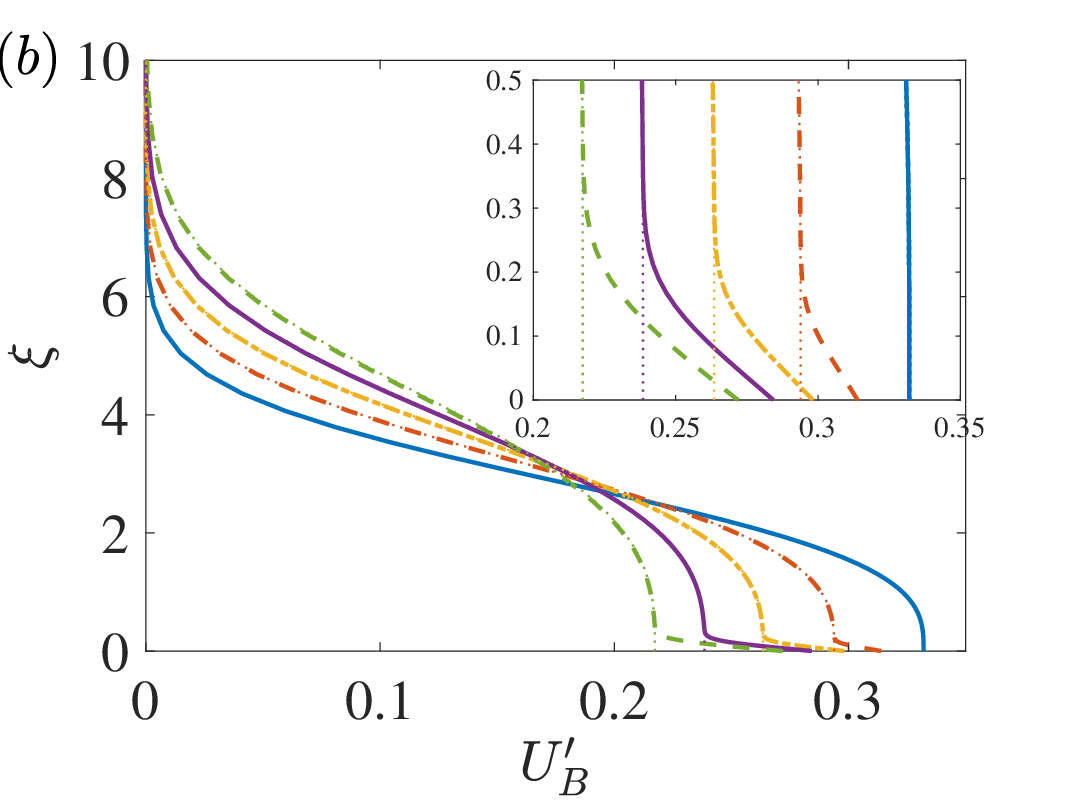}}
\centerline{
\includegraphics[width=0.5\textwidth]{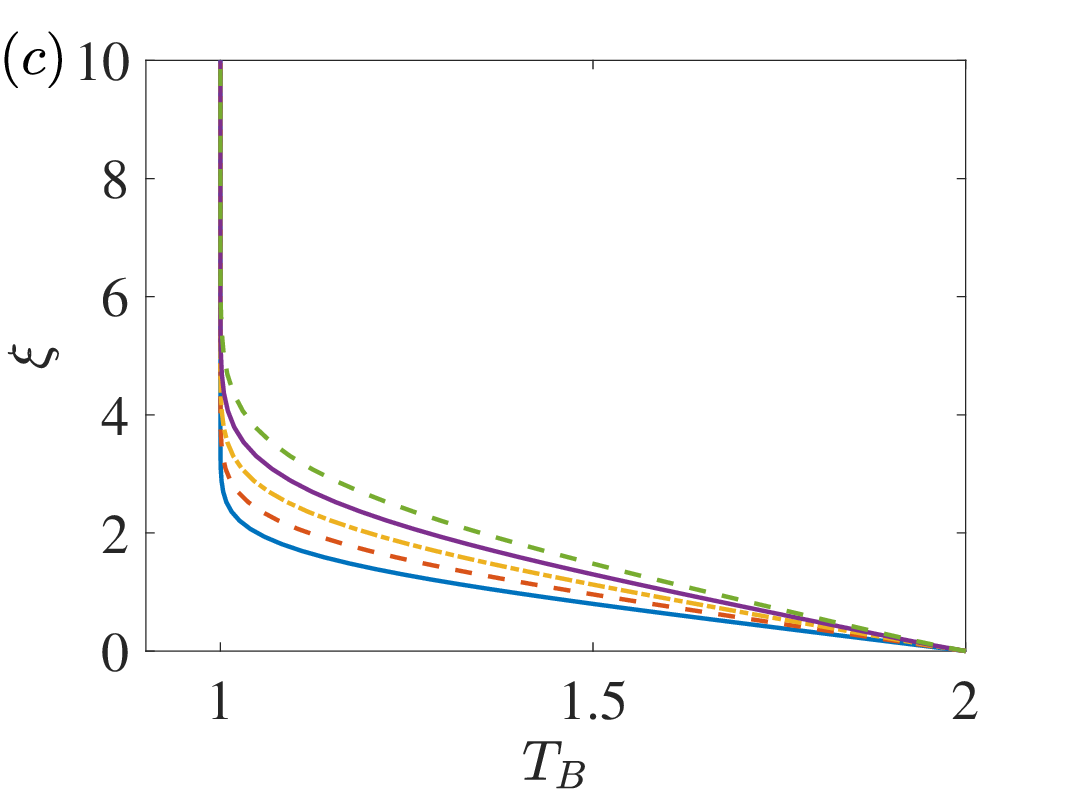}
\includegraphics[width=0.5\textwidth]{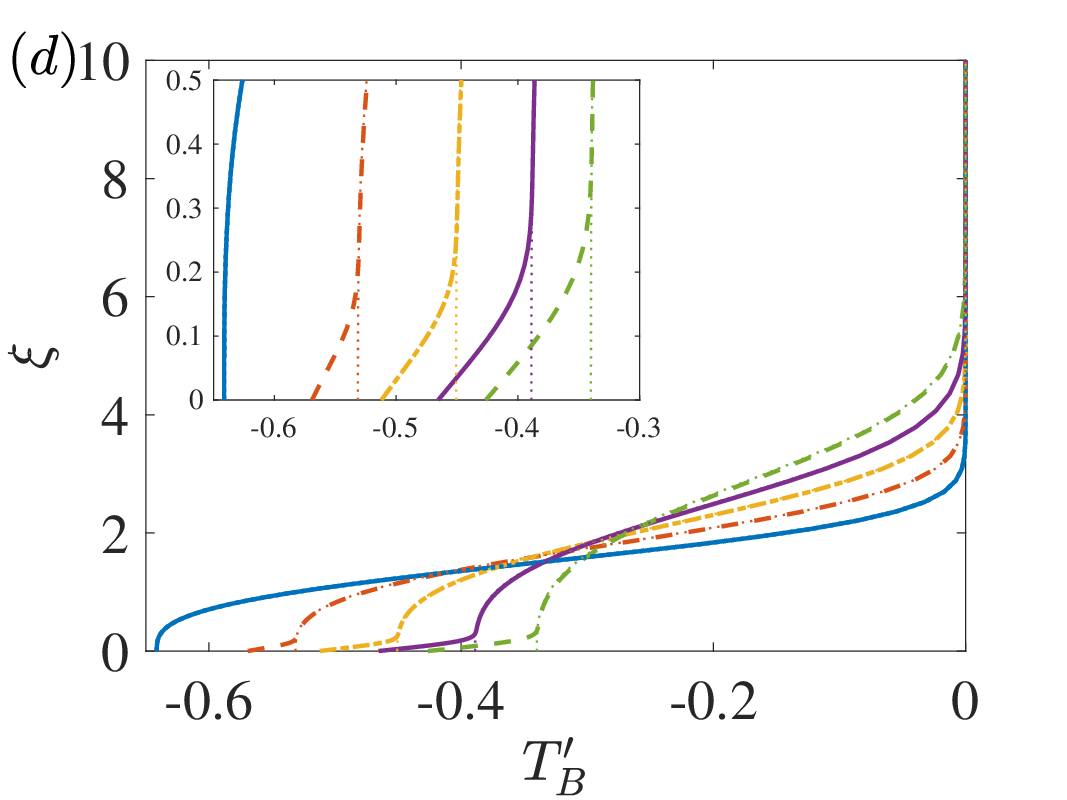}}
\centerline{
\includegraphics[width=0.5\textwidth]{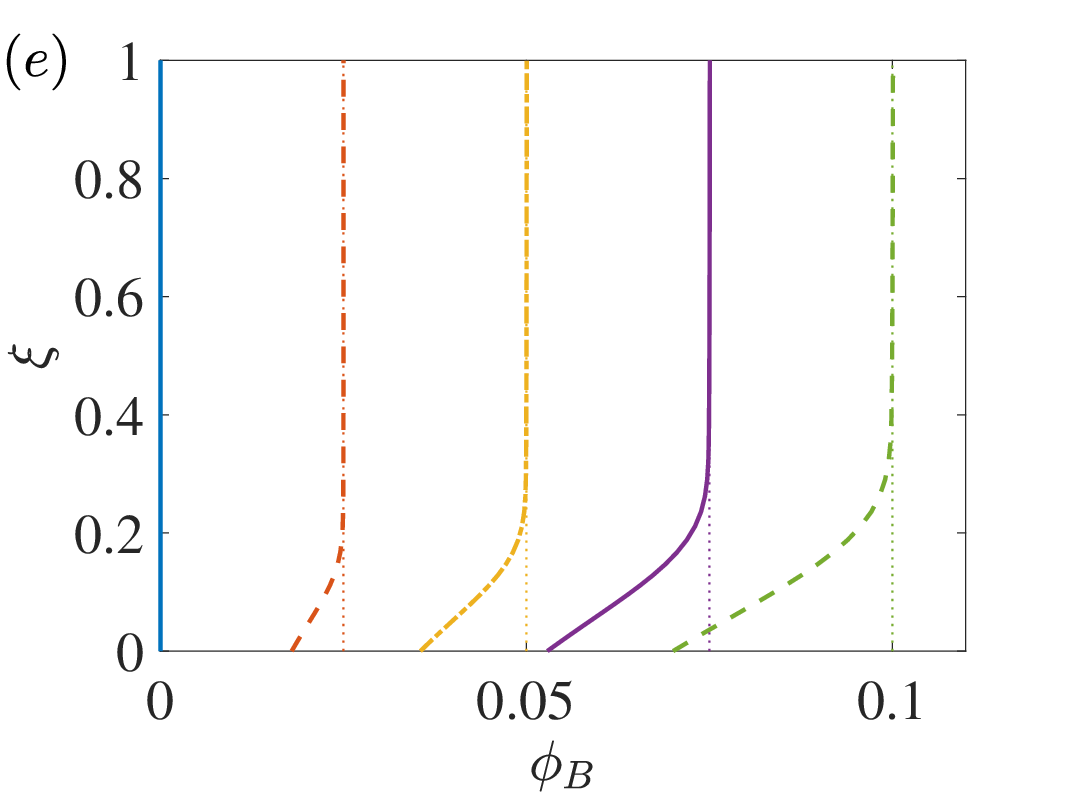}
\includegraphics[width=0.5\textwidth]{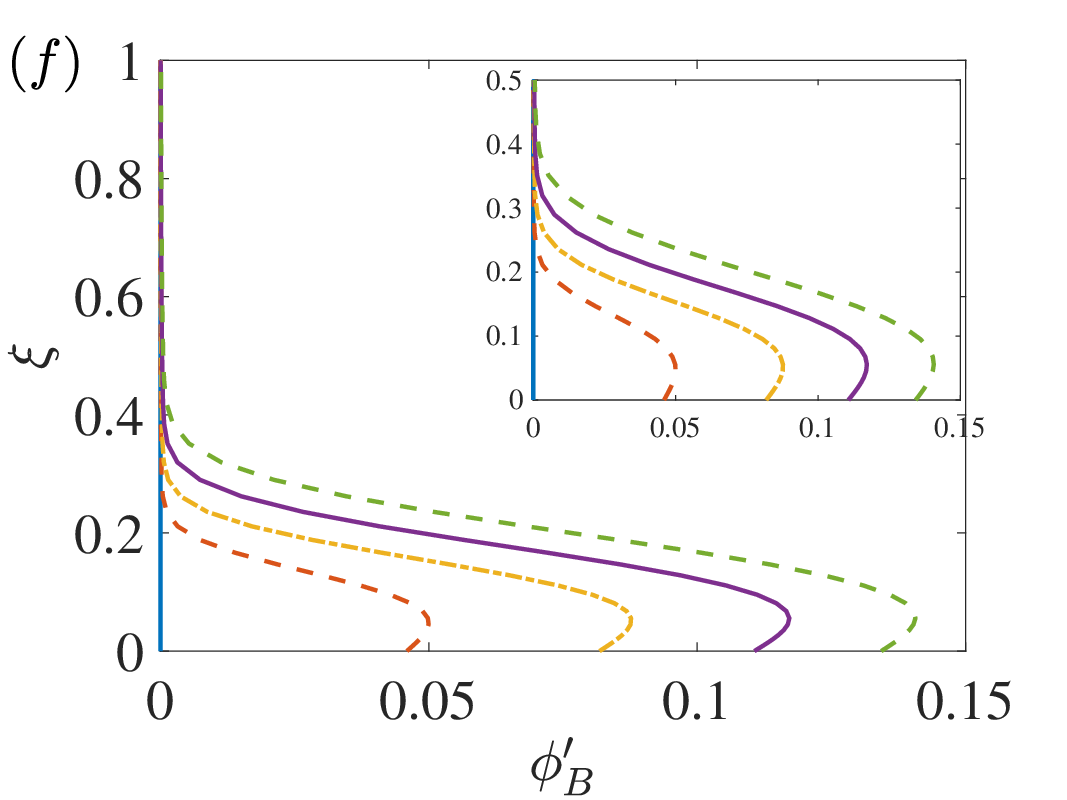}}
\caption{Steady base flow profiles for variable $\phi_{\infty}$ and $T_w = 2$, for copper (Cu) nanoparticles in water. (\emph{a}) Streamwise velocity $U_B=f'(\xi)$, (\emph{b}) $U_B'=f''(\xi)$, (\emph{c}) temperature $T_B=\theta(\xi)$, (\emph{d}) $T_B'=\theta'(\xi)$, (\emph{e}) nanoparticle volume concentration $\phi_B=\varphi(\xi)$, and (\emph{f}) $\phi'_B=\varphi'(\xi)$. Dotted lines depict the equivalent solutions in the instance $Le\rightarrow\infty$ and $Sc\rightarrow\infty$.}\label{Fig3}
\end{figure*}

On the left-hand side of figure~\ref{Fig3}, the steady streamwise velocity $U_B=f'(\xi)$, temperature $T_B=\theta(\xi)$, and nanoparticle volume concentration $\phi_B=\varphi(\xi)$ are plotted for five values of $\phi_{\infty}$ and the wall temperature $T_w=2$. Similar profiles are obtained for other values of $T_w$. The solid, dashed, and chain lines represent solutions of the full boundary-layer equations~\eqref{Gov4} for copper (Cu) nanoparticles in water (see table~\ref{Table1} for thermophysical properties). A thin concentration layer develops in the $\phi_{B}$ profile, consistent with the observations of \citet{Avramenko2011}, which alters the near-wall behaviour of the velocity and temperature profiles. This behaviour is most clearly illustrated on the right-hand side of figure~\ref{Fig3}, which plots the profiles $U_B' = f''(\xi)$, $T_B' = \theta'(\xi)$, and $\phi_B'=\varphi'(\xi)$. These profiles reveal that, in contrast to the standard Blasius flow, $U_B'$ does not approach a constant as $\xi\rightarrow 0$.

When Brownian motion and thermophoresis are neglected (i.e., $Sc\rightarrow\infty$ and $Le\rightarrow\infty$), the concentration layer disappears with $\phi_B=\phi_{\infty}$ everywhere (see the vertical dotted lines in figure~\ref{Fig3}(\emph{e})). In this limit, the standard Blasius flow structure is recovered, with $U_B'$ approaching a constant near the wall, as indicated by the dotted lines in figure~\ref{Fig3}(\emph{b}). 

Table \ref{Table2} compares the base flow properties on the plate surface for varying $\phi_{\infty}$ and $T_w=2$. The differences between the results obtained with and without Brownian motion and thermophoresis are negligible for $\phi_{\infty}<10^{-3}$, but grow, due to the impact of the concentration layer, at larger $\phi_{\infty}$.

\setlength{\tabcolsep}{8pt}
\begin{table}
\begin{center}
\begin{tabular}{cccc}
        $\phi_{\infty}$ & $U_{B}'(0)=f''(0)$ & $T_{B}'(0)=\theta'(0)$ & $\phi_B(0)=\varphi(0)$ \\ \hline 
        $0$ & $0.332057$ ($0.332057$) & $-0.641309$ ($-0.641309$) & $0.000000$ ($0.000000$) \\
        $10^{-6}$ & $0.332057$ ($0.332056$) & $-0.641307$ ($-0.641304$) & $0.000001$ ($0.000001$) \\  
        $10^{-5}$ & $0.332049$ ($0.332040$) & $-0.641276$ ($-0.641257$) & $0.000007$ ($0.000010$) \\
        $10^{-4}$ & $0.331979$ ($0.331884)$ & $-0.640981$ ($-0.640790$) & $0.000073$ ($0.000100$) \\
        $10^{-3}$ & $0.331273$ ($0.330335$) & $-0.638043$ ($-0.636146$) & $0.000726$ ($0.001000$) \\
        $10^{-2}$ & $0.324446$ ($0.315631$) & $-0.610189$ ($-0.592849$) & $0.007229$ ($0.010000$) \\
        $10^{-1}$ & $0.271857$ ($0.217365$) & $-0.426271$ ($-0.339946$) & $0.070052$ ($0.100000$)
\end{tabular}
\caption{Base flow properties on $\xi=0$ for variable $\phi_{\infty}$ and $T_w=2$, where a prime denotes differentiation with respect to the similarity variable $\xi$. Solutions based on copper (Cu) nanoparticles in water, while the results in brackets correspond to the solutions obtained in the absence of Brownian motion and thermophoresis.}
\label{Table2}
\end{center}
\end{table}

\begin{figure*}
\centerline{\includegraphics[width=0.75\textwidth]{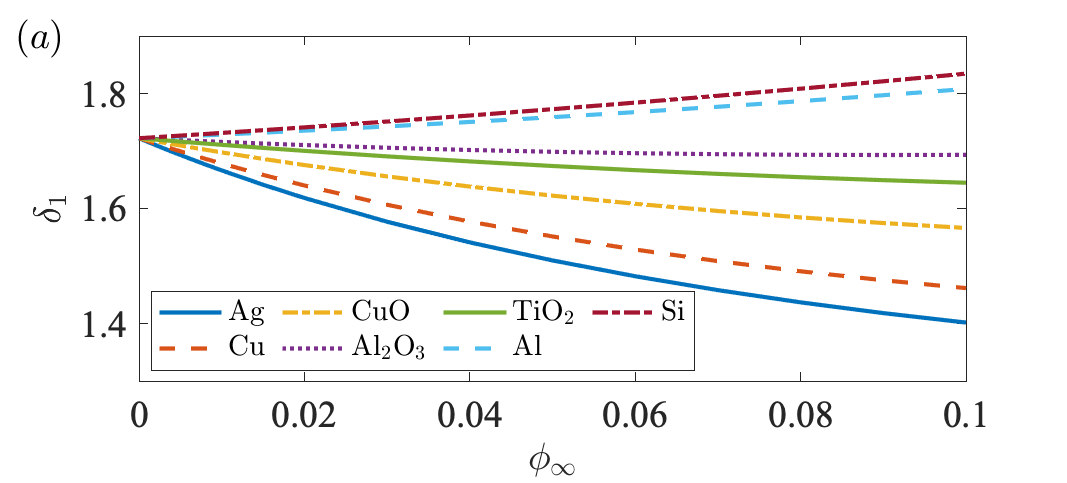}}
\centerline{\includegraphics[width=0.75\textwidth]{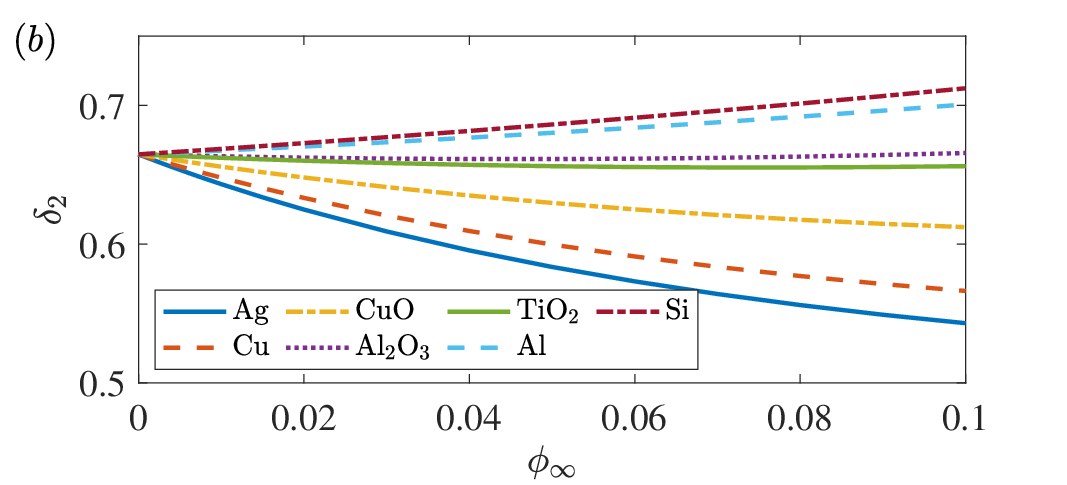}}
\centerline{\includegraphics[width=0.75\textwidth]{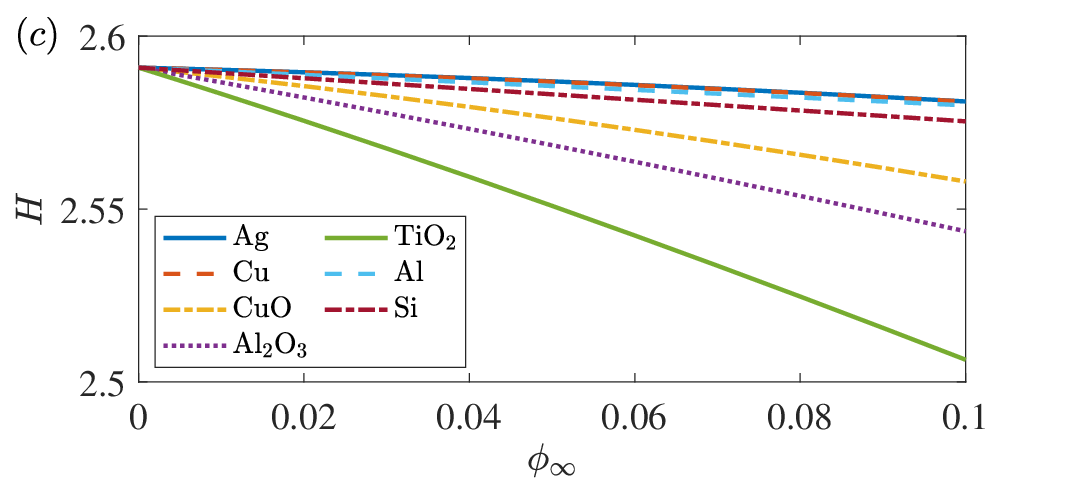}}
\caption{(\emph{a}) Displacement thickness $\delta_1$, (\emph{b}) momentum thickness $\delta_2$, and (\emph{c}) shape factor $H$ as functions of the free-stream nanoparticle volume concentration $\phi_{\infty}$, for different nanoparticle materials.}\label{Fig4}
\end{figure*}

Since the base flow profiles in figure~\ref{Fig3} are plotted against the density-weighted similarity variable $\xi$, a physically meaningful measure of the boundary-layer thickness is provided by the displacement thickness. The dimensional displacement thickness $\delta^*_1=x^*\delta_1/\Rey_x^{1/2}$ and momentum thickness $\delta^*_2=x^*\delta_2/\Rey_x^{1/2}$, for
\begin{subequations}
\begin{equation}
\delta_1 = \int_0^{\infty}\frac{1}{\rho(\xi)}-\frac{f'(\xi)}{\rho_{\infty}} \;\textrm{d}\xi \quad \textrm{and} \quad
\delta_2 = \int_0^{\infty}\frac{f'(\xi)}{\rho_{\infty}}\left(1-f'(\xi)\right) \;\textrm{d}\xi, \tag{\theequation \emph{a,b}} 
\end{equation}
\end{subequations}
are shown in figure~\ref{Fig4}, along with the shape factor $H=\delta^*_1/\delta^*_2$. Here, $\Rey_x=U_{\infty}^*x^*\rho_{bf}^*/\mu_{bf}^*$ and $\rho_{\infty}=\rho^*_{\infty}/\rho^*_{bf}$ denotes the dimensionless free-stream density. Results are plotted for all seven nanoparticle materials listed in table~\ref{Table1}. For all but two of these materials, both $\delta_1$ and $\delta_2$ decrease as $\phi_{\infty}$ increases. The most significant reductions occur for silver (Ag) and copper (Cu) nanoparticles, which have the highest densities (and the largest non-dimensional $\hat{\rho}$ values). In contrast, silicon (Si) and aluminium (Al) nanoparticles, which have the lowest densities (and the smallest values of $\hat{\rho}$), show an increase in $\delta_1$ and $\delta_2$ as $\phi_{\infty}$ increases. (Solutions corresponding to the case without Brownian motion and thermophoresis are nearly identical to those shown in figure~\ref{Fig4}.)

\begin{figure*}
\centerline{\includegraphics[width=0.75\textwidth]{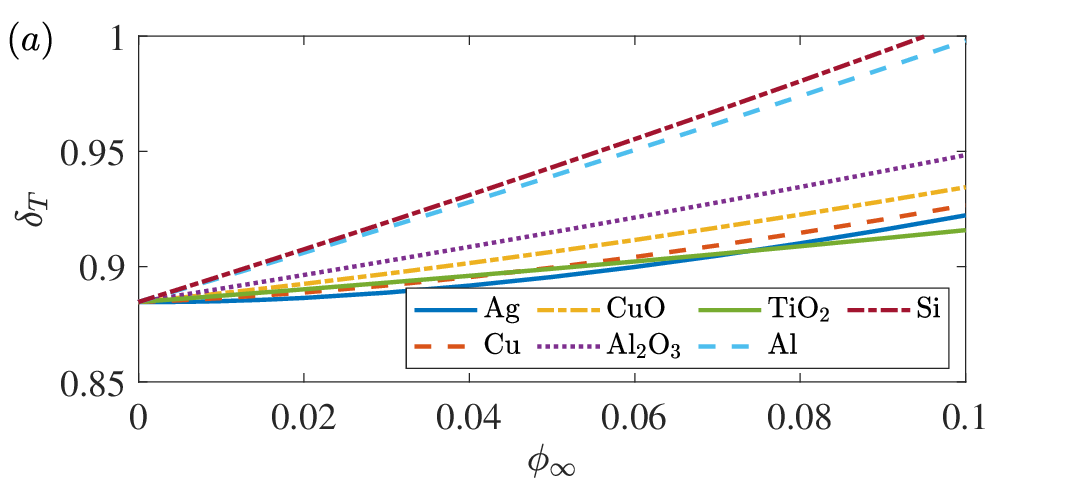}}
\centerline{\includegraphics[width=0.75\textwidth]{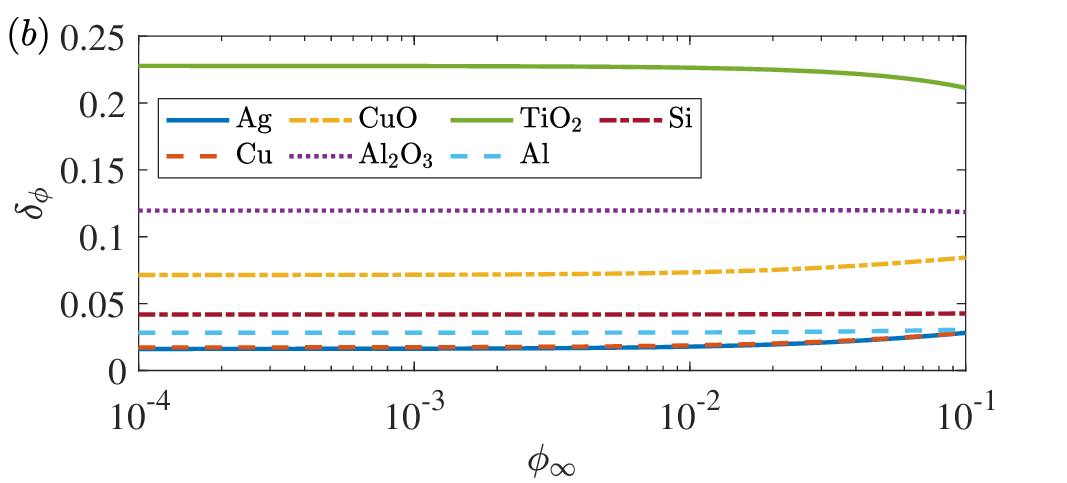}}
\caption{(\emph{a}) Thermal displacement thickness $\delta_1$ and (\emph{b}) concentration displacement thickness $\delta_{\phi}$ as functions of the free-stream nanoparticle volume concentration $\phi_{\infty}$, for different nanoparticle materials.}\label{Fig5}
\end{figure*}

The thermal displacement thickness $\delta^*_T=x^*\delta_T/\Rey_x^{1/2}$ and concentration displacement thickness $\delta^*_{\phi}=x^*\delta_{\phi}/\Rey_x^{1/2}$, for
\begin{subequations}
\begin{equation}
\delta_T = \int_0^{\infty}\frac{1}{\rho(\xi)}-\frac{\theta(\xi)-T_w}{\rho_{\infty}(1-T_w)} \;\textrm{d}\xi \quad \textrm{and} \quad
\delta_{\phi} = \int_0^{\infty}\frac{1}{\rho(\xi)}-\frac{\varphi(\xi)}{\rho_{\infty}\phi_{\infty}} \;\textrm{d}\xi, \tag{\theequation \emph{a,b}} 
\end{equation}
\end{subequations}
are plotted in figure~\ref{Fig5} as a function of $\phi_{\infty}$. In contrast to the displacement thickness $\delta_1$, the thermal displacement thickness $\delta_T$ increases with increasing $\phi_{\infty}$ for all seven nanoparticle materials. The most pronounced increases are observed for the less dense materials, aluminium (Al) and silicon (Si). On the other hand, the concentration displacement thickness $\delta_{\phi}$ (plotted on a semi-log scale along the horizontal axis) exhibits only minor variations across the range of $\phi_{\infty}$ shown. However, noticeable differences arise between the materials. Notably, titanium oxide (TiO$_2$) and alumina (Al$_2$O$_3$) exhibit larger values of $\delta_{\phi}$ than the other materials. This can be attributed to their respective $N_{\text{BT}}$ values being an order of magnitude smaller than those of the other materials (see table~\ref{Table1}). Thus, thermophoresis effects are more dominant than Brownian motion effects for these particular materials. Moreover, as $\phi_{\infty}$ approaches zero, $\delta_{\phi}$ tends toward a positive constant, indicating that $\varphi$ approaches a limiting solution. This behaviour will be examined in further detail in \S\ref{Limit_for_phi}.

\begin{figure*}
\centerline{\includegraphics[width=0.75\textwidth]{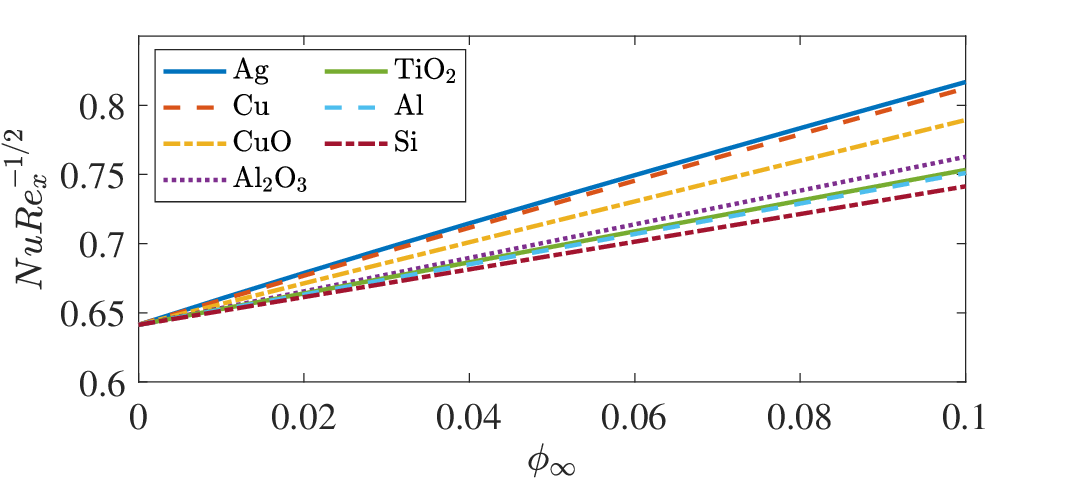}}
\caption{Scaled local Nusselt number $Nu\Rey_x^{-1/2}$ as a function of the free-stream nanoparticle volume concentration $\phi_{\infty}$, for different nanoparticle materials.}\label{Fig6}
\end{figure*}

Despite the thickening of the thermal boundary-layer, the local Nusselt number, defined as
\begin{equation}
Nu = \frac{\Rey_x^{1/2}\rho_w k_w\theta'(0)}{1-T_w},
\end{equation}
increases with increasing $\phi_{\infty}$, as shown in figure~\ref{Fig6}. Thus, all of the nanoparticles improve the heat transfer capabilities of the fluid. The most pronounced increases in $Nu$ are observed for denser materials with higher thermal conductivities and smaller specific heat capacities, such as silver (Ag) and copper (Cu) nanoparticles. Consequently, these materials have greater thermodynamic benefits.


\subsection{Asymptotic behaviour in the limit $\phi_{\infty}\rightarrow 0$}\label{Limit_for_phi}

The behaviour of the steady base flow is now examined in the limit as the free-stream nanoparticle volume concentration $\phi_{\infty}$ approaches zero. Similarity variables $f$, $\theta$, and $\varphi$ are expanded in powers of $\phi_{\infty}$, as
\begin{subequations}\label{LowPhiLimit}
\begin{align}
f(\xi) &= f_0(\xi) + \phi_{\infty}f_1(\xi) + O(\phi_{\infty}^2), \\ 
\theta(\xi) &= \theta_0(\xi) + \phi_{\infty}\theta_1(\xi) + O(\phi_{\infty}^2), \\ 
\varphi(\xi) &= \phi_{\infty}\varphi_1(\xi) + O(\phi_{\infty}^2),   
\end{align}
while the physical quantities $\mu$, $\rho$, $c$, and $k$ are of the form
\begin{equation}
(\mu,\rho,c,k)(\xi) = 1 +\phi_{\infty}(\mu_1,\rho_1,c_1,k_1)(\xi) + O(\phi_{\infty}^2).
\end{equation}
\end{subequations}

Substituting \eqref{LowPhiLimit} into equations \eqref{Momentum4} and \eqref{Energy4} and retaining the leading-order terms yields the Blasius boundary-layer equations for the velocity and temperature
\begin{subequations}\label{LowPhiLimit_Gov1}
\begin{equation}
2f_0'''+f_0f_0'' = 0 \quad \textrm{and} \quad 2\theta_0'' + \Pran f_0\theta_0' = 0,    \tag{\theequation \emph{a,b}}
\end{equation}
subject to the boundary conditions
\begin{equation}
f_0 = f'_0 = 0, \quad \theta_0 = T_w \quad \textrm{on} \quad \xi=0, \tag{\theequation \emph{c-e}} 
\end{equation}
\begin{equation}
f_0'\rightarrow 1, \quad \theta_0\rightarrow 1 \quad \textrm{as} \quad \xi\rightarrow\infty. 
\tag{\theequation \emph{f,g}} 
\end{equation}
\end{subequations}
Moreover, substituting \eqref{LowPhiLimit} into equation \eqref{Volume4} and equating terms of order $\phi_{\infty}$ gives the following second-order differential equation for $\varphi_1$
\begin{subequations}\label{LowPhiLimit_Gov2}
\begin{equation}
\theta_0\varphi_1'' + \left(\theta_0'+\frac{\theta_0'}{N_{\text{BT}}\theta_0} + \frac{Sc f_0}{2}\right)\varphi_1' 
+ \frac{1}{N_{\text{BT}}}\left(\frac{\theta_0''}{\theta_0} - \left(\frac{\theta_0'}{\theta_0}\right)^2 \right)\varphi_1 = 0,
\end{equation}
subject to the boundary conditions
\begin{equation}
\theta_0\varphi_1'+\frac{\varphi_1 \theta_0'}{N_{\textrm{BT}}\theta_0}=0
\quad \textrm{on} \quad \xi=0, 
\end{equation}    
\begin{equation}
\varphi_1\rightarrow 1 \quad \textrm{as} \quad \xi\rightarrow\infty.
\end{equation}
\end{subequations}

\begin{figure*}
\centerline{\includegraphics[width=0.75\textwidth]{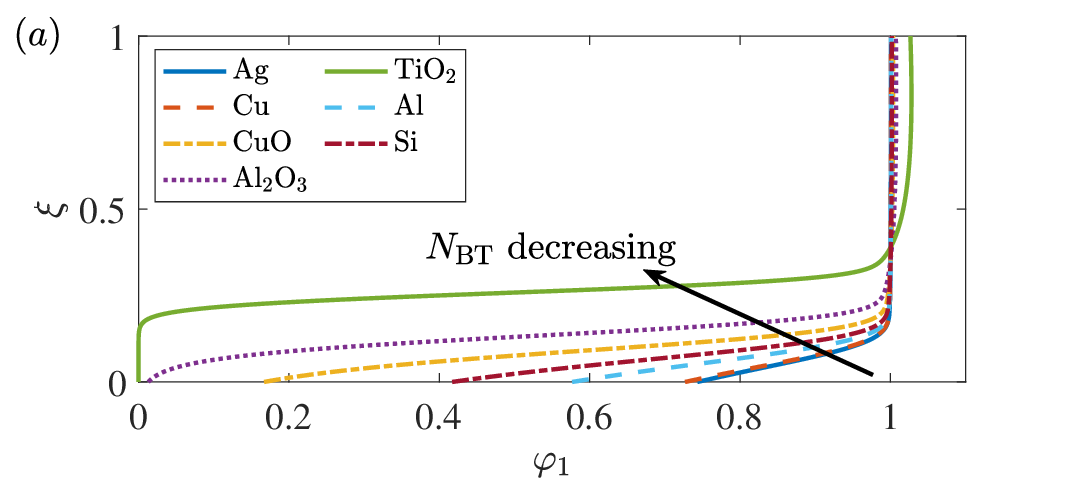}}
\centerline{\includegraphics[width=0.75\textwidth]{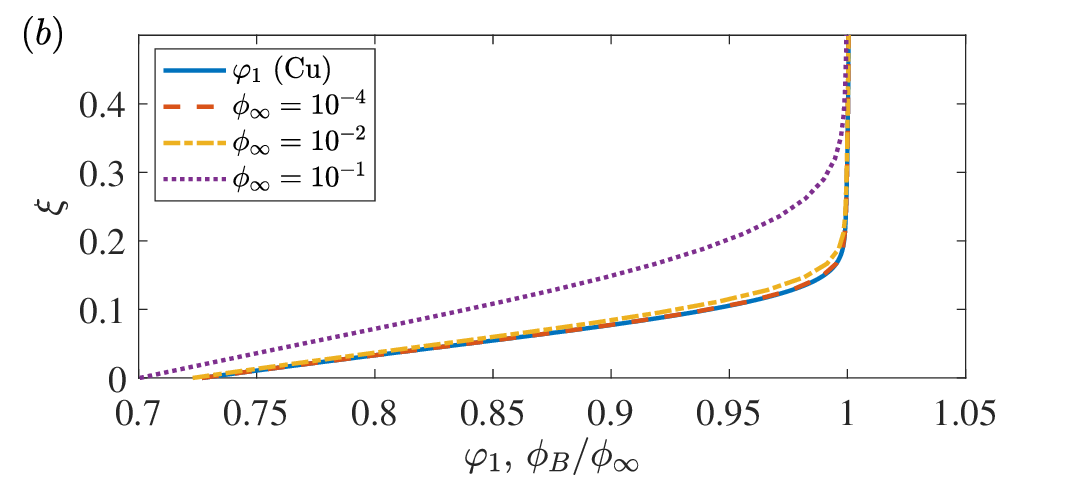}}
\centerline{\includegraphics[width=0.75\textwidth]{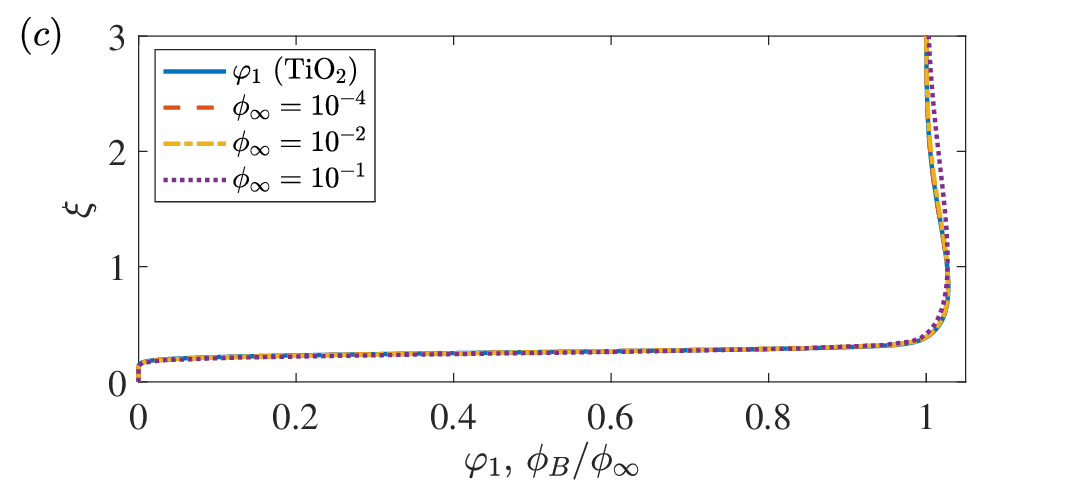}}
\caption{(\emph{a}) Scaled profile of the nanoparticle volume concentration $\varphi_1$ in the limit $\phi_{\infty}\rightarrow 0$, for different nanoparticle materials. (\emph{b, c}) Comparisons between the limiting solution $\varphi_1$ and numerical solutions $\phi_B/\phi_{\infty}$ for $\phi_{\infty}=10^{-4}$, $\phi_{\infty}=10^{-2}$, and $\phi_{\infty}=10^{-1}$, for copper (Cu) and titanium oxide (TiO$_2$) nanoparticles.}\label{Fig7}
\end{figure*}

Substituting the solution of \eqref{LowPhiLimit_Gov1} into \eqref{LowPhiLimit_Gov2} establishes the limiting solutions for $\varphi_1$, which are presented in figure~\ref{Fig7}(\emph{a}) for all seven nanoparticle materials given in table~\ref{Table1}. These solutions illustrate the influence of the Brownian motion to thermophoresis ratio $N_{\text{BT}}$ on the behaviour of the concentration layer. As $N_{\text{BT}}$ decreases, the concentration layer becomes thicker. Notably, the solution corresponding to titanium oxide (TiO$_2$), represented by the green solid line, exhibits an overshoot near the wall, where $\varphi_1>1$ before approaching the free-stream value for larger $\xi$ (beyond the range shown in figure~\ref{Fig7}(\emph{a})). Conversely, as $N_{\text{BT}}$ increases and Brownian motion dominates diffusion effects, the nanoparticle volume concentration $\varphi_1\rightarrow 1$ for all $\xi$, indicating a uniform concentration profile across the boundary layer. 

Figures~\ref{Fig7}(\emph{b}) and~\ref{Fig7}(\emph{c}) compare the limiting solution $\varphi_1$ and numerical solutions $\phi_B/\phi_{\infty}$ for $\phi_{\infty}\in[10^{-4},10^{-1}]$, for copper (Cu) and titanium oxide (TiO$_2$) nanoparticles, respectively. In both cases, the numerical solution converges to the limiting profile $\varphi_1$ as $\phi_{\infty}\rightarrow 0$. Indeed, significant deviations only emerge for $\phi_{\infty}=10^{-1}$.

\subsection{The concentration layer}

The base flow profiles in figures~\ref{Fig3} and~\ref{Fig7} reveal a thin concentration layer within the boundary layer, similar to the particle concentration layer reported by \citet{Pelekasis1995} for the flow of a well-mixed particle suspension past a flat plate. As $Sc\rightarrow\infty$, the concentration layer narrows. Since $U_B\sim Y$ as $Y\rightarrow 0$, the following transformations are introduced to balance the diffusion and convection terms in equation~\eqref{Volume3}:
\begin{subequations}\label{ConTransformation}
\begin{equation}
Y = {Sc}^{-1/3}\bar{Y}, 
\quad 
U_B = {Sc}^{-1/3}\bar{U}_B, 
\quad V_B = {Sc}^{-2/3}\bar{V}_B,
\tag{\theequation \emph{a-c}}
\end{equation}
\end{subequations} 
which gives the re-scaled concentration equation
\begin{equation}\label{ConVolume1}
\frac{\partial(\phi_B \bar{U}_B)}{\partial x} + \frac{\partial(\phi_B \bar{V}_B)}{\partial \bar{Y}} = \frac{\partial}{\partial \bar{Y}}\left(T_B\frac{\partial\phi_B}{\partial \bar{Y}}+\frac{\phi_B}{N_{\textrm{BT}}T_B}\frac{\partial T_B}{\partial \bar{Y}}\right).
\end{equation}
Thus, the concentration layer has a characteristic thickness of $O({\Rey}^{-1/2}{Sc}^{-1/3})$. 

Substituting \eqref{ConTransformation} into \eqref{Continuity3}-\eqref{Energy3}, with $Le\rightarrow\infty$ and
\begin{equation}
\phi_B=\phi_{\infty}+\frac{\psi(x,\bar{Y})}{{Sc}^{1/3}},
\end{equation}
gives to leading-order
\begin{subequations}\label{ConGov1}
\begin{equation}
\frac{\partial \bar{U}_B}{\partial x} + \frac{\partial \bar{V}_B}{\partial\bar{Y}} = 0, 
\quad
\frac{\partial^2\bar{U}_B}{\partial\bar{Y}^2} = 0, 
\quad 
\frac{\partial^2T_B}{\partial\bar{Y}^2} = 0.  
\tag{\theequation \emph{a-c}}
\end{equation}
\end{subequations}
The leading-order term in the concentration equation~\eqref{ConVolume1} is also given by~(\ref{ConGov1}\emph{a}). 
Thus,
\begin{subequations}
\begin{equation}
\bar{U}_B = \frac{\hat{\lambda}\bar{Y}}{x^{1/2}}, 
\quad 
\bar{V}_B = \frac{\hat{\lambda}\bar{Y}^2}{4x^{3/2}}, 
\quad 
T_B = T_w + \frac{\hat{\sigma}\bar{Y}}{{Sc}^{1/3}x^{1/2}},
\tag{\theequation \emph{a-c}}
\end{equation}
\end{subequations}
for $\hat{\lambda}=\rho_{w}f''(0)$ and $\hat{\sigma}=\rho_{w}\theta'(0)$.

The next order term in the concentration equation~\eqref{ConVolume1} is given as 
\begin{subequations}
\begin{equation}
\bar{U}_B\frac{\partial\psi}{\partial x} + \bar{V}_B\frac{\partial\psi}{\partial\bar{Y}} = T_w\frac{\partial^2\psi}{\partial\bar{Y}^2},    
\end{equation}
with boundary conditions
\begin{equation}
\frac{\partial\psi}{\partial\bar{Y}} +\frac{\phi_{\infty}\hat{\sigma}}{N_{\text{BT}}T_w^2x^{1/2}}=0 
\quad \textrm{on} \quad \bar{Y}=0
\end{equation}
and
\begin{equation}
\psi\rightarrow 0 \quad \textrm{as} \quad \bar{Y}\rightarrow \infty.    
\end{equation}
\end{subequations}

\begin{figure*}
\centerline{\includegraphics[width=0.75\textwidth]{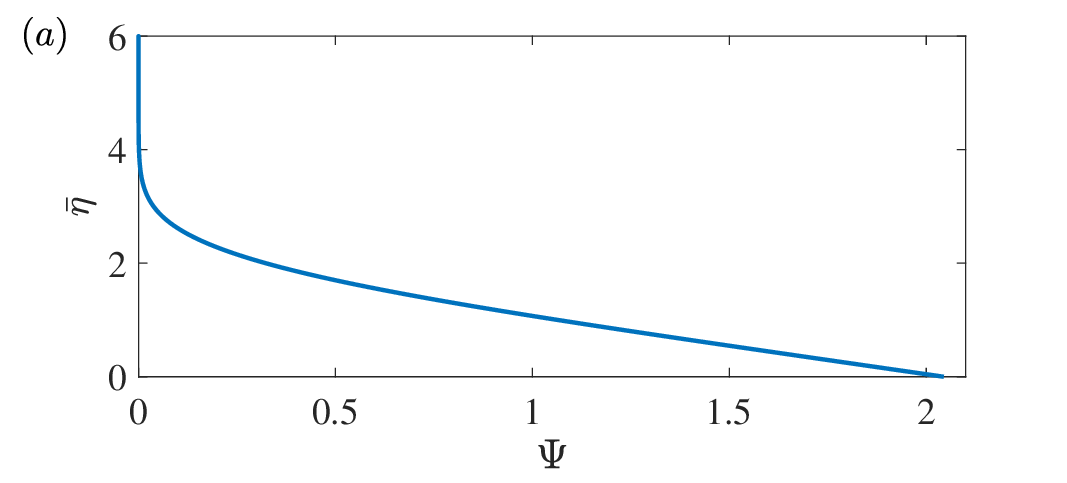}}
\centerline{
\includegraphics[width=0.5\textwidth]{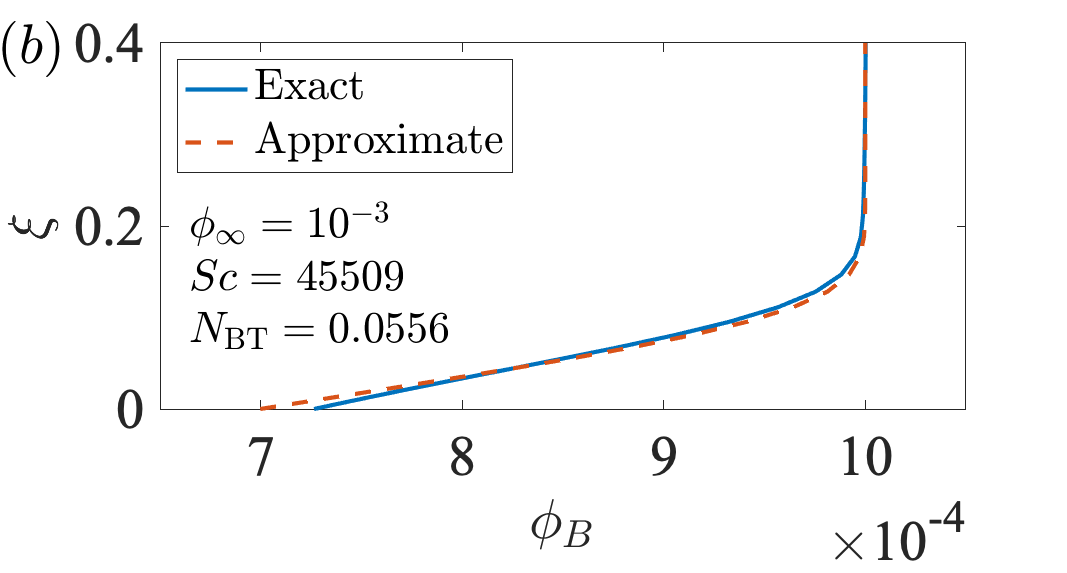}
\includegraphics[width=0.5\textwidth]{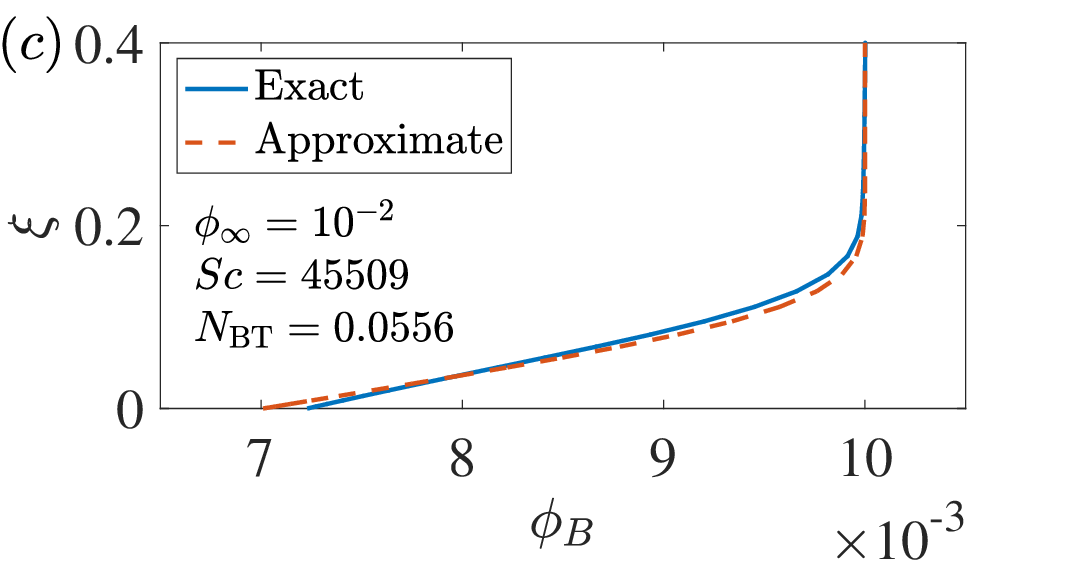}}
\centerline{
\includegraphics[width=0.5\textwidth]{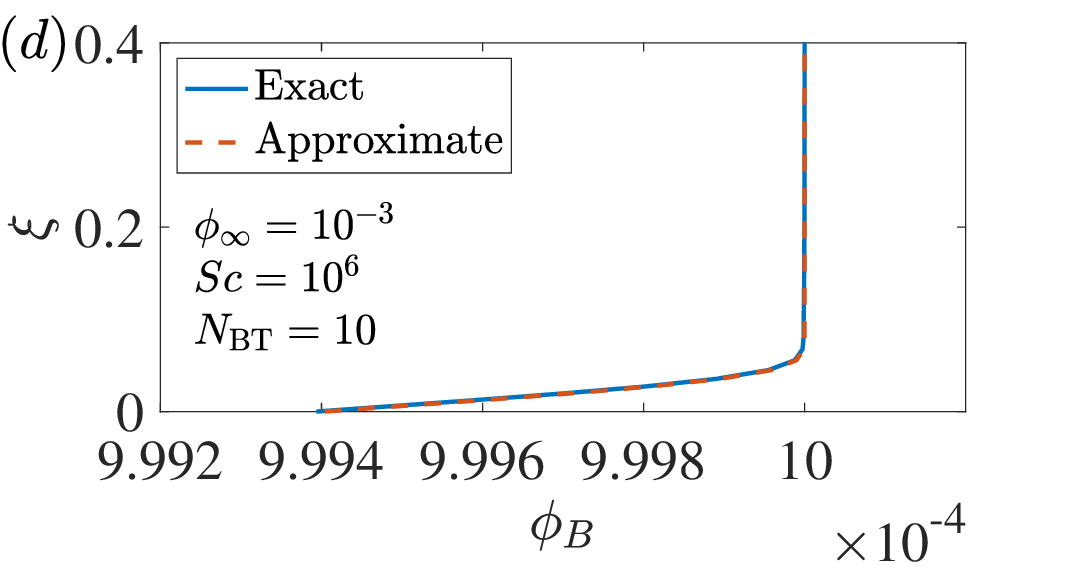}
\includegraphics[width=0.5\textwidth]{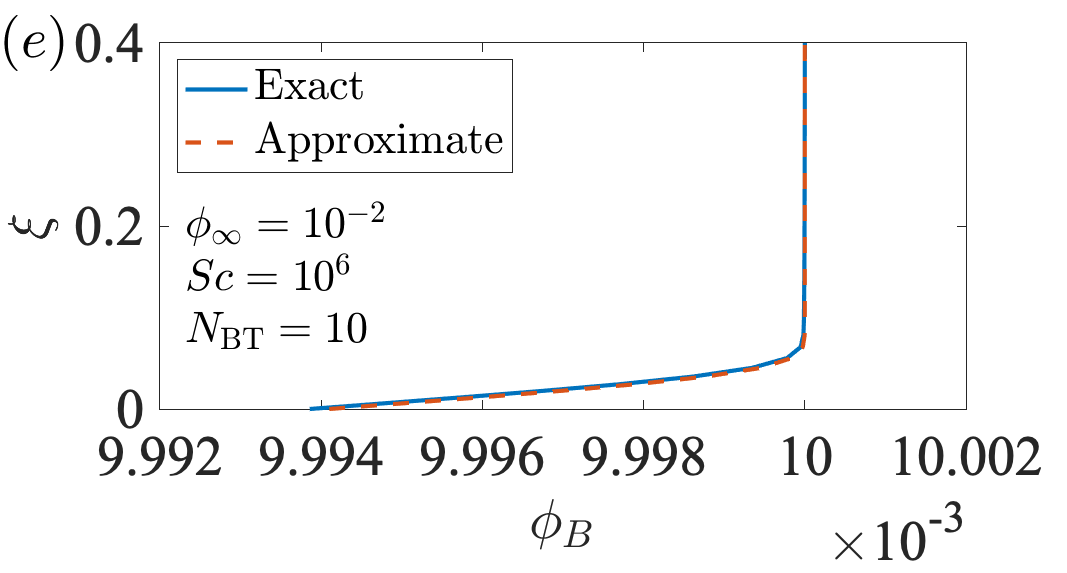}}
\caption{(\emph{a}) Similarity solution $\Psi$ for the nanoparticle volume concentration, as given by~\eqref{Gamma_Solution}. (\emph{b-e}) Nanoparticle volume concentration profiles $\phi_B$ given by the exact solution to equations~\eqref{Gov4} (solid blue lines) and the approximate solution~\eqref{Approx_Sol_Concentration} (dashed red), for copper (Cu) nanoparticles.}\label{Fig8}
\end{figure*}

Introducing the similarity transformation
\begin{subequations}
\begin{equation}
\psi(x,\bar{Y})
=\frac{\phi_{\infty}\hat{\sigma}\Psi(\bar{\eta})}{N_{\textrm{BT}}\hat{\lambda}^{1/3}T_w^{5/3}}, 
\end{equation}
for 
\begin{equation}
\bar{\eta} = \left(\frac{\hat{\lambda}}{T_w}\right)^{1/3}\frac{\bar{Y}}{x^{1/2}},
\end{equation}
\end{subequations}
gives the similarity equation
\begin{subequations}\label{Psi_Simsolution}
\begin{equation}
\frac{\textrm{d}^2\Psi}{\textrm{d}\bar{\eta}^2}+\frac{\bar{\eta}^2}{4}\frac{\textrm{d}\Psi}{\textrm{d}\bar{\eta}}=0,   
\end{equation}
with boundary conditions
\begin{equation}
\frac{\textrm{d}\Psi}{\textrm{d}\bar{\eta}}=-1 \quad \textrm{on} \quad \bar{\eta}=0
\end{equation}
and
\begin{equation}
\Psi\rightarrow 0 \quad \textrm{as} \quad \bar{\eta}\rightarrow \infty.
\end{equation}
\end{subequations}
The solution for $\Psi$ is given in terms of the upper incomplete Gamma function $\Gamma$:
\begin{equation}\label{Gamma_Solution}
\Psi(\bar{\eta}) = \left(\frac{2}{3}\right)^{2/3}\Gamma\left(\frac{1}{3},\frac{\bar{\eta}^3}{12}\right),
\end{equation}
and is plotted in figure~\ref{Fig8}(\emph{a}). At the wall, $\Psi(0) \approx 2.0444$. Hence, to a first approximation, the nanoparticle volume concentration is given by
\begin{equation}\label{Approx_Sol_Concentration}
\phi_B = \phi_{\infty}\left(1+\frac{\hat{\sigma}\Psi(\bar{\eta})}{N_{\textrm{BT}}\hat{\lambda}^{1/3}T_w^{5/3}{Sc}^{1/3}}
\right).
\end{equation}

Figures~\ref{Fig8}(\emph{b}) and~\ref{Fig8}(\emph{c}) compare the exact nanoparticle volume concentration profiles $\phi_B$, obtained by solving~\eqref{Gov4}, with the approximate solution given by~\eqref{Approx_Sol_Concentration}, for copper (Cu) nanoparticles and $T_w=2$. Results are plotted for $\phi_{\infty}=10^{-3}$ and $\phi_{\infty}=10^{-2}$. In both cases, the approximate solution is qualitatively similar to the exact solution, with only minor differences near the wall, corresponding to a maximum relative error of about 3\%. Such small differences are to be expected since $N_{\textrm{BT}}Sc^{1/3}\sim O(1)$ for the parameter settings used in figures ~\ref{Fig8}(\emph{b}) and~\ref{Fig8}(\emph{c}). For materials with smaller $N_{\textrm{BT}}$ values, such as alumina (Al$_2$O$_3$) and titanium oxide (TiO$_2$), the approximation is less accurate, and higher-order terms are required to improve the solution. However, by increasing both $Sc$ and $N_{\textrm{BT}}$, as is modelled in figures~\ref{Fig8}(\emph{d}) and~\ref{Fig8}(\emph{e}), the agreement between the exact and approximate solutions improves significantly, with the maximum relative error reduced to 0.001\%.


\section{Linear stability analysis}

\subsection{Linearised stability equations}

The linear stability equations are derived by decomposing the total velocity, pressure, temperature, and nanoparticle volume concentration fields as
\begin{subequations}\label{Perturbation1}
\begin{equation}
\begin{alignedat}{3}
u &{}={} U_B + \epsilon \tilde{u}, \quad 
&v&{}={} {\Rey}^{-1/2}V_B + \epsilon \tilde{v}, \quad
&w&{}={} \epsilon \tilde{w}, \\
p &{}={} \epsilon \tilde{p}, \quad
&T&{}={} T_B+\epsilon \tilde{T}, \quad
&\phi&{}={} \phi_B+\epsilon \tilde{\phi},
\end{alignedat}
\tag{\theequation \emph{a-f}}
\end{equation}
\end{subequations}
for perturbations $\tilde{\boldsymbol{q}}=(\tilde{\boldsymbol{u}}, \tilde{p}, \tilde{T}, \tilde{\phi})$, with $\tilde{\boldsymbol{u}}=(\tilde{u}, \tilde{v}, \tilde{w})$ and $\epsilon\ll 1$. Similarly, 
\begin{subequations}\label{Perturbation2}
\begin{equation}
\begin{alignedat}{3}
\rho &= \rho_B+\epsilon\tilde{\rho}, \quad
&\rho c &{}={} (\rho c)_B+\epsilon\tilde{\rho}\tilde{c}, \quad
&c &{}={} c_B+\epsilon\tilde{c}, \\
\mu &{}={} \mu_B+\epsilon\tilde{\mu}, \quad
&k &{}={} k_B+\epsilon\tilde{k}.
\end{alignedat}
\tag{\theequation \emph{a-e}}
\end{equation}
\end{subequations}
Here, base flow quantities $\boldsymbol{Q}_B=(U_B, V_B, T_B, \phi_B)$ depend on $x$ and $y$, while perturbations $\tilde{\boldsymbol{q}}$ are functions of $\boldsymbol{x}$ and $t$. Substituting \eqref{Perturbation1} and \eqref{Perturbation2} into \eqref{Gov2}, and linearising in $\epsilon$, gives the following linear stability equations
\begin{subequations}\label{LinearNavierStokes1}
\begin{equation}
\rho_B\nabla\cdot\tilde{\boldsymbol{u}} + \frac{\partial\tilde{\rho}}{\partial t}+U_B\frac{\partial\tilde{\rho}}{\partial x}+\rho_{B,y}\tilde{v} = 
g_1(V_B,\boldsymbol{Q}_{B,x}),
\end{equation}
\begin{multline}
\rho_B\left(\frac{\partial\tilde{u}}{\partial t}+U_B\frac{\partial \tilde{u}}{\partial x}+U_{B,y}\tilde{v}\right) = - \frac{\partial\tilde{p}}{\partial x}
+ \frac{1}{\Rey}\left(\mu_B\bigg(\nabla^2\tilde{u} + \frac{1}{3}\frac{\partial}{\partial x}\nabla\cdot\tilde{\boldsymbol{u}}\right) \\
+ \mu_{B,y}\left(\frac{\partial\tilde{v}}{\partial x} + \frac{\partial\tilde{u}}{\partial y}\right) + U_{B,yy}\tilde{\mu}  + U_{B,y}\frac{\partial\tilde{\mu}}{\partial y}\Bigg) 
+ g_2(V_B,\boldsymbol{Q}_{B,x}),
\end{multline}
\begin{multline}
\rho_B\left(\frac{\partial\tilde{v}}{\partial t}+U_B\frac{\partial \tilde{v}}{\partial x}\right) = - \frac{\partial\tilde{p}}{\partial y}
+ \frac{1}{\Rey}\left(\mu_B\bigg(\nabla^2\tilde{v} + \frac{1}{3}\frac{\partial}{\partial y}\nabla\cdot\tilde{\boldsymbol{u}}\right) \\ 
+ \frac{2\mu_{B,y}}{3}\left(2\frac{\partial\tilde{v}}{\partial y} - \left(\frac{\partial\tilde{u}}{\partial x}+\frac{\partial\tilde{w}}{\partial z}\right)\right) + U_{B,y}\frac{\partial\tilde{\mu}}{\partial x}\Bigg) 
+ g_3(V_B,\boldsymbol{Q}_{B,x}),
\end{multline}
\begin{multline}
\rho_B\left(\frac{\partial\tilde{w}}{\partial t}+U_B\frac{\partial \tilde{w}}{\partial x}\right) = - \frac{\partial\tilde{p}}{\partial z}
+ \frac{1}{\Rey}\Bigg(\mu_B\left(\nabla^2\tilde{w} + \frac{1}{3}\frac{\partial}{\partial z}\nabla\cdot\tilde{\boldsymbol{u}}\right)
\\
+ \mu_{B,y}\left(\frac{\partial\tilde{v}}{\partial z} + \frac{\partial\tilde{w}}{\partial y}\right)\Bigg)
+ g_4(V_B,\boldsymbol{Q}_{B,x}),
\end{multline}
\begin{multline}
\rho_BT_B\left(\frac{\partial\tilde{c}}{\partial t}+U_B\frac{\partial\tilde{c}}{\partial x}+c_{B,y}\tilde{v}\right)
+ (\rho c)_B\left(\frac{\partial\tilde{T}}{\partial t}+U_B\frac{\partial\tilde{T}}{\partial x}+T_{B,y}\tilde{v}\right) \\ 
= \frac{1}{\Rey \Pran}\left(\frac{\partial}{\partial y}\left(k_B\frac{\partial\tilde{T}}{\partial y}+T_{B,y}\tilde{k}\right) + k_B\widehat{\nabla}^2\tilde{T}\right) \\
+ \frac{1}{\Rey \Pran Le}\left(T_{B,y}\mathcal{A} + \mathcal{B}\frac{\partial\tilde{T}}{\partial y}\right) 
+ g_5(V_B,\boldsymbol{Q}_{B,x}),
\end{multline}
\begin{multline}
\phi_B\nabla\cdot\tilde{\boldsymbol{u}}+\frac{\partial\tilde{\phi}}{\partial t}+U_B\frac{\partial\tilde{\phi}}{\partial x}+\phi_{B,y}\tilde{v} \\ = 
\frac{1}{\Rey Sc}\left(\frac{\partial\mathcal{A}}{\partial y}+T_B\widehat{\nabla}^2\tilde{\phi} + \frac{\phi_B}{N_{\textrm{BT}}T_B}\hat{\nabla}^2\tilde{T} \right) 
+ g_6(V_B,\boldsymbol{Q}_{B,x}),
\end{multline}
\end{subequations}
where functions $g_{\star}$ depend on the wall-normal velocity $V_B$ and $x$-derivatives of the base flow $\boldsymbol{Q}_B$, and
\[\mathcal{A} = T_{B}\frac{\partial\tilde{\phi}}{\partial y}+\phi_{B,y}\tilde{T}+\frac{1}{N_{\textrm{BT}}T_B}\left(\phi_B\frac{\partial\tilde{T}}{\partial y}+T_{B,y}\tilde{\phi}-\frac{\phi_BT_{B,y}}{T_B}\tilde{T}\right),\]
\[\mathcal{B} = \phi_{B,y}T_B+\frac{\phi_BT_{B,y}}{N_{\textrm{BT}}T_B},\]
and
\[\widehat{\nabla}^2=\frac{\partial^2}{\partial x^2}+\frac{\partial^2}{\partial z^2}.\]
(The exact form of the functions $g_{\star}$ are given in appendix~\ref{AppendixB}.) The corresponding boundary conditions are given as
\begin{subequations}\label{LinearStabilityBoundaryConditions1}
\begin{equation}
\tilde{u}=\tilde{v}=\tilde{w}=\tilde{T}=\mathcal{A}=0 \quad \textrm{on} \quad y=0, \tag{\theequation \emph{a-e}}
\end{equation}
and
\begin{equation}
\tilde{u}\rightarrow 0, \; \tilde{v}\rightarrow 0, \; \tilde{w}\rightarrow 0, \; \tilde{p}\rightarrow 0, \; \tilde{T}\rightarrow 0, \; \tilde{\phi}\rightarrow 0 \quad \textrm{as} \quad y\rightarrow\infty. \tag{\theequation \emph{f-k}}  
\end{equation}
\end{subequations}

The length scale $L^*$ used in the subsequent linear stability analysis is based on the displacement thickness $\delta_1^*$, to give the Reynolds number
\begin{equation}\label{ReynoldsNumberDisplacement}
R=\frac{U_{\infty}^*\delta_1^*\rho_{bf}^*}{\mu_{bf}^*},
\end{equation}
which ensures consistency with earlier investigations~\citep{Mack1984,Schmid2001}. This gives the following relationships $R = \delta_1\Rey_x^{1/2}$ and $R = \delta_1(x\Rey)^{1/2}$. Consequently, $\Rey$ in the system of equations~\eqref{LinearNavierStokes1} is replaced with $R$. 

Additionally, the parallel flow approximation is imposed, where the flow is assumed to be in the $x$-direction and depends only on the wall-normal $y$-direction, i.e., $g_{\star}=0$. Subsequently, perturbations $\tilde{\boldsymbol{q}}$ are  decomposed into the normal mode form
\begin{equation}
\tilde{\boldsymbol{q}}(\boldsymbol{x},t) = \breve{\boldsymbol{q}}(y)\exp{(\textrm{i}(\alpha x+\beta z-\omega t))} + \textrm{c.c},
\end{equation}
(and similarly for quantities $\tilde{\rho}$, $\tilde{\mu}$, etc.) for a streamwise wavenumber $\alpha\in\mathbb{R}$, spanwise wavenumber $\beta\in\mathbb{R}$, and frequency $\omega\in\mathbb{C}$. Here, $\textrm{c.c}$ denotes the complex conjugate. Consequently, equations \eqref{LinearNavierStokes1} become
\begin{subequations}\label{LinearNavierStokes2}
\begin{equation}
\rho_B\left(\textrm{i}\left(\alpha\breve{u}+\beta\breve{w}\right)+\textrm{D}\breve{v} \right) + \textrm{i}\left(\alpha U_B-\omega\right)\breve{\rho}+\rho_{B,y}\breve{v} = 0,
\end{equation}
\begin{multline}
\rho_B\left(\textrm{i}\left(\alpha U_B-\omega\right)\breve{u}+U_{B,y}\breve{v}\right) = -\textrm{i}\alpha\breve{p}
+ \frac{1}{R}\Bigg(\mu_B\bigg(\left(\textrm{D}^2 - \left(\alpha^2+\beta^2\right)\right)\breve{u} \\
+ \frac{\textrm{i}\alpha}{3}\left(\textrm{i}\left(\alpha\breve{u}+\beta\breve{w}\right)+\textrm{D}\breve{v}\right)\bigg) + \mu_{B,y}\left(\textrm{i}\alpha\breve{v} + \textrm{D}\breve{u}\right) + \left(U_{B,yy}+U_{B,y}\textrm{D}\right)\breve{\mu}\Bigg),
\end{multline}
\begin{multline}
\textrm{i}\rho_B\left(\alpha U_B-\omega\right)\breve{v} = - \textrm{D}\breve{p} + \frac{1}{R}\Bigg(\mu_B\bigg(\left(\textrm{D}^2 - \left(\alpha^2+\beta^2\right)\right)\breve{v} \\
+ \frac{\textrm{D}}{3}\left(\textrm{i}\left(\alpha\breve{u}+\beta\breve{w}\right) + \textrm{D}\breve{v}\right)\bigg) + \frac{2\mu_{B,y}}{3}\left(2\textrm{D}\breve{v} - \textrm{i}\left(\alpha\breve{u}+\beta\breve{w}\right)\right) + \textrm{i}\alpha U_{B,y}\breve{\mu}\Bigg),
\end{multline}
\begin{multline}
\textrm{i}\rho_B\left(\alpha U_B-\omega\right)\breve{w} = -\textrm{i}\beta\breve{p} + \frac{1}{R}\Bigg(\mu_B\bigg(\left(\textrm{D}^2-\left(\alpha^2+\beta^2\right)\right)\breve{w} \\
+ \frac{\textrm{i}\beta}{3}\left(\textrm{i}\left(\alpha\breve{u}+\beta\breve{w}\right)+\textrm{D}\breve{v}\right)\bigg) + \mu_{B,y}\left(\textrm{i}\beta\breve{v} + \textrm{D}\breve{w}\right)\Bigg),
\end{multline}
\begin{multline}
\rho_BT_B\left(\textrm{i}\left(\alpha U_B-\omega\right)\breve{c}+c_{B,y}\breve{v}\right)
+ (\rho c)_B\left(\textrm{i}\left(\alpha U_B-\omega\right)\breve{T}+T_{B,y}\breve{v}\right) \\ 
= \frac{1}{R \Pran}\left(\textrm{D}\left(k_B\textrm{D}\breve{T}+T_{B,y}\breve{k}\right) - \left(\alpha^2+\beta^2\right) k_B\breve{T}\right)
+ \frac{1}{R \Pran Le}\left(T_{B,y}\mathcal{A} + \mathcal{B}\textrm{D}\breve{T}\right),
\end{multline}
\begin{multline}\label{LinearNavierStokes2f}
\phi_B\left(\textrm{i}\left(\alpha\breve{u}+\beta\breve{w}\right)+\textrm{D}\breve{v}\right)+\textrm{i}\left(\alpha U_B-\omega\right)\breve{\phi}+\phi_{B,y}\breve{v}
\\ 
= \frac{1}{R Sc}\left(\textrm{D}\mathcal{A}-\left(\alpha^2+\beta^2\right)\left(T_B\breve{\phi} + \frac{\phi_B}{N_{\textrm{BT}}T_B}\breve{T}\right) \right),
\end{multline}
\end{subequations}
where $\textrm{D}=\textrm{d}/\textrm{d}y$. The exact form of the perturbed quantities, including $\breve{\rho}$, $\breve{\mu}$, etc., are given in Appendix~\ref{AppendixB}.


\subsection{Numerical methods}

A temporal linear stability analysis was conducted using the Chebyshev collocation method developed by~\citet{Trefethen2000}. Derivatives in the $y$-direction were approximated using Chebyshev matrices, with $N$ Chebyshev mesh points mapped from the semi-infinite physical domain $y \in [0, \infty)$ onto the computational interval $\zeta \in [1,-1]$ via the coordinate transformation
\begin{equation}\label{Cheb-Phys-Transformation}
y = \frac{l(1-\zeta)}{1+\zeta},
\end{equation} 
where $l$ is a stretching parameter.

The linear stability equations \eqref{LinearNavierStokes2} were transformed into the following eigenvalue problem
\begin{equation}
\mathsfbi{A}\breve{\boldsymbol{q}}^T= \omega \mathsfbi{B}\breve{\boldsymbol{q}}^T,
\end{equation} 
\label{eigenvalue}
where $\mathsfbi{A}$ and $\mathsfbi{B}$ are $6N\times 6N$ matrices. The frequencies $\omega$ and the corresponding linear perturbations $\breve{\boldsymbol{q}}$ were then computed using the \texttt{eig} command in \textsc{Matlab}.

\setlength{\tabcolsep}{8pt}
\begin{table}
\begin{center}
\begin{tabular}{ccccc}
        $N$ & $l$ & $\omega \; (\phi_{\infty}=10^{-4})$ & $\omega \; (\phi_{\infty}=10^{-2})$ & $\omega \; (\phi_{\infty}=10^{-1})$ \\ \hline
        $32$ & $2$ & $0.11931-\textrm{i}0.00035$ & $0.11987-\textrm{i}0.00019$ & $0.11141 +\textrm{i}0.00308$ \\
        $64$ & $2$ & $0.11928-\textrm{i}0.00029$ & $0.11844+\textrm{i}0.00008$ & $0.11384 +\textrm{i}0.00151$ \\
        $96$ & $2$ & $0.11929-\textrm{i}0.00028$ & $0.11845+\textrm{i}0.00009$ & $0.11382 +\textrm{i}0.00153$ \\
        $128$ & $2$ & $0.11929-\textrm{i}0.00028$ & $0.11845+\textrm{i}0.00009$ & $0.11382 +\textrm{i}0.00153$ \\
        $32$ & $3$ & $0.11926-\textrm{i}0.00036$ & $0.12153-\textrm{i}0.00029$ & $0.11162 +\textrm{i}0.00313$ \\
        $64$ & $3$ & $0.11928-\textrm{i}0.00029$ & $0.11845+\textrm{i}0.00008$ & $0.11384 +\textrm{i}0.00151$ \\
        $96$ & $3$ & $0.11929-\textrm{i}0.00028$ & $0.11845+\textrm{i}0.00009$ & $0.11383 +\textrm{i}0.00152$ \\
        $128$ & $3$ & $0.11929-\textrm{i}0.00028$ & $0.11845+\textrm{i}0.00009$ & $0.11383 +\textrm{i}0.00151$ \\
        $32$ & $4$ & $0.11925-\textrm{i}0.00033$ & $0.11823+\textrm{i}0.00088$ & $0.11229 -\textrm{i}0.00199$ \\
        $64$ & $4$ & $0.11928-\textrm{i}0.00029$ & $0.11844+\textrm{i}0.00008$ & $0.11382 +\textrm{i}0.00153$ \\
        $96$ & $4$ & $0.11929-\textrm{i}0.00028$ & $0.11844+\textrm{i}0.00008$ & $0.11383 +\textrm{i}0.00152$ \\
        $128$ & $4$ & $0.11929-\textrm{i}0.00028$ & $0.11845+\textrm{i}0.00009$ & $0.11383 +\textrm{i}0.00152$
\end{tabular}
\caption{Frequencies $\omega=\omega_r+\textrm{i}\omega_i$ for variable $N$ and $l$, for $R=500$, $\alpha=0.3$, $\beta=0$, $T_w=2$, and $\phi_{\infty}=10^{-4}$, $\phi_{\infty}=10^{-2}$, and $\phi_{\infty}=10^{-1}$.}
\label{Table3}
\end{center}
\end{table}

Table~\ref{Table3} presents the frequency $\omega$ corresponding to the TS wave for varying values of $N$ and $l$, for copper (Cu) nanoparticles and free-stream nanoparticle volume concentrations $\phi_{\infty}\in[10^{-4},10^{-1}]$. In each case, the Reynolds number $R=500$, the streamwise wavenumber $\alpha=0.3$, the spanwise wavenumber $\beta=0$, and the wall temperature $T_w=2$. The results are identical to four decimal places for all $l$ considered when $N\geq 64$. Therefore, for the remainder of this investigation, $N=96$ Chebyshev mesh points were used with the mapping parameter $l=2$.


\subsection{Numerical results}

In the following linear stability analysis, unless stated otherwise, the nanofluid is composed of copper (Cu) nanoparticles dispersed in a base fluid of water. In addition, the wall temperature $T_w=2$.

\subsubsection{Eigenspectrum}

\begin{figure*}
\centerline{\includegraphics[width=0.5\textwidth]{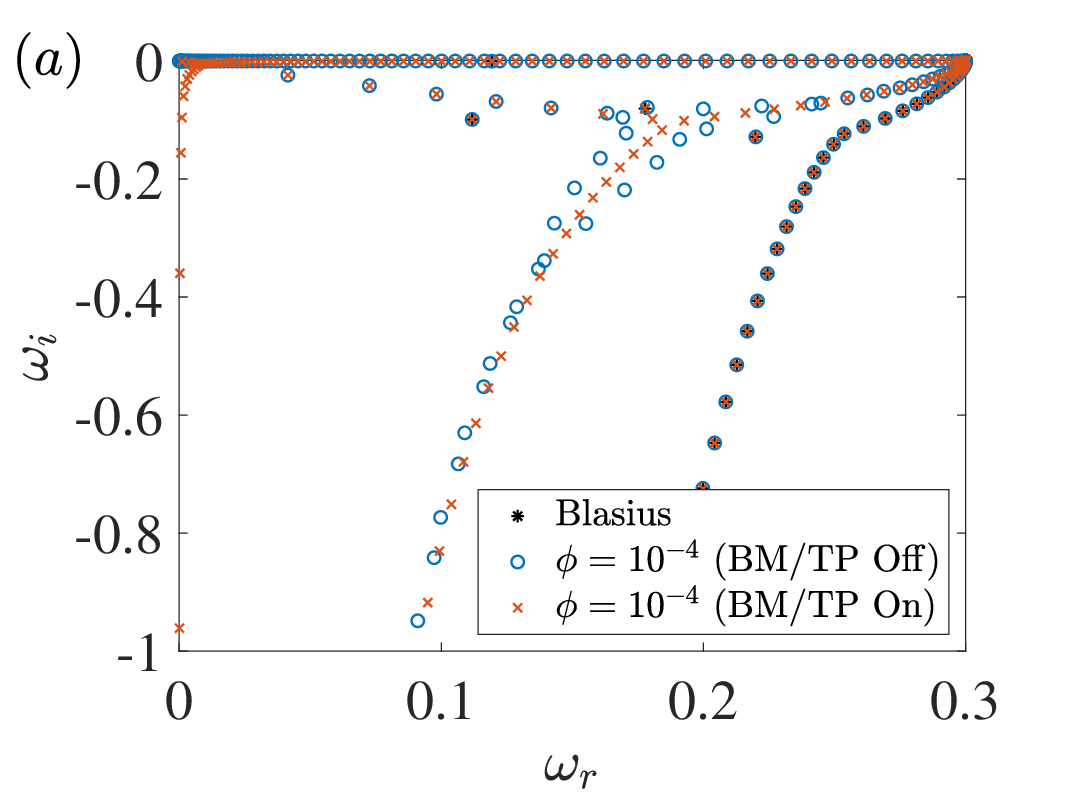}\includegraphics[width=0.5\textwidth]{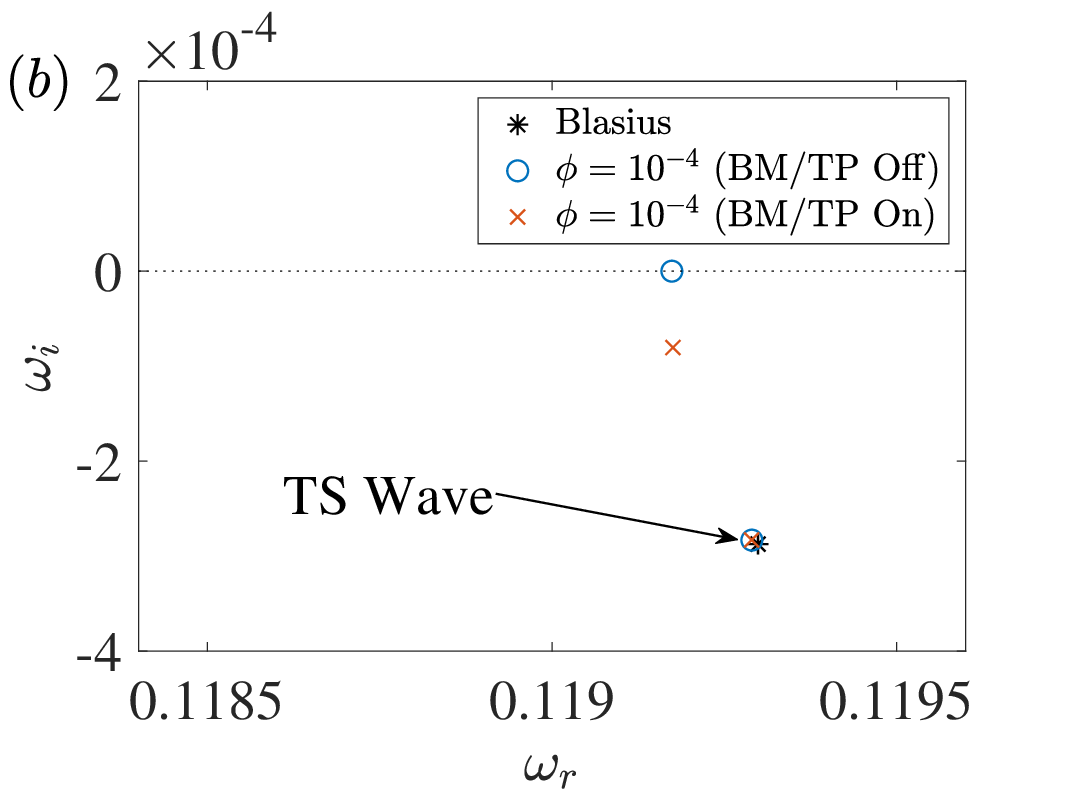}}
\centerline{\includegraphics[width=0.5\textwidth]{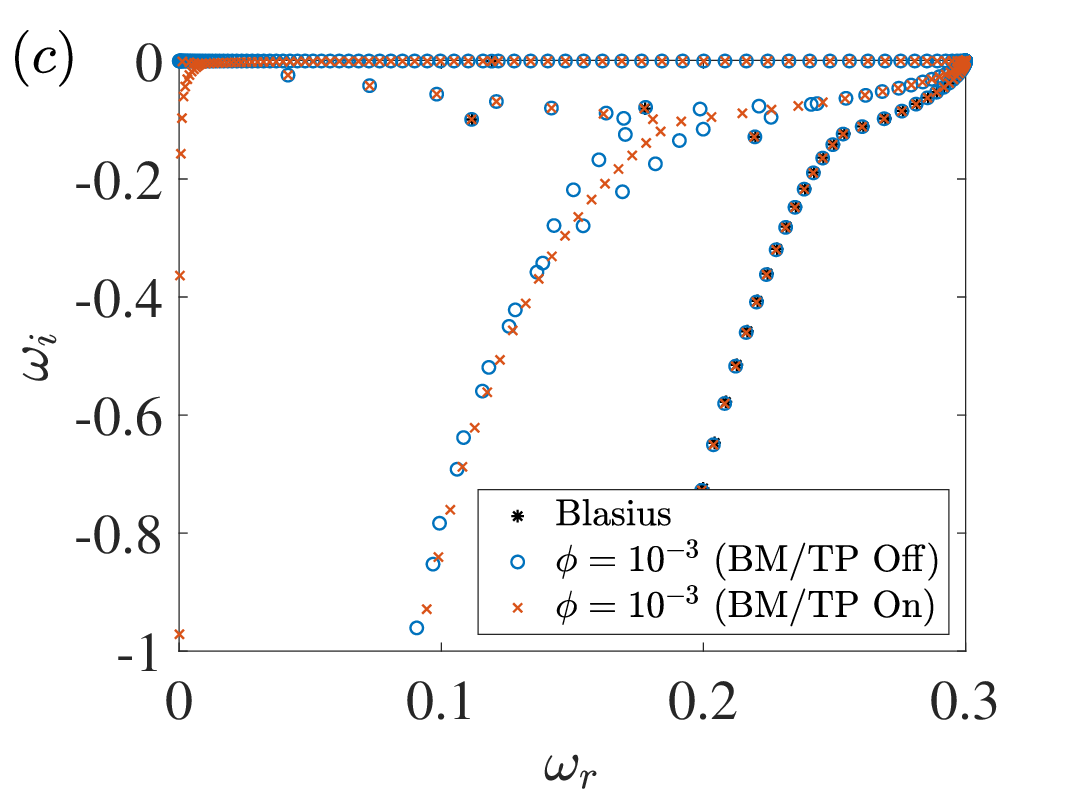}\includegraphics[width=0.5\textwidth]{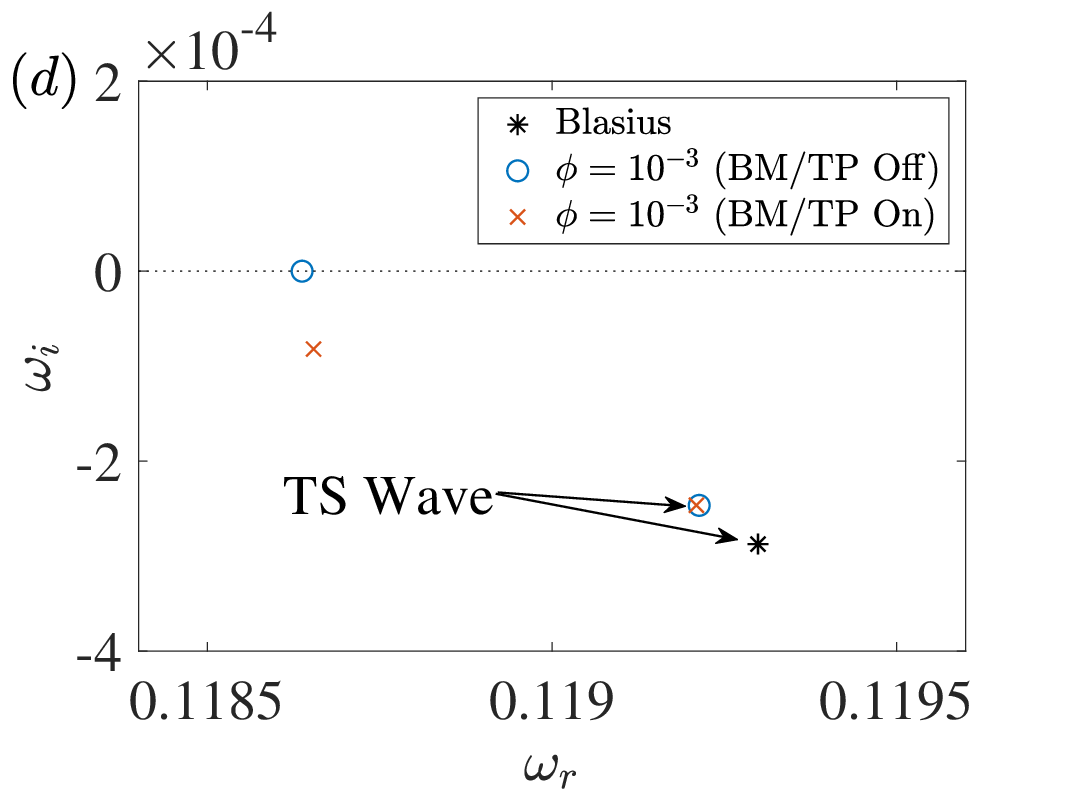}}
\centerline{\includegraphics[width=0.5\textwidth]{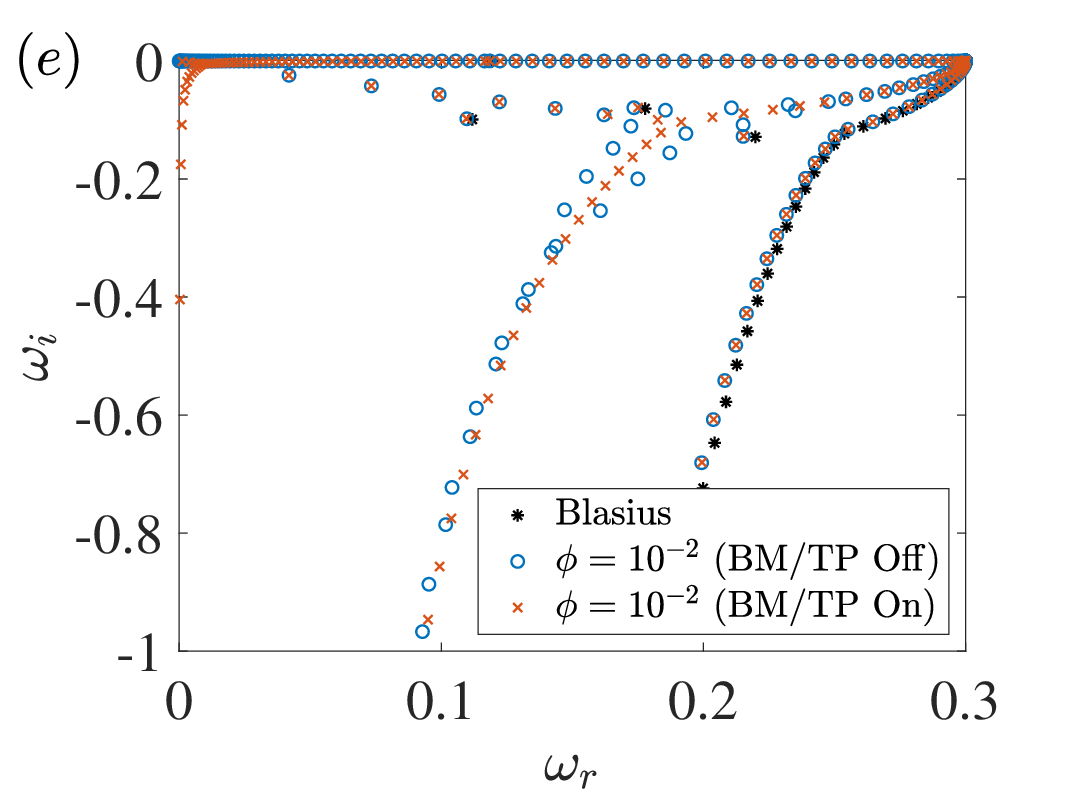}\includegraphics[width=0.5\textwidth]{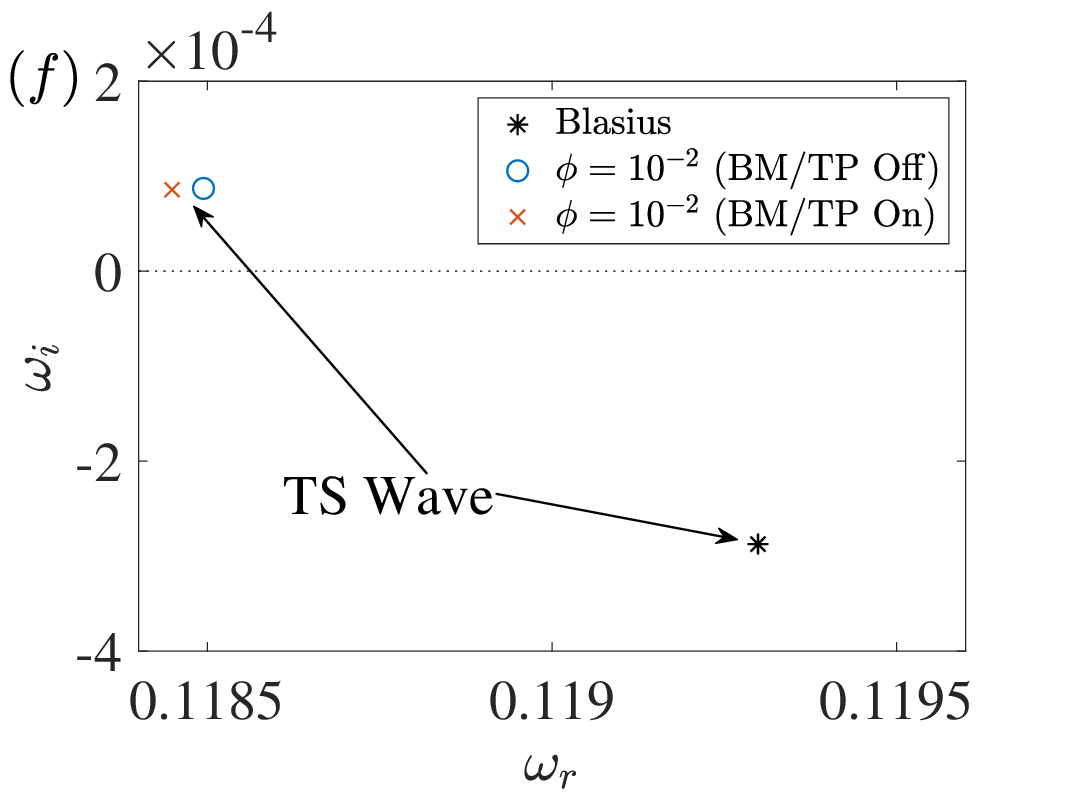}}
\caption{Eigenspectrum in the $(\omega_r,\omega_i)$-plane for $R=500$, $\alpha=0.3$, $\beta=0$, $T_w=2$, and (\emph{a},\emph{b}) $\phi_{\infty}=10^{-4}$, (\emph{c},\emph{d}) $\phi_{\infty}=10^{-3}$, and (\emph{e},\emph{f}) $\phi_{\infty}=10^{-2}$. Black asterisk markers represent solutions of the Blasius flow, while blue circles and red crosses represent solutions of the nanofluid flow without (BM/TP Off) and with (BM/TP On) Brownian motion and thermophoresis.}\label{Fig9}
\end{figure*}

Figure~\ref{Fig9} presents a representative eigenspectrum in the complex $\omega$-plane for the parameter settings $R=500$, $\alpha=0.3$, and $\beta=0$, and three values of $\phi_{\infty}$. For the standard Blasius flow without nanoparticles, these conditions are linearly stable. The left-hand plots display the eigenspectrum on a large scale, while the right-hand plots provide a zoomed-in view. The blue circular markers correspond to solutions where Brownian motion and thermophoresis are ignored, whereas the red crosses indicate the corresponding solutions when these effects are included. The black star markers represent the eigenspectrum for the Blasius flow without nanoparticles, where the nanoparticle volume concentration equations have been removed from the analysis. 

Consistent with previous studies~\citep{Mack1976, GroschSalwen1978, SalwenGrosch1981, Schmid2001}, the eigenspectrum consists of multiple branches. A discrete set of modes are located on the A-branch~\citep{Mack1976} in the upper left-hand corner of figures~\ref{Fig9}(\emph{a,c,e}). This branch contains the TS wave, which is highlighted in the right-hand plots and discussed further below. Additionally, the eigenspectrum features three continuous branches, each associated with different governing equations. (The eigenspectrum shown is a discrete representation of the continuous spectrum, with the resolution governed by the number of Chebyshev mesh points $N$.) The first two branches, approximately aligned with the vertical axis, are associated with the momentum and energy equations, respectively. As the number of Chebyshev mesh points $N$ increases, these two branches shift to the right toward the vertical line $\omega_r\rightarrow\alpha$, although their qualitative behaviour is unchanged. The third continuous branch, associated with the nanoparticle volume concentration equation, runs parallel to the real $\omega$-line but with a negative imaginary part. Like the other two continuous branches, this branch also shifts to the right as $N$ increases, but at a significantly slower rate due to the size of the Schmidt number $Sc$. Notably, when Brownian motion and thermophoresis are neglected, this branch is located along the real $\omega$-line (i.e., $\omega_i=0$), as expected, since equation \eqref{LinearNavierStokes2f} simplifies to 
\begin{equation*}
\left(\alpha U_B-\omega\right)\breve{\phi} = 0
\end{equation*}
in this case.

The zoomed-in plots on the right-hand side of figure~\ref{Fig9} focus on the behaviour of the frequency $\omega$ of the TS wave as the free-stream nanoparticle volume concentration $\phi_{\infty}$ increases. For $\phi_{\infty}=10^{-4}$, the value of $\omega$ closely matches that of the Blasius flow without nanoparticles, with linearly stable conditions, as the imaginary part of $\omega$ is negative. However, as $\phi_{\infty}$ increases, a noticeable shift occurs. At $\phi_{\infty}=10^{-3}$, the frequency $\omega$ shifts slightly to the left and upward in the $\omega$-plane, remaining linearly stable but less stable than the standard Blasius flow. With a further increase to $\phi_{\infty}=10^{-2}$, $\omega$ moves into the upper half-plane, where a positive imaginary part indicates linearly unstable behaviour. Thus, for the given flow conditions, the nanofluid destabilises the TS wave. Furthermore, the differences in $\omega$ obtained with and without the effects of Brownian motion and thermophoresis are minimal, with only slight variations in the real component and no discernible changes in the imaginary component. (In addition to the frequency $\omega$ of the TS wave, eigenspectra from the branch arising from the nanoparticle volume concentration equation are also shown in figures~\ref{Fig9}(\emph{b,d}), further illustrating how this branch aligns with the real $\omega$-axis.)  

\begin{figure*}
\centerline{\includegraphics[width=0.75\textwidth]{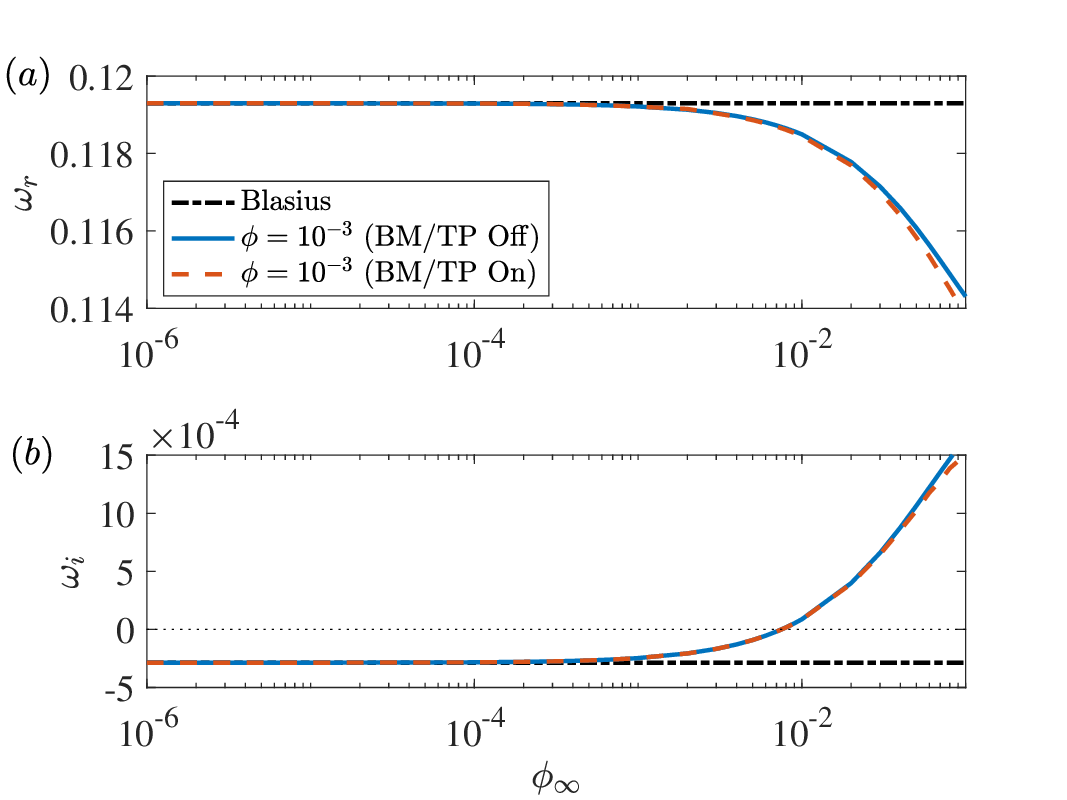}}
\caption{Frequency $\omega=\omega_r+\textrm{i}\omega_i$ as a function of $\phi_{\infty}$ for $R=500$, $\alpha=0.3$, $\beta=0$, and $T_w=2$. (\emph{a}) Real part and (\emph{b}) imaginary part. The solid blue and dashed red lines represent solutions of the nanofluid flow without (BM/TP Off) and with (BM/TP On) Brownian motion and thermophoresis. The horizontal chain lines indicate the corresponding solutions for the Blasius flow without nanoparticles.}\label{Fig10}
\end{figure*}

Figure~\ref{Fig10} further illustrates the variation of the frequency $\omega$ of the TS wave as the free-stream nanoparticle volume concentration $\phi_{\infty}$ increases, for the same conditions as given in figure~\ref{Fig9}. The plots show the evolution of both the real and imaginary components of $\omega$ with increasing $\phi_{\infty}$, supporting the trend observed in figure~\ref{Fig9}. As more nanoparticles are added to the base fluid, the TS wave becomes increasingly destabilised, with the imaginary part of $\omega$ shifting from negative to positive values near $\phi_{\infty}=0.008$, signalling the onset of linear instability. Additionally, solutions demonstrate that the effects of Brownian motion and thermophoresis are negligible, since the differences between cases without (solid blue lines) and with (dashed red) these effects are minimal, with only slight variations in the real part of $\omega$ and no significant impact on the imaginary part.


\subsubsection{Three-dimensional instabilities}

\begin{figure*}
\centerline{\includegraphics[width=0.75\textwidth]{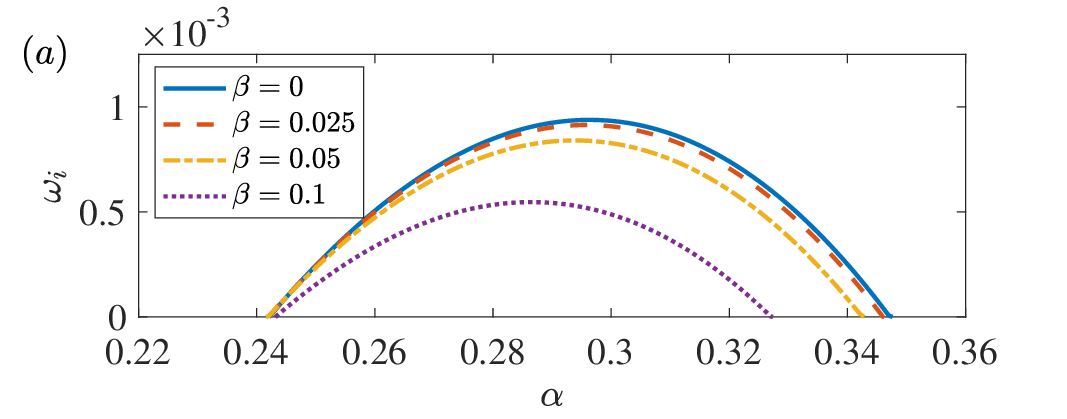}}
\centerline{\includegraphics[width=0.75\textwidth]{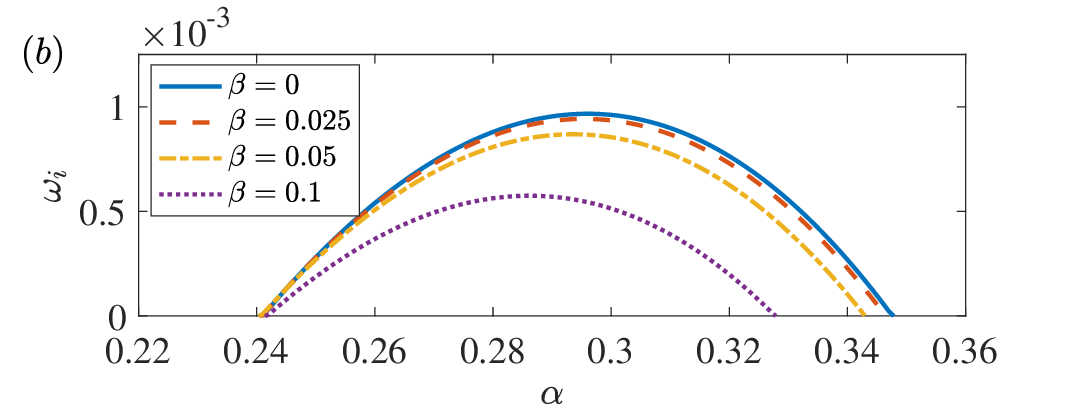}}
\centerline{\includegraphics[width=0.75\textwidth]{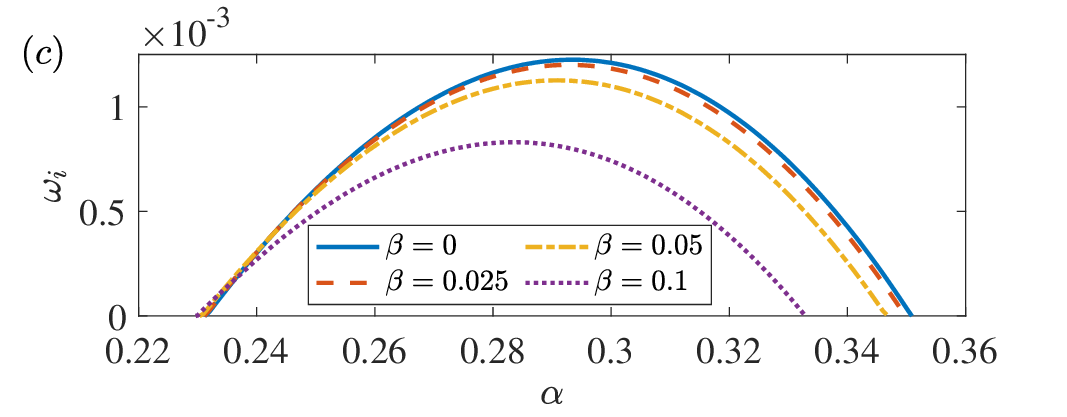}}
\caption{Temporal growth rate $\omega_i$ as a function of the streamwise wavenumber $\alpha$ for $R=600$, $T_w=2$, $\beta\in[0,0.1]$, and (\emph{a}) $\phi_{\infty}=10^{-4}$, (\emph{b}) $\phi_{\infty}=10^{-3}$, and (\emph{c}) $\phi_{\infty}=10^{-2}$.}\label{Fig11}
\end{figure*}

Although Squire’s theorem cannot be applied directly to the full linear stability equations \eqref{LinearNavierStokes2}, it is applicable to the simplified linear stability equations that neglect Brownian motion and thermophoresis. Since these diffusion effects have a minimal impact on both the base flow and the linear stability calculations, we conclude that Squire's theorem is approximately valid for the full equations. Consequently, it is sufficient to limit the stability analysis to two-dimensional instabilities.

This conclusion is supported by the results shown in figure~\ref{Fig11}, which plots the temporal growth rate $\omega_i$ as a function of the streamwise wavenumber $\alpha$, for the Reynolds number $R=600$, spanwise wavenumbers $\beta\in[0,0.1]$, and nanoparticle volume concentrations $\phi_{\infty}\in[10^{-4},10^{-2}]$. The results indicate that $\omega_i$ decreases as $\beta$ increases, confirming that two-dimensional instabilities are more unstable than three-dimensional instabilities. Therefore, based on this and further observations, the remainder of this study focuses on two-dimensional disturbances by setting $\beta = 0$.


\subsubsection{Conditions for neutral stability}

\begin{figure*}
\centerline{\includegraphics[width=0.75\textwidth]{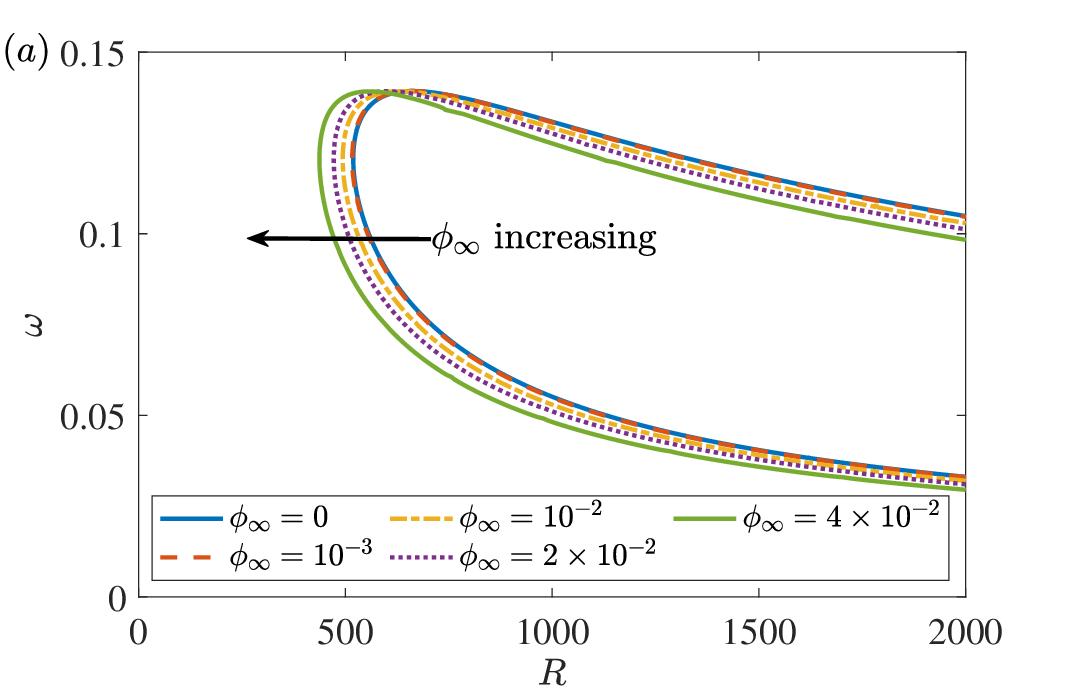}}
\centerline{\includegraphics[width=0.75\textwidth]{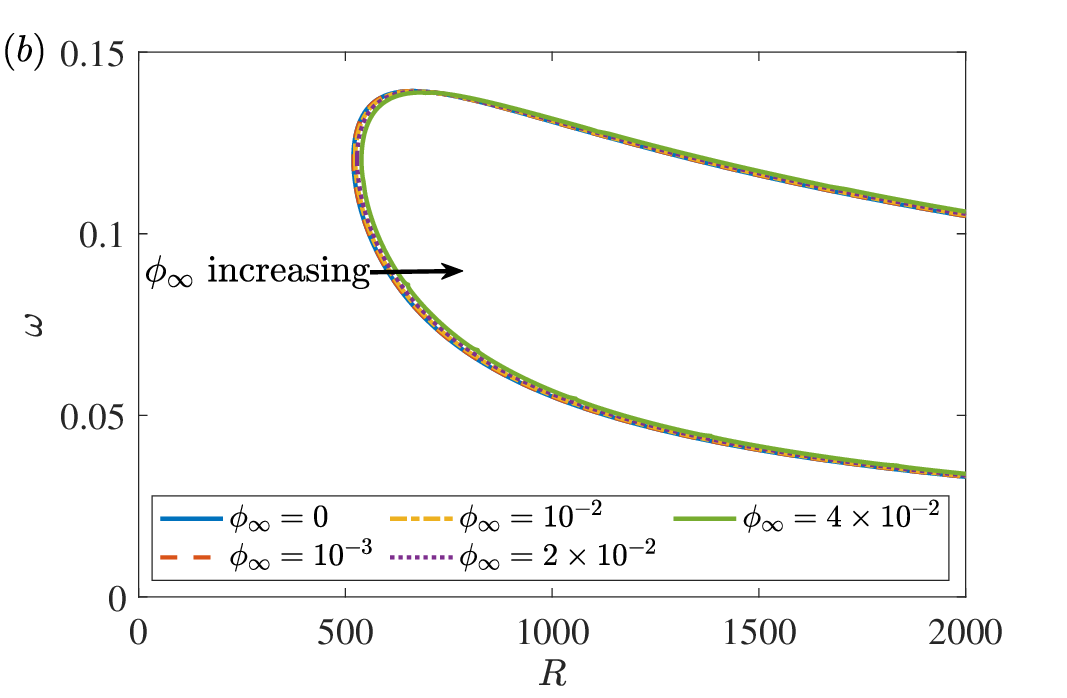}}
\caption{Neutral stability curves in the $(R,\omega)$-plane for variable $\phi_{\infty}$, $\beta=0$, $T_w=2$, and (\emph{a}) copper (Cu) nanoparticles and (\emph{b}) aluminium (Al) nanoparticles.}\label{Fig12}
\end{figure*}

The neutral conditions $(\omega, R)$ for linear instability were computed using streamwise wavenumber increments of $\Delta a=10^{-4}$. To accurately trace the frequency $\omega$ associated with the TS wave within the complex $\omega$-plane, small Reynolds number steps $\Delta R=0.01$ were used. This ensured that the TS frequency was correctly identified, minimising interference with the eigenspectra found on the branch due to the nanoparticle volume concentration equation. The critical Reynolds number for the Blasius flow, in the absence of nanoparticles, was obtained as $R_c\approx 519.4$ for a streamwise wavenumber $\alpha_c\approx 0.304$, frequency $\omega_c\approx 0.121$, and phase speed $s_c=\omega_c/\alpha_c\approx 0.397$, in agreement with previous studies~\citep{Schmid2001}.

Neutral stability curves were obtained for freestream nanoparticle volume concentrations $\phi_{\infty}\in[0,4\times 10^{-2}]$, with solutions for the copper (Cu) nanoparticles shown in figure~\ref{Fig12}(\emph{a}). The destabilisation of the TS wave is further demonstrated, with neutral stability curves shifting horizontally to the left and smaller Reynolds numbers as $\phi_{\infty}$ increases. Notably, there is no discernible vertical variation in the neutral stability curves. Thus, while the critical Reynolds number $R_c$ shrinks, the corresponding frequency $\omega_c$, the streamwise wavenumber $\alpha_c$, and the phase velocity $s_c$, remain relatively constant for the range of $\phi_{\infty}$ considered.

A second set of neutral stability curves is shown in figure~\ref{Fig12}(\emph{b}), but for nanoparticles made of aluminium (Al). Like the copper (Cu) nanoparticles, there is no vertical variation as $\phi_{\infty}$ increases. However, a small stabilising effect is observed, with neutral curves shifting to the right and marginally larger Reynolds numbers $R$. Therefore, the type of material used for the nanoparticles plays a significant role in determining whether the TS wave is stabilised or destabilised. 

\begin{figure*}
\centerline{\includegraphics[width=0.75\textwidth]{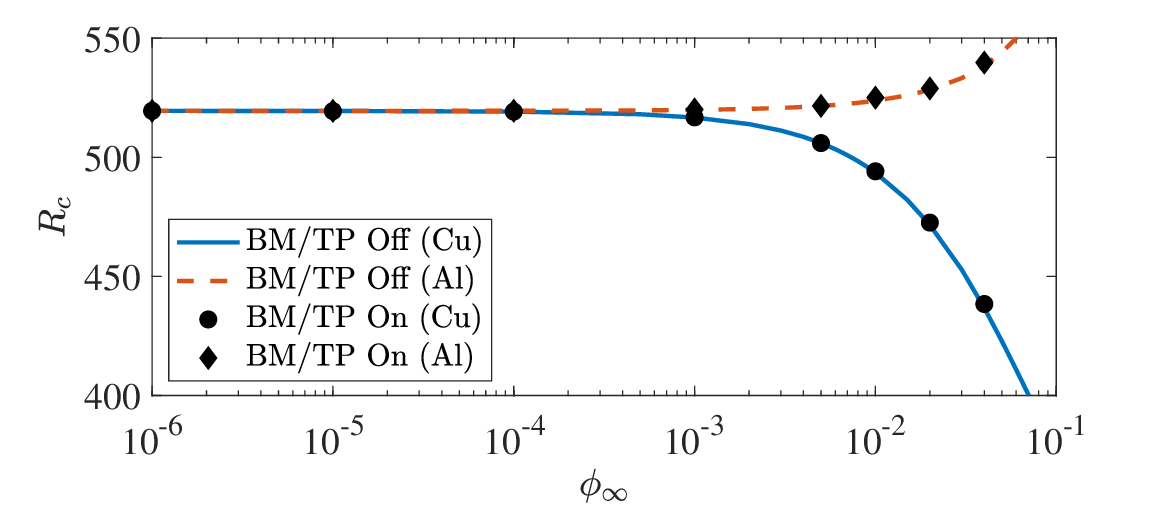}}
\caption{Critical Reynolds number $R_c$ as a function of $\phi_{\infty}$, for copper (Cu) nanoparticles (solid blue line and circular markers) and aluminium (Al) nanoparticles (dashed red line and diamond markers) in a base fluid of water without (BM/TP Off) and with (BM/TP On) Brownian motion and thermophoresis.}\label{Fig13}
\end{figure*}

Figure~\ref{Fig13} presents further evidence of the stabilising benefits of aluminium (Al) nanoparticles compared to the destabilising effects of copper (Cu) nanoparticles. The circular (Cu) and diamond (Al) markers indicate the critical Reynolds numbers $R_c$ obtained from the full linear stability equations~\eqref{LinearNavierStokes2}, with a noticeable reduction in $R_c$ for copper (Cu) nanoparticles and a small increase for aluminium (Al) nanoparticles. Additionally, the critical Reynolds number $R_c$ for these two types of nanoparticles is plotted when Brownian motion and thermophoresis are neglected, as represented by the solid blue and dashed red curves. In this case, the critical Reynolds number $R_c=\mu\widehat{R}_c/\rho$, where $\widehat{R}_c\approx 519.4$ is the critical Reynolds number for the Blasius flow without nanoparticles. Thus, using the definition for density $\rho$ and the Brinkman dynamic viscosity $\mu$, given by \eqref{Density} and \eqref{NondimBrinkman} respectively, the critical Reynolds for the nanofluid flow is approximated as
\begin{equation}\label{CriticalRePrediction}
R_c = \frac{519.4}{(1-\phi_{\infty})^{2.5}(1+(\hat{\rho}-1)\phi_{\infty})}.   
\end{equation}
Unsurprisingly, the results with and without Brownian motion and thermophoresis are nearly identical. Thus, the impact of these diffusion effects on the linear stability of the nanofluid flow are negligible. Table~\ref{Table4} lists critical Reynolds numbers $R_c$ at select $\phi_{\infty}$ values for both copper (Cu) and aluminium (Al) nanoparticles. 


\begin{table}
\begin{center}
\begin{tabular}{ccccc}
\multirow{3}{*}{$\phi_{\infty}$} & 
\multicolumn{1}{c}{Copper (Cu)} & 
\multicolumn{1}{c}{Aluminium (Al)} \\ \\
& $R_c$ & $R_c$ \\ \hline
$0$ & $519.4$ & $519.4$  \\
$10^{-6}$ & $519.4$ ($519.4$) & $519.4$ ($519.4$) \\
$10^{-5}$ & $519.3$ ($519.3$) & $519.5$ ($519.5$) \\
$10^{-4}$ & $519.2$ ($519.2$) & $519.5$ ($519.5$) \\
$10^{-3}$ & $516.7$ ($516.7$) & $520.1$ ($519.9$) \\
$10^{-2}$ & $493.7$ ($493.6$) & $523.9$ ($523.8$) \\
$2\times 10^{-2}$ & $471.9$ ($471.6$) & $528.6$ ($528.3$) \\
$4\times 10^{-2}$ & $437.5$ ($436.7$) & $539.2$ ($538.5$)
\end{tabular}
\caption{Critical Reynolds numbers $R_c$ for copper (Cu) and aluminium (Al) nanoparticles in a base fluid of water, while the results in brackets correspond to the solutions obtained in the absence of Brownian motion and thermophoresis.}
\label{Table4}
\end{center}
\end{table}

\begin{figure*}
\centerline{\includegraphics[width=0.75\textwidth]{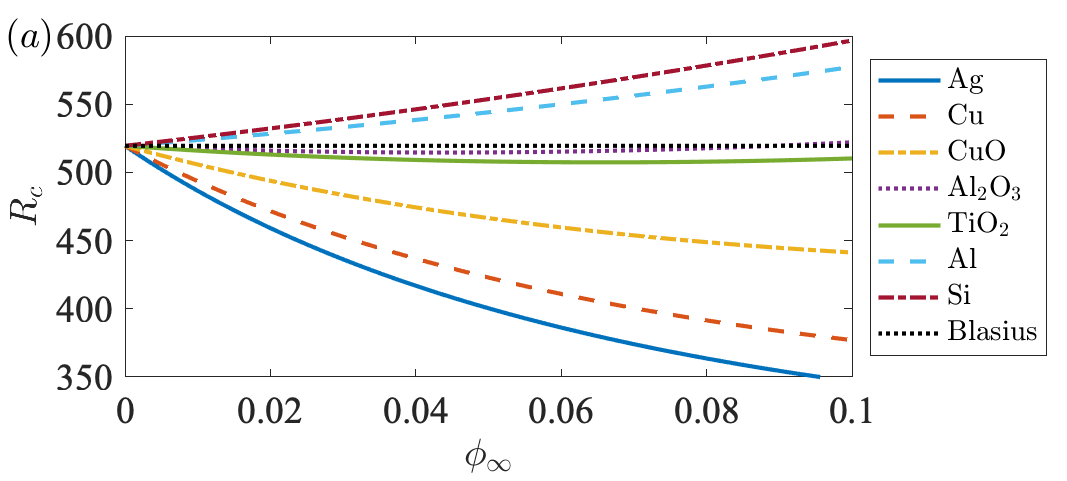}}
\centerline{\includegraphics[width=0.75\textwidth]{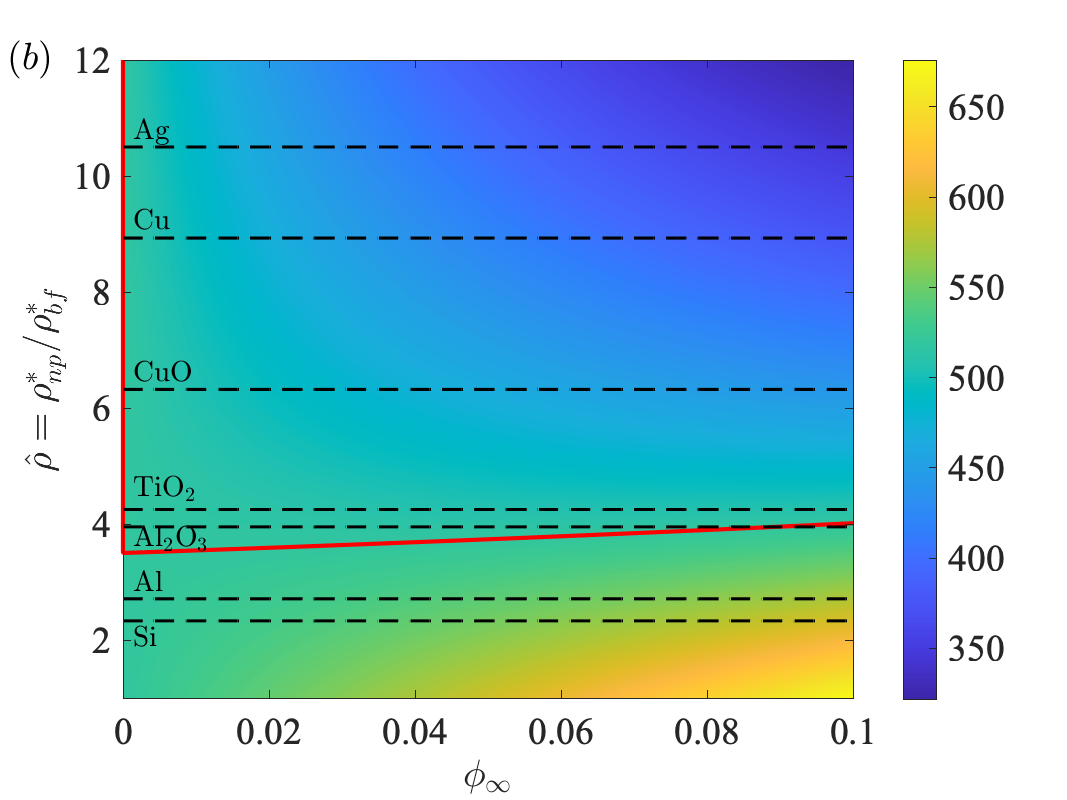}}
\caption{Plots of the critical Reynolds number $R_c$ for the seven nanoparticle materials tabulated in table~\ref{Table1} in a base fluid of water, with the dynamic viscosity $\mu$ based on the \citet{Brinkman1952} model \eqref{NondimBrinkman}. ($a$) $R_c$ as a function of $\phi_{\infty}$. ($b$) Contours of $R_c$ in the ($\phi_{\infty},\hat{\rho}$)-plane, where the solid red contour represents the contour level $R_c=519.4$, matched to the critical conditions for the Blasius flow without nanoparticles.}\label{Fig14}
\end{figure*}

Consequently, the critical Reynolds number $R_c$ is governed by the dynamic viscosity $\mu$ and the density $\rho$ of the nanofluid, which are in turn influenced by the free-stream nanofluid volume concentration $\phi_{\infty}$ and the ratio of densities $\hat{\rho}$. Figure~\ref{Fig14} illustrates $R_c$ as approximated by equation~\eqref{CriticalRePrediction}. In the first plot, figure~\ref{Fig14}(\emph{a}), $R_c$ is plotted as a function of $\phi_{\infty}$ and demonstrates the influence of both $\phi_{\infty}$ and the material used for the nanoparticles. Denser materials with larger $\hat{\rho}$ ratios, like silver (Ag) and copper (Cu), have a destabilising effect, while lighter materials, like silicon (Si) and aluminium (Al), stabilise the flow. On the other hand, alumina (Al$_2$O$_3$) exhibits a marginally destabilising effect at small $\phi_{\infty}$, with a stabilising benefit realised for large $\phi_{\infty}$ (for $\phi_{\infty}\gtrapprox 0.09$). 

Figure~\ref{Fig14}(\emph{b}) further demonstrates the impact of nanofluids on the onset of linear instability, with $R_c$ plotted in the $(\phi_{\infty},\hat{\rho})$-plane. The solid red contour corresponds to $R_c=519.4$ (i.e., the onset of linear instability in the standard Blasius flow), with solutions illustrating the negative impact of most nanoparticle materials, except silicon (Si) and aluminium (Al), on the hydrodynamic stability of the flow. More specifically, for a base fluid of water, only nanoparticles with a density ratio $\hat{\rho}\lessapprox 3.5$ are stabilising.


\section{Asymptotic analysis}

\begin{figure*}
\centering
\begin{tikzpicture}[scale=1]
\tikzstyle{every node}=[font=\normalsize]
\draw [thick] (1,0) -- (10,0);
\draw [] (0,1) -- (0,3.5) -- (2.5,3.5);
\draw [] (4,3.5) -- (10,3.5) -- (10,0);
\draw [>=latex,<->] (0,1) -- (0,0) -- (1,0);
\node [below] at (1,0) {$x^{*}$};
\node [left] at (0,1) {$y^{*}$};
%
\draw [blue] (2.5,0) 	-- (2.5,3.5);
\draw [blue] (4,1.5) 	-- (4,3.5);
\draw [blue] (2.5,0) arc (-90:0:1.5);
\draw [blue, >=latex,->] (2.5,3.5) 	-- (4,3.5);
\draw [blue, >=latex,->] (2.5,3) 		-- (4,3);
\draw [blue, >=latex,->] (2.5,2.5) 	-- (4,2.5);
\draw [blue, >=latex,->] (2.5,2) 		-- (4,2);
\draw [blue, >=latex,->] (2.5,1.5) 	-- (4,1.5);
\draw [blue, >=latex,->] (2.5,1) 		-- ({2.5+sqrt(2)},1);
\draw [blue, >=latex,->] (2.5,0.5) 	-- ({2.5+sqrt(1.25)},0.5);
\node [blue, above] at (3.25,0.5) {$U_{B}$};
\node [blue, above] at (3.25,3) {$U_{B}=1$};
\draw [thick, dashed] (5,0) 	-- (5,3.5);
\draw [thick, dashed] (8,0) 	-- (8,3.5);
\draw [thick, dashed] (5,0.5) 	-- (8,0.5);
\draw [thick, dashed] (5,1.5) 	-- (8,1.5);
\draw [>=latex,<->] (6,0) 	-- (6,0.5);
\node at (5.5,0.25)  {$O(\varepsilon^{5})$};
\draw [>=latex,<->] (6.5,0) 	-- (6.5,1.5);
\node at (6,1)  {$O(\varepsilon^{4})$};
\draw [>=latex,<->] (7,0) 	-- (7,3.5);
\node at (6.5,2.5)  {$O(\varepsilon^{3})$};
\node at (7.5,0.25) {$3$};
\node at (7.5,1) {$2$};
\node at (7.5,2.5) {$1$};
\draw [>=latex,<->] (5,-0.5) 	-- (8,-0.5);
\node at (6.5,-0.75)  {$O(\varepsilon^{3})$};
\clip (0,0) rectangle (10,4);
\end{tikzpicture}
\caption{Diagram of the triple deck structure of the lower-branch of the neutral stability curve for $\varepsilon={\Rey}^{-1/8}$. Regions $1$, $2$, and $3$ correspond to the upper, main, and lower decks, respectively.}\label{Fig15}
\end{figure*}
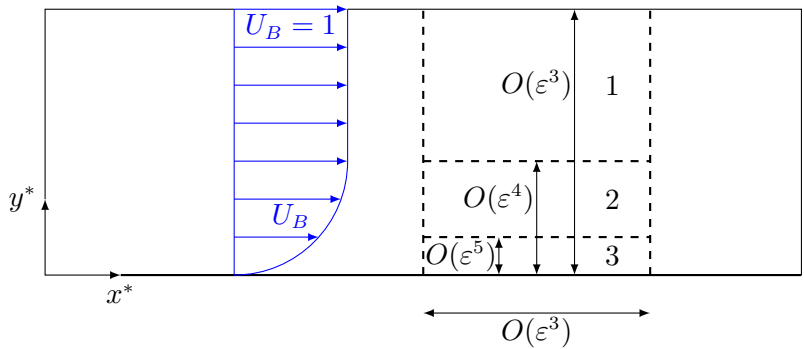

To describe the lower-branch structure of the neutral stability curve, we follow the approach of \citet{Smith1979} and assume a large Reynolds number $\Rey$. Consequently, linear disturbances on the lower branch are governed by a triple deck structure with a main deck of thickness $O({\Rey}^{-1/2})$, an upper deck of thickness $O({\Rey}^{-3/8})$, and a lower deck of thickness $O({\Rey}^{-5/8})$, with streamwise length $O({\Rey}^{-3/8})$ and frequency $O({\Rey}^{-1/4})$. A diagram of the triple deck structure is shown in figure~\ref{Fig15} for $\varepsilon={\Rey}^{-1/8}$. In addition, 
\begin{subequations}
\begin{equation}
x=1+\varepsilon^3 X 
\quad \textrm{and} \quad 
t=\varepsilon^2\hat{t},
\tag{\theequation \emph{a,b}}
\end{equation}
\end{subequations}
while linear disturbances are taken to be proportional to 
\begin{subequations}
\begin{equation}
E=\exp\left(\textrm{i}\left(\Theta(X) - \omega\hat{t}\right)\right),
\tag{\theequation \emph{a}}
\end{equation}
for
\begin{equation}
\frac{d\theta}{dX} = \alpha_1(x) +\varepsilon\alpha_2(x) + \cdots 
\quad \textrm{and} \quad 
\omega = \omega_1+\varepsilon\omega_2+\cdots.
\tag{\theequation \emph{b,c}}
\end{equation}
\end{subequations}
%


\subsection{The main deck}

Here $y=\varepsilon^4 y_2$, for $y_2=O(1)$, where perturbations $\tilde{\boldsymbol{q}}=(\tilde{u},\tilde{v},\tilde{p},\tilde{T},\tilde{\phi})$ are expanded as
\begin{subequations}\label{main_deck_expansions}
\begin{equation}
\begin{alignedat}{2}
\tilde{u} &{}={} \left(u_2 + O(\varepsilon)\right)E, \quad 
&\tilde{v} &{}={} \left(\varepsilon v_2+O(\varepsilon^2)\right)E, \\
\tilde{p} &{}={} \left(\varepsilon p_2+O(\varepsilon^2)\right)E, \quad
&\tilde{T} &{}={} \left(T_2+O(\varepsilon)\right)E, \\ 
\tilde{\phi} &{}={} \left(\phi_2+O(\varepsilon)\right)E, 
\end{alignedat}
\tag{\theequation \emph{a-e}} 
\end{equation}
\end{subequations}
where $u_2=u_2(x,y_2)$, etc. Similar expansions are given for the perturbed quantities $\tilde{\mu}$, $\tilde{\rho}$, $\tilde{c}$, and $\tilde{k}$. In addition, the nanoparticle volume concentration $\phi_B\sim\phi_{\infty}$. 

Substituting \eqref{main_deck_expansions} into the linear stability equations \eqref{LinearNavierStokes1} and collecting the leading-order terms, gives the solution
%
%
%
%
%
%
%
%
%
%
%
\begin{subequations}\label{main_sol}
\begin{equation}
u_2 = A(x)U_{B,y_2}, \quad
v_2 = -\textrm{i}\alpha_1 A(x) U_B, 
\quad \textrm{and} \quad
p_2 = p_2(x), 
\tag{\theequation \emph{a-c}}
\end{equation}
where $p_2(x)$ and $A(x)$ are unknown, slowly varying, amplitude functions, representing pressure and negative displacement perturbations, respectively. Similarly, 
\begin{equation}
T_2  = A(x)T_{B,y_2} 
\quad \textrm{and} \quad
\phi_2 = 0. 
\tag{\theequation \emph{d,e}}
\end{equation}
\end{subequations}
%


\subsection{The upper deck}

Here $y=\varepsilon^3 y_1$, for $y_1=O(1)$. To match with the main deck, perturbations are expanded as
\begin{subequations}\label{upper_deck_expansions}
\begin{equation}
\begin{alignedat}{2}
\tilde{u} &{}={} \left(\varepsilon u_1+O(\varepsilon^2)\right)E, \quad 
&\tilde{v} &{}={} \left(\varepsilon v_1+O(\varepsilon^2)\right)E, \\
\tilde{p} &{}={} \left(\varepsilon p_1+O(\varepsilon^2)\right)E, \quad
&\tilde{T} &{}={} \left(\varepsilon T_1+O(\varepsilon^2)\right)E, \\ 
\tilde{\phi} &{}={} \left(\varepsilon \phi_1+O(\varepsilon^2)\right)E, 
\end{alignedat}
\tag{\theequation \emph{a-e}} 
\end{equation}
\end{subequations}
where $u_1=u_1(x,y_1)$ etc. Similar expansions are again given for the perturbed quantities $\tilde{\mu}$, $\tilde{\rho}$, $\tilde{c}$, and $\tilde{k}$. In addition, the base flow is effectively given by the uniform free-stream conditions
\begin{subequations}\label{upper_deck_baseflow}
\begin{equation}
\begin{alignedat}{3}
U_B &{}\approx{} 1, \quad 
&V_B &{}\approx{} 0, \quad 
&T_B &{}\approx{} 1, \\ 
\phi_B &{}\approx{} \phi_{\infty}, \quad 
&c_B &{}\approx{} c_{\infty}, \quad
&\rho_B &{}\approx{} \rho_{\infty}.
\end{alignedat}
\tag{\theequation \emph{a-f}} 
\end{equation}
\end{subequations}

Substituting \eqref{upper_deck_expansions} and \eqref{upper_deck_baseflow} into the linear stability equations \eqref{LinearNavierStokes1}, gives 
%
%
%
%
%
%
%
%
%
%
%
\begin{equation}\label{upper_p_eq}
\left(\frac{\partial^2}{\partial y_1}-\alpha_1^2\right)p_1=0,
\end{equation}
with the bounded solution as $y_1\rightarrow\infty$ given by
\begin{subequations}\label{upper_sol}
\begin{equation}\label{upper_p_sol}
p_1 = P_1(x) \textrm{e}^{-\alpha_1 y_1},
\tag{\theequation \emph{a}}
\end{equation}
where $P_1(x)$ is an unknown function of $x$ and $\alpha_1>0$. Moreover,
\begin{equation}
u_1 = -\frac{P_1(x)\textrm{e}^{-\alpha_1 y_1}}{\rho_{\infty}}, 
\quad
v_1 = -\frac{\textrm{i}P_1(x) \textrm{e}^{-\alpha_1 y_1}}{\rho_{\infty}}, 
\quad
T_1 = 0,
\quad \textrm{and} \quad
\phi_1 = 0.
\tag{\theequation \emph{b-e}}
\end{equation}
\end{subequations}

Continuity of pressure requires
\begin{equation}
P_1(x)=p_2(x) 
\quad \textrm{as} \quad y_1\rightarrow 0.
\end{equation}
Similarly, continuity of the wall-normal velocity $\tilde{v}$ between the main deck solution (\ref{main_sol}\emph{b}) and the upper deck solution (\ref{upper_sol}\emph{c}) yields the condition
\begin{equation}\label{A_p}
\alpha_1 A(x) = \frac{p_2(x)}{\rho_{\infty}}.
\end{equation}


\subsection{The lower deck}

Recall that the concentration layer has a characteristic thickness of $O({\Rey}^{-1/2}{Sc}^{-1/3})$. By setting ${Sc}^{-1/3}\sim {\Rey}^{-1/8}$, the lower deck coincides with the concentration layer. 

To match with the main deck, in the lower deck $y=\varepsilon^5 y_3$, for $y_3=O(1)$.
Perturbations in the lower deck are then expanded as
\begin{subequations}\label{lower_exps}
\begin{equation}
\begin{alignedat}{2}
\tilde{u} &{}={} \left(u_3+O(\varepsilon)\right)E, \quad 
&\tilde{v} &{}={} \left(\varepsilon^2 v_3+O(\varepsilon^3)\right)E, \\
\tilde{p} &{}={} \left(\varepsilon p_3+O(\varepsilon^2)\right)E, \quad
&\tilde{T} &{}={} \left(T_3+O(\varepsilon)\right)E, \\ 
\tilde{\phi} &{}={} \left(\phi_3+O(\varepsilon)\right)E, 
\end{alignedat}
\tag{\theequation \emph{a-e}} 
\end{equation}
\end{subequations}
where $u_3=u_3(x,y_3)$ etc. As before, similar expansions are introduced for the perturbed quantities $\tilde{\mu}$, $\tilde{\rho}$, $\tilde{c}$, and $\tilde{k}$. 

In the main deck, the base velocity behaves as $U_B\sim\lambda y_2$ as $y_2\rightarrow 0$, where $\lambda = U_{B,y_2}|_{y_2=0} (\equiv \rho_wf''(0)/x^{1/2})$, and consequently from~(\ref{main_sol}\emph{a}) and~(\ref{main_sol}\emph{b})
\begin{subequations}\label{Maindeckuatlowerend}
\begin{equation}
u_2 \rightarrow \lambda A(x) 
\quad \textrm{and} \quad
v_2 \rightarrow -\textrm{i}\alpha_1 \lambda A(x)y_2 
\quad \textrm{as} \quad y_2 \rightarrow 0.
\tag{\theequation \emph{a,b}}
\end{equation}
\end{subequations}
Therefore, within the lower deck, the base flow is given by
\begin{subequations}\label{lower_basic}
\begin{equation}
\begin{alignedat}{2}
U_B &{}={} \varepsilon\lambda y_3 + O(\varepsilon^2), \quad
&V_B &{}={} -\tfrac{1}{2}\varepsilon^2\lambda_x y_3^2 + O(\varepsilon^3), \\
T_B &{}={} T_{w} + \varepsilon\sigma y_3 + O(\varepsilon^2), \quad
&\phi_B &{}={} \phi_{\infty} + \varepsilon\psi(x,y_3) + O(\varepsilon^2),
\end{alignedat}
\tag{\theequation \emph{a-d}} 
\end{equation}
\end{subequations}
where $\sigma = T_{B,y_2}|_{y_2=0} (\equiv \rho_w\theta'(0)/x^{1/2})$.

Substituting \eqref{lower_exps} and \eqref{lower_basic} into the linear stability equations \eqref{LinearNavierStokes1} gives
%
%
%
%
%
%
%
%
%
%
\begin{equation}
p_3=p_2(x),
\end{equation}
to match with the pressure in the main deck, and
\begin{subequations}
\begin{equation}\label{LowerdeckuatUpperend}
u_3=B(x)\int_{\chi_0}^\chi \Ai(\grave{\chi}) \;\textrm{d}\grave{\chi},
\end{equation}
\begin{equation}\label{lower_B_p_1}
p_2 = -\frac{\omega_1\rho_{\infty}}{\alpha_1}\frac{B(x)\Ai'(\chi_0)}{\chi_0},
\end{equation}
\end{subequations}
where $B$ is an unknown, amplitude function, $\Ai$ is the Airy function, and
\[
\chi = \left(\frac{\textrm{i}\alpha_1\lambda\rho_{\infty}}{\mu_{\infty}}\right)^{1/3}\left(y_3 -
\frac{\omega_1}{\alpha_1\lambda}\right), 
\]
for $\chi_0=\chi|_{y_3=0}$.

Matching the streamwise velocity $\tilde{u}$ between the main deck solution (\ref{Maindeckuatlowerend}\emph{a}) and the lower deck solution \eqref{LowerdeckuatUpperend}, gives
\begin{equation}\label{B_A}
B(x)\int_{\chi_0}^\infty \Ai(\chi) \; \textrm{d} \chi = \lambda A(x).
\end{equation}
Eliminating $A$, $B$, and $p_2$ from equations \eqref{A_p}, \eqref{lower_B_p_1}, and \eqref{B_A} yields the leading-order eigenrelation
\begin{equation}\label{eigenrelation}
%
\frac{\Ai'(\chi_0)}{\int_{\chi_0}^\infty \Ai(\chi) \; \textrm{d}\chi} = \left(\frac{\textrm{i}\alpha_1\lambda\rho_{\infty}}{\mu_{\infty}}\right)^{1/3}\frac{\alpha_1}{\lambda^{2}},
\end{equation}
which, following the parameter scaling
\begin{subequations}\label{alpha_omega_scales}
\begin{equation}
\alpha_1 = \lambda^{5/4}\left(\frac{\mu_{\infty}}{\rho_{\infty}}\right)^{1/4}\overline{\alpha}
\quad \textrm{and} \quad
%
\omega_1 = \lambda^{3/2}\left(\frac{\mu_{\infty}}{\rho_{\infty}}\right)^{1/2}\overline{\omega},
\tag{\theequation \emph{a,b}}
\end{equation}
\end{subequations}
becomes
\begin{subequations}
\begin{equation}\label{eigenrelation_scaled}
\frac{\Ai'(\chi_0)}
{\int_{\chi_0}^\infty \Ai(\chi) \; \textrm{d}\chi}
=\textrm{i}^{1/3}\overline{\alpha}^{4/3}
\quad \textrm{for} \quad
\chi_0=-\textrm{i}^{1/3}\frac{\overline{\omega}}
{\overline{\alpha}^{2/3}}.
\tag{\theequation \emph{a,b}}
\end{equation}
\end{subequations}

For neutral stability, $\alpha_1, \alpha_2$ etc. must be real, requiring $\chi_0 = -2.298 \textrm{i}^{1/3}$ and 
\begin{equation}\label{eigenrelation_result}
\frac{\Ai'(\chi_0)}
{\int_{\chi_0}^\infty \Ai(\chi) \; \textrm{d}\chi}
=1.001\textrm{i}^{1/3}.
\end{equation}
%
Consequently, the neutral values of $\alpha_1$ and $\omega_1$ are given as
\begin{subequations}\label{alpha_omega_x}
\begin{align}
\alpha_1 & = 1.001\hat{\lambda}^{5/4}\left(\frac{\mu_{\infty}}{\rho_{\infty}}\right)^{1/4}x^{-5/8},
\label{alpha_approx} \\
\omega_1 & = 2.299\hat{\lambda}^{3/2}\left(\frac{\mu_{\infty}}{\rho_{\infty}}\right)^{1/2}x^{-3/4},
\label{omega_approx}
\end{align}
\end{subequations}
where $\hat{\lambda}=\rho_wf''(0)$. This gives the leading-order approximation for the frequency of the lower branch in terms of the Reynolds number $R$:
\begin{equation}\label{comp_final}
\omega_N \sim 2.299[\delta_1\hat{\lambda}]^{3/2}\left(\frac{\mu_{\infty}}{\rho_{\infty}}\right)^{1/2}R^{-1/2}.
\end{equation}
Notably, in the limit $Sc\rightarrow\infty$, $\delta_1\hat{\lambda}\approx 0.572$ across all nanoparticle materials and $\phi_{\infty}$. Thus, $2.299[\delta_1\hat{\lambda}]^{3/2}\approx 0.994$.

\begin{figure*}
\centerline{\includegraphics[width=0.75\textwidth]{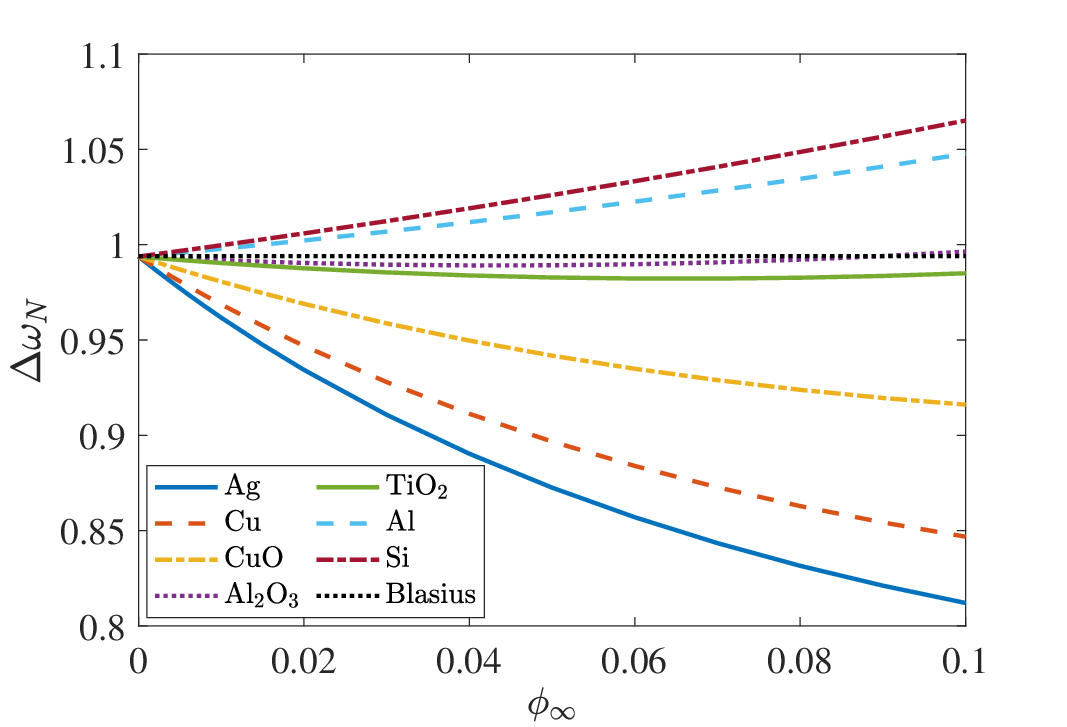}}
\caption{Gradient $\Delta\omega_N=0.994[\mu_{\infty}/\rho_{\infty}]^{1/2}$ of the lower branch~\eqref{comp_final} as a function of $\phi_{\infty}$ for different nanoparticle materials.}
\label{Fig16}
\end{figure*}

Figure~\ref{Fig16} depicts the gradient of the frequency $\omega_N$, defined as $\Delta\omega_N=0.994[\mu_{\infty}/\rho_{\infty}]^{1/2}$, as a function of $\phi_{\infty}$ for all seven nanoparticle materials listed in table~\ref{Table1}. A gradient $\Delta\omega_N<0.994$ indicates a destabilising effect, while $\Delta\omega_N>0.994$ corresponds to stabilising behaviour. The solutions are qualitatively similar to and consistent with the linear stability results shown in figure~(\ref{Fig14}\emph{a}): less dense materials are stabilising and denser materials are destabilising.

\section{Conclusions}

A linear stability study has been conducted on the nanofluid boundary-layer flow over a flat plate, extending the earlier work of~\citet{Buongiorno2006},~\citet{Avramenko2011},~\citet{MacDevette2014}, and~\citet{Turkyilmazoglu2020}. The model employs a two-phase flow formulation that incorporates the effects of Brownian motion and thermophoresis, with all quantities scaled on the base fluid characteristics, providing a physically consistent approach for investigating stability trends. Although the influence of Brownian motion and thermophoresis is relatively weak, a thin concentration layer with a characteristic thickness of $O({\Rey}^{-1/2}{Sc}^{-1/3})$ develops within the boundary layer, which modifies the near-wall velocity and temperature fields. The concentration layer disappears when Brownian motion and thermophoresis are ignored, with the nanoparticle volume concentration $\phi$ uniform throughout the boundary layer.

In terms of thermodynamic performance, all seven materials modelled herein establish an increasing Nusselt number $Nu$, with greater benefits obtained for denser materials like silver (Ag) and copper (Cu).

Despite the emergence of a thin concentration layer, numerical and asymptotic stability calculations show that Brownian motion and thermophoresis have a negligible impact on the onset of the TS wave. In fact, linear stability characteristics and the onset of TS waves can be accurately predicted using solutions to the Blasius flow without nanoparticles. The Reynolds number for the nanofluid is given as
\begin{equation}
\Rey = \frac{\mu\widehat{\Rey}}{\rho},   
\end{equation}
for the Blasius flow Reynolds number $\widehat{\Rey}$. Consequently, the stability of the nanofluid boundary-layer flow is governed by the density $\rho$ and viscosity $\mu$ of the nanofluid. In particular, the density ratio $\hat{\rho}=\rho^*_{np}/\rho^*_{bf}$ is critical to determining whether the nanofluid is stabilising or destabilising. Denser nanoparticle materials, such as silver (Ag) and copper (Cu), significantly destabilise the TS wave. In contrast, a small stabilising effect is achieved by lighter materials, like aluminium (Al) and silicon (Si). This observation differs from the one-phase flow study conducted by~\citet{Turkyilmazoglu2020}, which predicted the opposite outcome. However, in Turkyilmazoglu's investigation, physical quantities were scaled on the characteristics of the nanofluid rather than the base fluid, leading to a Reynolds number that varied with the type of nanoparticle material and volume concentration. 

The results presented above are based on a nanofluid with water as the base fluid. Replacing water with a less dense fluid, like ethanol, would increase the density ratio $\hat{\rho}$ for all materials. While this change would enhance the thermal benefits of the nanofluid, it would lead to a further destabilisation of the TS wave, even for those nanofluids composed of lighter materials like aluminium (Al) and silicon (Si).

Another key factor influencing the stability of nanofluids is the choice of viscosity model. In the above study, the~\citet{Brinkman1952} model~\eqref{Brinkman} was used to represent the dynamic viscosity of the nanofluid. However, alternative models can lead to significantly different results. For instance, the correlations due to~\citet{PakCho1998} and~\citet{Maiga2004} (see equations~\eqref{PakCho}) predict larger increases in viscosity as the nanoparticle volume concentration $\phi$ increases. Assuming these models can be applied to the boundary-layer flow on a flat plate, stability calculations indicate a strong stabilising effect for all nanoparticle materials, in contrast to the destabilising trends observed for the Brinkman model. Therefore, selecting an experimentally validated viscosity model is essential for accurately determining the stability of nanofluid flows.

Future investigations into nanofluid boundary-layer flows could include non-parallel effects and nonlinear stability effects by using parabolised stability equations, following the approach of~\citet{Bertolotti1992}. Additionally, the analysis may be applied to other geometries, including rotating disk boundary layers and wall jets, as considered by~\citet{Turkyilmazoglu2020}. However, based on the above observations, we anticipate the stability of such flows will still be well-approximated by the base flow without nanoparticles, unless Brownian motion and thermophoresis play a more dominant role.


\backsection[Acknowledgements]{JSBG is grateful to the Sydney Mathematics Research Institute (SMRI) for financial support and for hosting  a trip to the SMRI as a visiting researcher during July-August 2023 which helped facilitate this research project.}

\appendix

\section{On ignoring Brownian motion and thermophoresis}\label{AppendixA}

When the effects of Brownian motion and thermophoresis are ignored, the continuity equation for the nanoparticles, given by equation~\eqref{Volume1}, reduces to the form
\begin{equation}\label{VolumeB1}
\frac{\partial\phi}{\partial t^*}+\phi\nabla^*\cdot\boldsymbol{u}^*+\nabla^*\phi\cdot\boldsymbol{u}^* = 0.
\end{equation}
In addition, the continuity equation~\eqref{Continuity1} can be re-written in the form
\begin{equation}\label{ContinuityB1a}
\frac{\partial\rho^*}{\partial t^*}+\rho^*\nabla^*\cdot\boldsymbol{u}^*+\nabla^*\rho^*\cdot\boldsymbol{u}^* = 0,
\end{equation}
which on using the definition for density~\eqref{DimDensity} becomes
\begin{equation}\label{ContinuityB1b}
\left(\rho_{np}^*-\rho_{bf}^*\right)\left(\frac{\partial\phi}{\partial t^*}+\nabla^*\phi\cdot\boldsymbol{u}^* \right)+\rho^*\nabla^*\cdot\boldsymbol{u}^* = 0.
\end{equation}
Subsequently, combining \eqref{VolumeB1} and \eqref{ContinuityB1b} gives 
\begin{equation}
\left(\left(\rho_{bf}^*-\rho_{np}^*\right)\phi+\rho^*\right)\nabla^*\cdot\boldsymbol{u}^* = 0,  
\end{equation}
which implies the flow is incompressible
\begin{equation}
\nabla^*\cdot\boldsymbol{u}^* = 0  
\end{equation}
and the continuity equation for the nanoparticles \eqref{VolumeB1} reduces to 
\begin{equation}\label{VolumeB2}
\frac{\partial\phi}{\partial t^*}+\nabla^*\phi\cdot\boldsymbol{u}^* = 0.
\end{equation}
Consequently, the Prandtl scaling in \S\ref{BLEquations} gives
\begin{equation}
\phi'=0 \quad \textrm{with} \quad \phi\rightarrow\phi_{\infty} \quad \textrm{as} \quad y\rightarrow\infty.
\end{equation}
Thus, $\phi=\phi_{\infty}$ for all $y$, i.e., $\phi$ is a constant. Hence, base flow quantities, including the viscosity $\mu$, density $\rho$, specific heat capacity $c$, and thermal conductivity $k$ are constant. 

On coupling the scalings~\eqref{Scaling} with the following substitutions
\[p=\rho\hat{p}, \quad T=1+(T_w-1)\widehat{T}, \quad \widehat{\Rey} = \frac{\rho}{\mu}\Rey, \quad \widehat{\Pran}=\frac{\mu c}{k}\Pran,\]
transforms the non-dimensional governing equations~\eqref{Gov2} into the form
\begin{subequations}\label{GovB1}
\begin{equation}
\nabla\cdot\boldsymbol{u} = 0,    
\end{equation}
\begin{equation}
\frac{\partial\boldsymbol{u}}{\partial t} + (\boldsymbol{u}\cdot\nabla)\boldsymbol{u} = -\nabla \hat{p} + \frac{1}{\widehat{\Rey}}\nabla^2\boldsymbol{u},
\end{equation}
\begin{equation}
\frac{\partial\widehat{T}}{\partial t} + (\boldsymbol{u}\cdot\nabla)\widehat{T} = \frac{1}{\widehat{\Rey}\widehat{\Pran}}\nabla^2 \widehat{T},
\end{equation}
\end{subequations}
for boundary conditions
\begin{subequations}\label{BoundaryConditionsB1}
\begin{equation}
\boldsymbol{u}=0 \quad \textrm{and} \quad \widehat{T} = 1 \quad \textrm{on} \quad y=0, \tag{\theequation \emph{a,b}}
\end{equation}
\end{subequations}
and
\begin{subequations}\label{BoundaryConditions5}
\begin{equation}
\begin{alignedat}{3}
u&{}\rightarrow{} 1, \qquad
&v{}\rightarrow{}& 0, \qquad 
&w{}\rightarrow{}& 0, \\
\hat{p}&{}\rightarrow{} 0, \qquad 
&\widehat{T}\rightarrow{}& 0, \qquad 
&\phi{}\rightarrow{}&\phi_{\infty} 
\qquad \textrm{as} \quad y\rightarrow\infty.
\end{alignedat}
\tag{\theequation \emph{a-f}}
\end{equation}
\end{subequations}
Subsequently, applying the Prandtl transformation for $\widehat{\Rey}\rightarrow\infty$ establishes the Blasius boundary-layer equations \eqref{Gov6}, with an equivalent set of linear stability equations for the Reynolds number $\widehat{\Rey}$. Thus, when Brownian motion and thermophoresis are neglected, the linear stability of the nanofluid flow reduces to the Blasius flow, with the nanofluid Reynolds number given as $\Rey=\mu\widehat{\Rey}/\rho$.


\section{Base flow and perturbation quantities}\label{AppendixB}

\subsection{Terms in equations \eqref{LinearNavierStokes1}}

The functions $g_{\star}$ in the linear stability equations \eqref{LinearNavierStokes1} are given as 
\begin{subequations}
\begin{equation}
g_1(V_B,\boldsymbol{Q}_{B,x}) = -\left(\rho_{B,x}\tilde{u}+U_{B,x}\tilde{\rho} +{\Rey}^{-1/2}\frac{\partial}{\partial y}\left(V_B\tilde{\rho}\right)\right), 
\end{equation}
\begin{multline}
g_2(V_B,\boldsymbol{Q}_{B,x}) = 
\frac{1}{\Rey}\bigg(\mu_{B,x}\left(\frac{4}{3}\frac{\partial\tilde{u}}{\partial x} - 
\frac{2}{3}\left(\frac{\partial\tilde{v}}{\partial y} +
\frac{\partial\tilde{w}}{\partial z}\right)\right) 
\\
+\frac{\partial}{\partial x}\left(\left(\frac{4}{3}U_{B,x}-\frac{2}{3}{\Rey}^{-1/2}V_{B,y}\right)\tilde{\mu}\right)
+{\Rey}^{-1/2}\frac{\partial}{\partial y}\left(V_{B,x}\tilde{\mu}\right)\bigg) 
\\
-(U_{B,x}\left(\rho_B\tilde{u}+U_B\tilde{\rho}\right)
-{\Rey}^{-1/2}V_B\left(\rho_B\frac{\partial\tilde{u}}{\partial y}+U_{B,y}\tilde{\rho}\right),
\end{multline}
\begin{multline}
g_3(V_B,\boldsymbol{Q}_{B,x}) = 
\frac{1}{\Rey}\bigg(\mu_{B,x}\left(\frac{\partial\tilde{v}}{\partial x} + \frac{\partial\tilde{u}}{\partial y}\right) 
+\frac{\partial}{\partial y}\left(\left(\frac{4}{3}{\Rey}^{-1/2}V_{B,y}-\frac{2}{3}U_{B,x}\right)\tilde{\mu}\right)
\\
+{\Rey}^{-1/2}\frac{\partial}{\partial x}\left(V_{B,x}\tilde{\mu}\right)\bigg) 
-{\Rey}^{-1/2}\bigg(V_{B,x}\left(\rho_B\tilde{u}+U_B\tilde{\rho}\right)
\\
+V_B\left(\rho_B\frac{\partial\tilde{v}}{\partial y}
-{\Rey}^{-1/2}V_{B,y}\tilde{\rho}\right) +\rho_BV_{B,y}\tilde{v}\bigg),
\end{multline}
\begin{multline}
g_4(V_B,\boldsymbol{Q}_{B,x}) = 
\frac{1}{\Rey}\left(\mu_{B,x}\left(\frac{\partial\tilde{w}}{\partial x} + \frac{\partial\tilde{u}}{\partial z}\right)
-\frac{2}{3}\frac{\partial}{\partial z}\left(\left(U_{B,x}+{\Rey}^{-1/2}V_{B,y}\right)\tilde{\mu}\right)\right)
\\
-{\Rey}^{-1/2}\rho_BV_B\frac{\partial\tilde{w}}{\partial y}, 
\end{multline}
\begin{multline}
g_5(V_B,\boldsymbol{Q}_{B,x}) = 
\frac{1}{\Rey \Pran}\left(k_{B,x}\frac{\partial\tilde{T}}{\partial x} + T_{B,x}\frac{\partial\tilde{k}}{\partial x}+T_{B,xx}\tilde{k}\right)
+\frac{1}{\Rey\Pran  Le}\Bigg(T_{B,x}\phi_{B,x}\tilde{T} 
\\ + T_B\left(T_{B,x}\frac{\partial\tilde{\phi}}{\partial x}+\phi_{B,x}\frac{\partial\tilde{T}}{\partial x}\right)  
+\frac{1}{N_{BT}T_B}\left(2\phi_BT_{B,x}\frac{\partial\tilde{T}}{\partial x} +T_{B,x}^2\left(\tilde{\phi}-\frac{\phi_B\tilde{T}}{T_B}\right)\right)\Bigg)
\\
- \left(\rho_BT_Bc_{B,x}\tilde{u}+U_Bc_{B,x}\left(T_B\tilde{\rho}+\rho_B\tilde{T}\right)+(\rho c)_BT_{B,x}\tilde{u}+U_BT_{B,x}\tilde{\rho}\tilde{c}
\right)
\\
- {\Rey}^{-1/2}V_B\left(\rho_BT_B\frac{\partial\tilde{c}}{\partial y}+c_{B,y}\left(T_B\tilde{\rho}+\rho_B\tilde{T}\right)+(\rho c)_B\frac{\partial\tilde{T}}{\partial y}+T_{B,y}\tilde{\rho}\tilde{c}
\right),
\end{multline}
\begin{multline}
g_6(V_B,\boldsymbol{Q}_{B,x}) = \frac{1}{\Rey Sc}\left(T_{B,x}\frac{\partial\tilde{\phi}}{\partial x}+\phi_{B,xx}\tilde{T}+\phi_{B,x}\frac{\partial\tilde{T}}{\partial x}
\right)
\\
+ \frac{1}{\Rey Sc N_{BT}}\Bigg(\left(\frac{\phi_{B,x}}{T_B}-\frac{\phi_BT_{B,x}}{T_B^2}\right)\frac{\partial\tilde{T}}{\partial x}
-T_{B,xx}\left(\frac{\tilde{\phi}}{T_B}-\frac{\phi_B\tilde{T}}{T_B^2}\right)
\\
+T_{B,x}\left(\frac{1}{T_B}\frac{\partial\tilde{\phi}}{\partial x} - \frac{T_{B,x}}{T_B^2}\tilde{\phi} - \frac{\phi_B}{T_B^2}\frac{\partial\tilde{T}}{\partial x} + \left(\frac{2\phi_BT_{B,x}}{T_B^3} - \frac{\phi_{B,x}}{T_B^2}\right)\tilde{T}
\right)
\Bigg)
\\
-\phi_{B,x}\tilde{u} - U_{B,x}\tilde{\phi} - {\Rey}^{-1/2}\left(V_B\frac{\partial\tilde{\phi}}{\partial y}+V_{B,y}\tilde{\phi}\right)
.
\end{multline}
\end{subequations}

\subsection{Terms in equations \eqref{LinearNavierStokes2}}

The base flow quantities in the system of equations \eqref{LinearNavierStokes2} are given as
\begin{subequations} 
\begin{equation}
\begin{alignedat}{2}
\rho_B&{} = 1+(\hat{\rho}-1)\phi_B, \quad
&\rho_{B,y}{} =& (\hat{\rho}-1)\phi_{B,y}, \\
(\rho c)_B&{} = 1+(\hat{\rho}\hat{c}-1)\phi_B, \quad
& {} & \\
c_B&{} = \frac{(\rho c)_B}{\rho_B}, \quad
&c_{B,y}{} =& \frac{\hat{\rho}(\hat{c}-1)\phi_{B,y}}{\rho_B^2}, \\
\mu_B&{} = \frac{1}{(1-\phi_B)^{2.5}}, \quad
&\mu_{B,y}{} =& \frac{2.5\mu_B\phi_{B,y}}{1-\phi_B}, \\
k_B&{} = \frac{\hat{k}+2+2(\hat{k}-1)\phi_B}{\hat{k}+2-(\hat{k}-1)\phi_B}, \quad
&k_{B,y}{} =& \mathcal{K}\phi_{B,y}, \\
\end{alignedat}
\tag{\theequation \emph{a-i}}
\end{equation}
\end{subequations}
and the perturbation quantities are given as 
\begin{subequations}
\begin{equation}
\begin{alignedat}{2}
\breve{\rho} &= (\hat{\rho}-1)\breve{\phi}, \quad
& {} & \\
(\breve{\rho}\breve{c})&{} = (\hat{\rho}\hat{c}-1)\breve{\phi}, \quad
&\breve{c}{} =& \frac{\hat{\rho}(\hat{c}-1)\breve{\phi}}{\rho_B^2}, \\
\breve{\mu}&{} = \frac{2.5\mu_B\breve{\phi}}{1-\phi_B}, \quad
&\textrm{D}\breve{\mu}{} =& \frac{2.5\mu_B}{1-\phi_B}\left(\textrm{D}+\frac{3.5\phi_{B,y}}{1-\phi_B}\right)\breve{\phi}, \\
\breve{k}{}& = \mathcal{K}\breve{\phi}, \quad
&\textrm{D}\breve{k}{} =& \mathcal{K}\left(\textrm{D} + \frac{2(\hat{k}-1)\phi_{B,y}}{\hat{k}+2-(\hat{k}-1)\phi_B}\right)\breve{\phi},
\end{alignedat}
\tag{\theequation \emph{a-g}}
\end{equation}
\end{subequations}
where
\[\mathcal{K}=\frac{3(\hat{k}-1)(\hat{k}+2)}{(\hat{k}+2-(\hat{k}-1)\phi_B)^2}.\]

\bibliographystyle{jfm}
\bibliography{JFM_Nanofluids}

\end{document}